\documentclass[usenatbib]{mn2e}
\usepackage{amsmath,graphicx}
\newcommand{\me}{\mathrm{e}}
\newcommand{\dif}{\mathrm{d}}
\begin{document}
\title[Formation of Star Clusters]{The Formation of Star Clusters I: 3D Simulations of Hydrodynamic Turbulence}
\author[Tilley, D. A. and Pudritz, R. E.]{David A. Tilley$^1$\thanks{E-mail: tilley@physics.mcmaster.ca (DAT); pudritz@physics.mcmaster.ca (REP)} and Ralph E. Pudritz$^1$\footnotemark[1] \\ $^1$ Department of Physics and Astronomy, McMaster University, Hamilton, Ontario, Canada L8S 4M1}
\maketitle
\begin{abstract}
  We present the results of a series of numerical simulations of compressible, self-gravitating hydrodynamic turbulence of cluster-forming clumps in molecular clouds.  We examine the role that turbulence has in the formation of gravitationally bound cores, studying the dynamical state, internal structure and bulk properties of these cores.  Complex structure in turbulent clumps is formed provided that the damping time of the turbulence, $t_\mathrm{damp}$ is longer than the gravitational free-fall time $t_\mathrm{ff}$ in a region.  We find a variety of density and infall velocity structures among the cores in the simulation, including cores that resemble the Larson-Penston collapse of an isothermal sphere ($\rho \propto r^{-2}$) as well as cores that resemble the McLaughlin-Pudritz collapse of logatropic spheres ($\rho \propto r^{-1}$).  The specific angular momentum profiles range between $j \propto r^{1} - r^{2}$.  The masses of the bound cores that form are well-fit by the turbulent mass spectrum of \citet{padoan02}, while the specific angular momentum distribution can be fit by a broken power law.  While our hydrodynamic simulations reproduce many of the observed properties of cores, we find an upper limit for the star formation efficiency (SFE) in clusters of 40-50 per cent.

\end{abstract}
\begin{keywords}
  Hydrodynamics -- turbulence -- stars: formation -- ISM: clouds -- ISM: evolution -- ISM: kinematics and dynamics
\end{keywords}

\section{INTRODUCTION}

Two of the most fundamental physical properties of stars that must be explained by a complete theory star formation are the distribution of their initial masses (the initial mass function or IMF) and their initial angular momenta (the initial angular momentum function - or IAMF).  The past decade has witnessed significant progress in measurement of the IMFs of stars in young star clusters. Infrared camera surveys of young stars within forming star clusters in molecular clouds reveal that the stellar mass distribution of stars is universal and follows the IMF of field stars (see, for example, the reviews of \citealt*{myers00}; \citealt{vazquez00}; \citealt{kroupa02}; \citealt{pudritz02}; \citealt*{lada03}).

Molecular cloud cores in which individual (or binary) stars are born have been intensively studied for more than a decade, and complete surveys of their physical properties have become available (e.g. \citealt*{jijina99}; \citealt{johnstone01}).  Millimetre and submillimetre surveys of the dense gas cores within cluster forming clumps have a core mass function or CMF (in which individual or binary stars will form) that is essentially the same as IMF (e.g. \citealt*{motte98}; \citealt{testi98}; \citealt{johnstone00}).  This suggests the powerful and simple hypothesis that the IMF is inherited from the mass spectrum of molecular cloud cores.  In such a picture, one-to-one correspondence between the IMF and the CMF naturally arises if the gravitational collapse of molecular cores into protostellar discs is followed by the efficient accretion of this material onto their central young stellar objects (YSOs).  Such an accretion scenario could in principle account for both low mass \citep*{shu87}  and high mass star formation (e.g. \citealt{mclaughlin97}; \citealt{mckee03})

Excellent quantitative information is also available about the IAMF of young stellar objects.  A recent study of 254 stars in the Orion nebula as an example, finds that their YSO rotation periods periods are statistically consistent with a uniform distribution \citep{stassun99}.  Molecular cloud cores are observed to have specific angular momenta that are broadly distributed around a characteristic value of $10^{21}$ cm$^2$ s$^{-1}$ - much larger than the average angular momenta of stars.  This is the observational basis for the well-known ``angular momentum problem'' for young stellar objects - that the total amount of angular momentum seen in molecular cloud cores exceeds that measured for the IAMF by several orders of magnitude.  Since YSOs acquire their initial spin from their parental cores, it is important to understand how the core angular momentum function (CAMF) arises.

These basic questions emphasize the need for theoretical and numerical work on the origin and evolution of molecular cloud cores.  A promising current idea is that molecular cloud cores result from the turbulent fragmentation of clouds and their clumps.  Cores are often observed to be arranged in long, filamentary structures, a feature that can be qualitatively reproduced in supersonic turbulence simulations as a network of shocks (\citealt*{porter94}; \citealt{burkert00}; \citealt{klessen00a}; \citealt{klessen01a}; \citealt*{ostriker01}; \citealt{padoan01}).  The statistical properties of the turbulence may naturally account for the distribution of masses and other properties of molecular cloud cores, such as their specific angular momenta \citep{burkert00,klessen01a}

This paper examines the origin, internal structure, and evolution of gravitationally bound structures that are produced in 3D hydrodynamic simulations of self-gravitating gas.  Our simulations follow the hydrodynamic evolution of an initially uniform density fluid that is perturbed by a spectrum of velocity fluctuations that damp out with time.  

Support against self-gravity in real molecular clouds involves a combination of turbulent and magnetic effects, as thermal and rotational support is negligible.  In purely hydrodynamical models for clouds, only the self-gravity and turbulence matters.  Unless feedback processes from star formation (eg. bipolar outflows -- see \citealt{matzner99}) can replenish the turbulent energy within a dense clump in a molecular cloud, turbulent energy will damp out with time.

We establish that the ratio $t_\mathrm{damp}/t_\mathrm{ff}$ determines the time it takes until gravitational collapse occurs; if $t_\mathrm{damp}/t_\mathrm{ff} < 1$ then turbulent fluctuations must damp before significant gravitational evolution can occur, and thus collapse is significantly delayed.  If $t_\mathrm{damp}/t_\mathrm{ff} > 1$, the fluid will experience gravitational collapse considerably earlier, while significant turbulent motions remain, such that there will be multiple gravitationally-bound cores.

We find that turbulent fragmentation of a system of shocks not only accounts for the observed CMF - as other numerical simulations have shown - but for their angular momentum distribution (CAMF) as well, as the shocks will typically collide at an oblique angle.  We also find that as long as clumps and cores remain turbulent, then their radial density profiles more closely resemble those of turbulent logatropes, analyzed by \citet{mclaughlin96}, than isothermal spheres.

We discuss the technical aspects of our simulations in Section \ref{simulations} where we also describe a new ``watershed'' core finding algorithm.  In Section \ref{energetics}, we follow the energetics of our turbulent system, examining in particular the role of damping of turbulent energy and the applicability of the virial theorem and Bonnor-Ebert equilibria.  We then go on to show the internal structures of cores (Section \ref{structure}) as well as statistical properties of the ensembles of cores such as the CMF and CAMF (Section \ref{statistics}).

\section{SIMULATIONS}\label{simulations}
	
\begin{table}
\begin{center}
\begin{tabular}{|c||c|c|c|c|c|}
\hline Run & $n_\mathrm{J}$ & $M_\mathrm{RMS}$ & $m_\mathrm{TOT} (m_\odot)^\star$ & $L (\mathrm{pc})^\star$ & $m_J$ ($m_\odot$)\\
\hline A2 & 1.1 & 2.0 & 12.5 & 0.10 & 11.6\\
A5 & 1.1 & 5.0 & 12.5 & 0.10 & 11.6\\
B2 & 4.6 & 2.0 & 105.1 & 0.32 & 22.9\\
B5 & 4.6 & 5.0 & 105.1 & 0.32 & 22.9\\
\hline
\end{tabular}
\caption{The parameters for each of the runs that are highlighted in this paper.  Each simulation was performed on 4 processors at a resolution of $256^3$ grid points, with a kinetic energy spectrum described by a $k^{-11/3}$ power law for wavelengths shorter than one-quarter of the length of one side of the simulation box.\label{table_sim_list}}
\end{center}
$^\star$ The original clump parameters drawn from \citet*{lada91}
\end{table}

\begin{figure}
\begin{center}
\includegraphics[width=84mm]{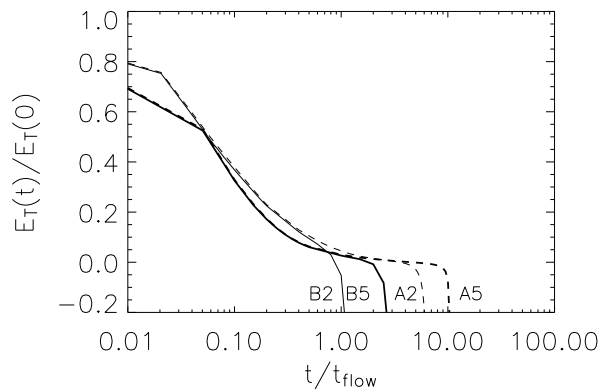}
\caption{Evolution of the total energy in our four simulations, normalized to the initial total energy of each run.  The abscissa measures the time evolution in terms of the flow crossing time of each run.  \label{energy_evolution}}
\end{center}
\end{figure}

	Our simulations are performed using the parallel implementation of the \textsc{zeus} code (\textsc{zeus-mp}) of \citet{stone92a}.  We use the parallelized FFTW libraries \citep{frigo98} to calculate the gravitational potential from the Poisson equation.  The simulations were performed using a Compaq AlphaServer SC40 at the SHARCNET supercomputing facility at McMaster University.  Our runs were performed at $256^3$ resolution on 4 processors.

	An initial fluctuating velocity field was generated by constructing a scalar field with a power spectrum with of the form $k^{-n/2}$, where $k=2\pi/\lambda$ is the wavenumber for a particular isotropic Fourier mode corresponding to wavelength $\lambda$, and $n$ is the three-dimensional power-law index for the fluctuation spectrum.  We use $n=11/3$, the power spectrum representing Kolmogorov turbulence.  For true turbulent motion, the phases of these plane waves are correlated, but the details of this correlation are not well understood.  We apply initial random phases to these plane waves, and give each plane wave a randomly-oriented direction in the computational grid.  We make the initial fluctuation spectrum divergence-free ($\nabla\cdot\mathbf{v}=0$) by taking the curl (or, in the momentum-space representation, taking the cross-product with the wavevector $-i\mathbf{k}$) of this vector field.  The net result for the momentum-space velocity field is
\begin{eqnarray}
  \delta \mathbf{v}_k & = & \left|\mathbf{k}\right|^{n/2}\me^{i\phi}\left(-i\mathbf{k}\times\mathbf{u}\right)/|\mathbf{k}\times\mathbf{u}|^2\label{pwrspectrumeq}
\end{eqnarray}
where $\mathbf{u}$ is the random unit vector giving the direction of the plane wave.  We also truncate our velocity spectrum at a maximum wavelength of $\lambda_\mathrm{max} = L/4$, where $L$ is the total length of one side of our periodic box.  This procedure strongly curtails any nonlinear effects that may arise from the periodic boundary conditions, such as large scale shock waves.  Our initial velocity field is then transformed into position-space, where it is normalized to the RMS Mach number for that simulation.  As we do not include any forcing of the velocity field during the simulation, the initial turbulent velocity field gradually decays away with time.

	Our initial density field was uniform, and normalized such that a specified number of Jeans masses $n_\mathrm{J}=m_{\mathrm{tot}}/m_\mathrm{J}$ were on the box.  The thermal pressure was determined via an isothermal equation of state.

	The \textsc{zeus} code evolves the hydrodynamic equations
\begin{eqnarray}
\frac{\partial\rho}{\partial t} + \nabla\cdot(\rho\boldsymbol{v}) & = & 0\label{eq_continuity}\\
\rho\frac{\partial\boldsymbol{v}}{\partial t} + \rho (\boldsymbol{v}\cdot\nabla)\boldsymbol{v} & = & - \nabla P - \rho \nabla\Phi \\
\nabla^2\Phi & = & 4\pi G \rho
\end{eqnarray}
(The \textsc{zeus} code can use the full magnetohydrodynamic equations, but for this paper we are not using magnetic fields, as we want to isolate the purely hydrodynamic effects.  We will consider magnetic fields in a future paper.)

\subsection{Observational basis for initial conditions}

	Our simulations explored a region of parameter space in $(n_\mathrm{J},M_\mathrm{RMS})$.  We performed many simulations but in this paper discuss a total of 4 different but representative simulations within this parameter space,  listed in Table \ref{table_sim_list}.  The values for $n_\mathrm{J}$ were chosen from the masses and sizes of cores in \citet*{lada91}.  These are cores detected in CS emission, corresponding to mean densities of $10^4 - 10^5 \;\mathrm{cm}^{-3}$.  Given their masses of $10-100\; \,m_\odot$ and radii of $0.1-0.3$ pc, and assuming a thermal temperatures of $20\;\mathrm{K}$ (typical for molecular cloud cores), these clumps contain a few ($1-10$) initial thermal Jeans masses each.  The \citet*{lada91} data set also contains typical linewidths for these cores of $1-2 \;\mathrm{ km s}^{-1}$.  For an isothermal fluid at $20\;\mathrm{K}$, for which the sound speed is $\sim 0.4 \;\mathrm{km s}^{-1}$, these linewidths correspond to flows of a few times the sound speed.  We adopt the values of $M_\mathrm{RMS} = v_\mathrm{RMS}/c_s =(2,5)$ for the RMS turbulent Mach number for the fluid.  We focus our analysis on four particular combinations of $n_\mathrm{J}$ and $M_\mathrm{RMS}$, namely $(n_\mathrm{J},M_\mathrm{RMS})$ = $(1.1,2.0)$ (Run A2), $(1.1,5.0)$ (Run A5), $(4.6,2.0)$ (Run B2), and $(4.6,5.0)$ (Run B5).

While the calculations were performed in dimensionless units, they can be scaled to real values by choosing two of the total mass $m_\mathrm{tot}$, the total length of one side of the box $L$, or the temperature $T$, via
\begin{eqnarray}
\frac{m_\mathrm{tot}}{m_\odot} & = & 119 \,n_\mathrm{J}^{2/3} \left(\frac{L}{\mathrm{pc}}\right) \left(\frac{T}{20 \mathrm{K}}\right)\label{massScaling}
\end{eqnarray}
It should be noted that Equation (\ref{massScaling}) cannot realistically be scaled to arbitrarily large or small values of $m_\mathrm{tot}$ or $L$, as real molecular clouds will only have $n_\mathrm{J}$ in the range of $1-5$ for a certain range of $L$.  

	Our simulations were allowed to run until the local Jeans length at some point within the computational grid was less than 4 pixels in length, a condition established to avoid artificial fragmentation effects \citep{truelove98}.  For our simulations this works out to a density threshold that we can express in terms of the number of Jeans masses on the grid, and the initial or mean density:
\begin{eqnarray}
\frac{\rho}{\rho_0} & = & \frac{N^2\pi}{80}\left(\frac{4\pi}{3}\right)^{1/3}n_\mathrm{J}^{-2/3} = 4148.6 n_\mathrm{J}^{-2/3}
\end{eqnarray}
where $N$ is the number of pixels along one edge of the box, and $\rho_0$ is the initial density of the fluid.

\subsection{Watershed algorithm for finding fluctuations}\label{watershedalgorithm}

	We use a watershed algorithm (e.g. \citealt{vincent91}; \citealt{mangan99}) to find fluctuations in our simulations.  We chose this algorithm because of its intuitive simplicity and speed of execution.  As we show in Appendix \ref{appendixa}, it is comparable to the commonly used \textsc{clumpfind} algorithm \citep*{williams94}, in the limit of infinitesimal contour levels.  The algorithm consists of the following procedure, iterated over the entire grid.

	At each initial cell, we find the local gradient vector and move to the next cell in that direction; we repeat this process until we reach a local maximum.  At each cell that we pass through in this path, we assign it an identification number $q$ for this path.  If we come to a cell that already has an identification number $q'$, we re-assign all of the cells on the current path $q$ to the identification number of the new path $q'$.  This way, all cells that are associated with a single local maximum are assigned to the same fluctuation.

	Our algorithm is extremely sensitive to small fluctuations in density, resulting in a large number of fluctuations found at the numerical resolution limit.   We therefore first smooth the data by averaging each cell with its nearest neighbours in each direction before running the fluctuation-finding algorithm.  This greatly reduces the number of fluctuations found with a radius of only a few pixels.  After the fluctuation-finding routine had been run, we used the original, unsmoothed data to calculate the fluctuation properties.

	When all of the grid cells have been assigned to fluctuations, we can look at the boundaries of each fluctuation (defined as a cell which has a 6-neighbour that belongs to a different fluctuation), and calculate the surface density (from which we can extract the surface pressure), and the mean radius of the fluctuation.  We can also calculate the mass, mean velocity and internal velocity dispersion of the fluctuations from the fluctuation data.

	We show in Appendix \ref{appendixa} that our watershed algorithm gives results that agree with the results returned by \textsc{clumpfind} when applied to the same numerical data.  Both the \textsc{clumpfind} algorithm and our watershed algorithm find a similar number of bound, self-gravitating fluctuations that we define as ``cores'' (see Section \ref{subsection_virial_stability}); however, the watershed algorithm finds many more unbound density fluctuations than \textsc{clumpfind}.  As the density resolution of \textsc{clumpfind} is increased, the number of unbound fluctuations increases, approaching the number found by the watershed algorithm.

As one of the goals of this paper is to examine what properties are needed for a density fragment to undergo gravitational collapse, we want to have a complete census of turbulent fluctuations.  The watershed algorithm presented here is more sensitive to these unbound fluctuations, so we consider this definition of a fluctuation to be better suited for our purposes than the definition provided by \textsc{clumpfind}.

\section{ENERGETICS}\label{energetics}

\begin{figure}
\begin{center}
\includegraphics[width=84mm]{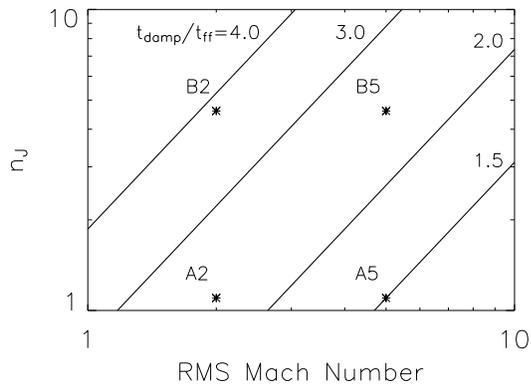}
\caption{The ratio of $t_\mathrm{damp}/t_\mathrm{ff}$ predicted by Equation (\ref{dampff}).  The solid lines, from right to left, represent ratios of $1.5, 2.0, 3.0$ and $4.0$.}\label{njvsmE}
\end{center}
\end{figure}

\begin{figure*}
\begin{center}
\includegraphics[width=84mm]{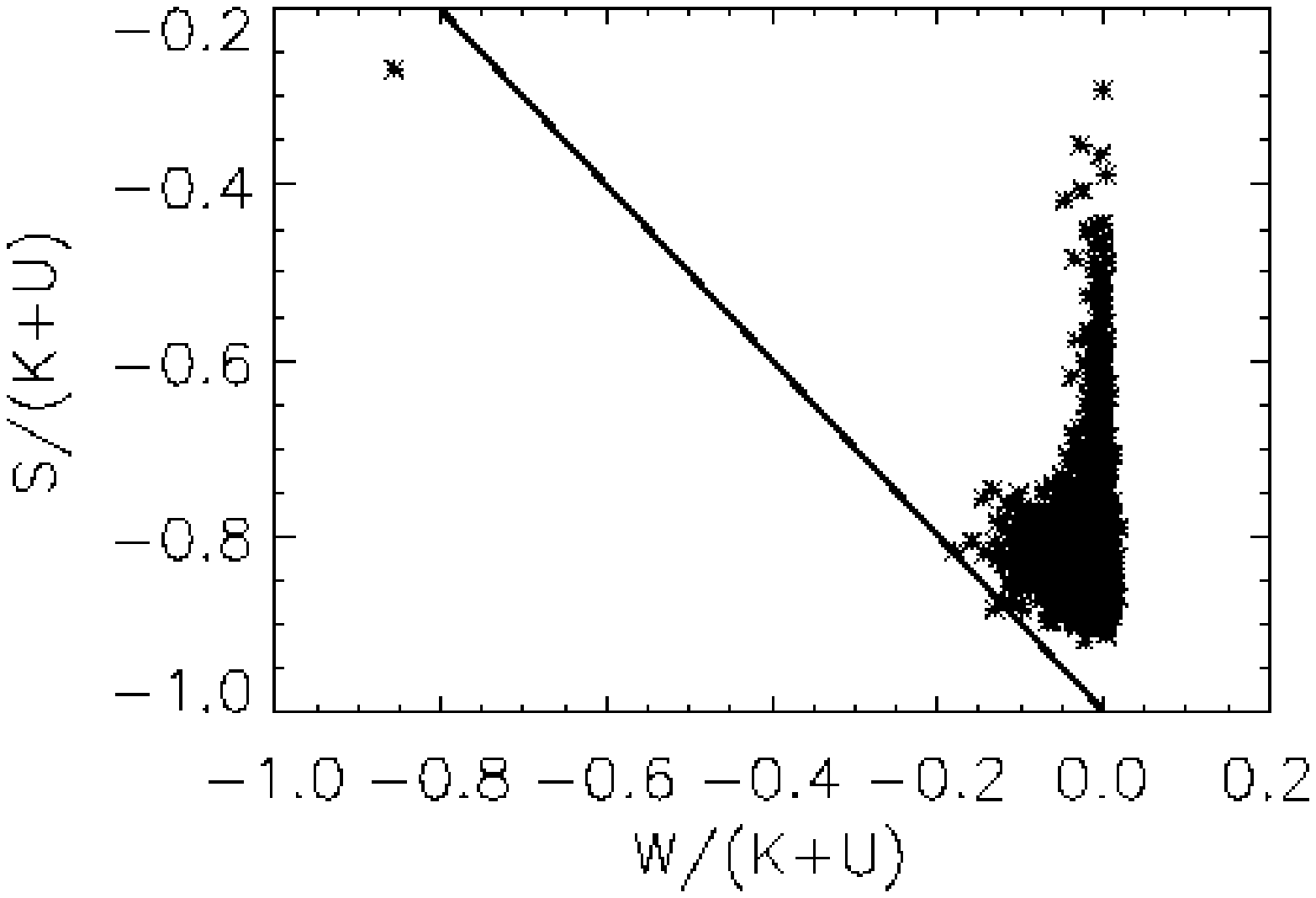} \includegraphics[width=84mm]{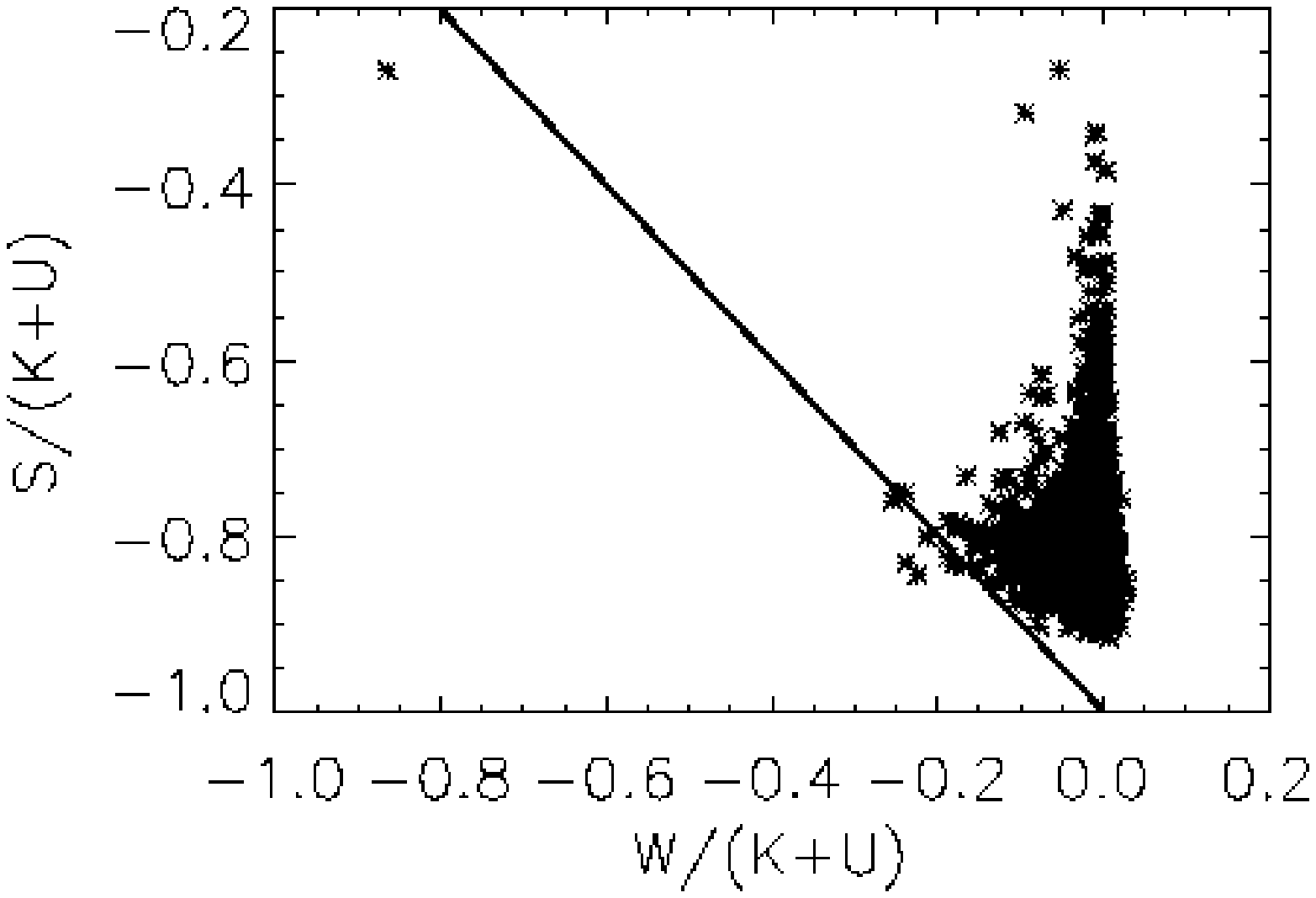}\\
\includegraphics[width=84mm]{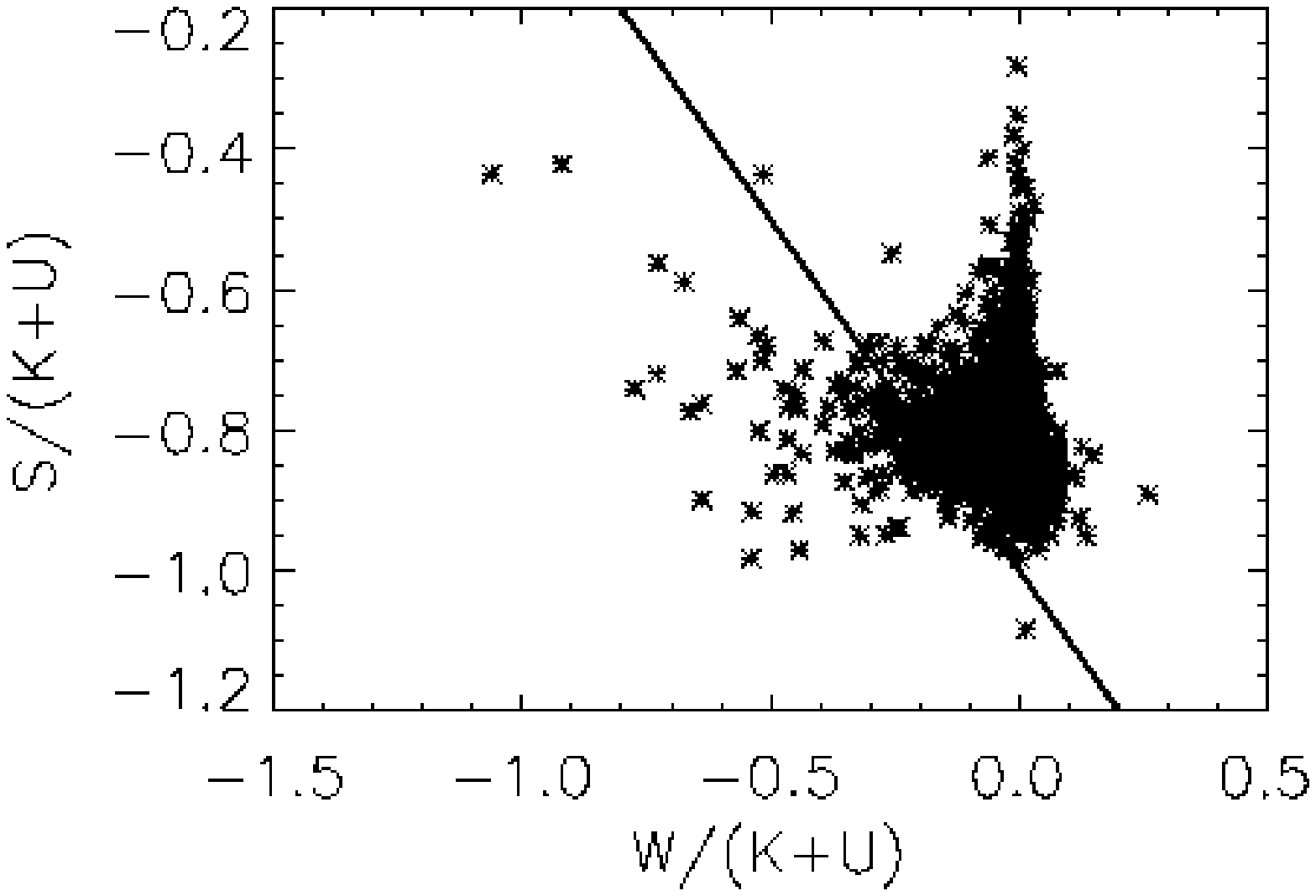} \includegraphics[width=84mm]{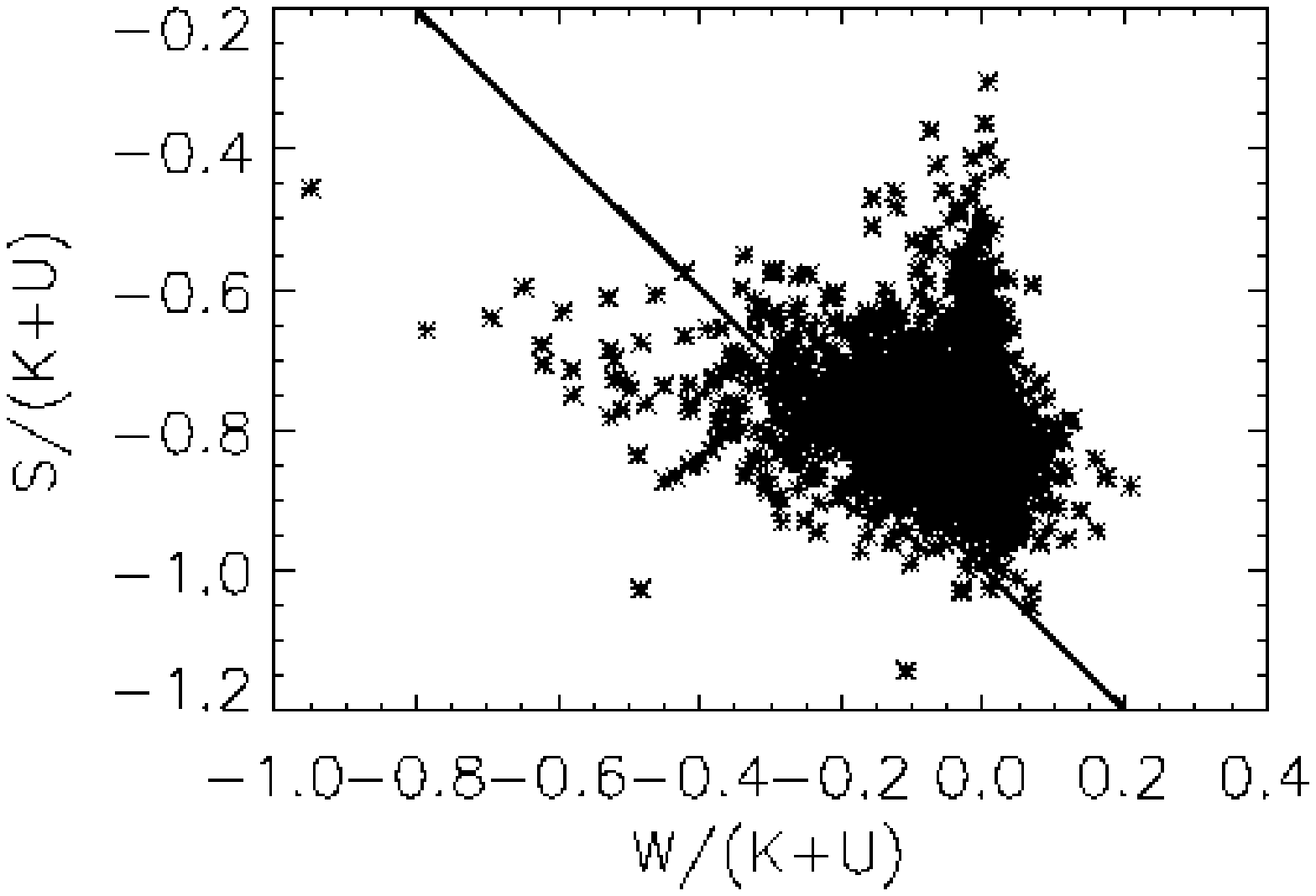}
\caption{The importance of the different terms in the virial equation.  S is the surface terms, W is the gravitational term, K is the kinetic energy term, and U is the internal energy term.  The solid line corresponds to virial equilibrium, $\ddot{I}'=0$.  In this figure and the following figures, run A2 is plotted in the upper-left, A5 in the upper-right, B2 in the lower-left, and B5 in the lower-right.\label{internal_dynamics}}
\end{center}
\end{figure*}

	Core formation within our simulations is driven by the energetics of self-gravity and turbulent decay.  We plot the evolution of total energy $E_\mathrm{T} = E_\mathrm{kin} + E_\mathrm{grav}$ in Fig. \ref{energy_evolution}.  As our simulations employ an isothermal equation of state, total energy is not conserved (in a real system, this would correspond to energy losses via radiative cooling).  The evolution of the systems can be divided into three phases:  an initial phase that is dominated by shocks forming in reaction to the initial conditions; an intermediate phase that primarily involves the dissipation of kinetic energy in the shocks; and a late phase that is dominated by gravitational contraction of cores.  

	The first phase lasts $\sim0.1\;t_\mathrm{flow}$ in each of our runs, marked in Fig. \ref{energy_evolution} by the rollover at early times.  

	Turbulent kinetic energy decays via the power-law $E_\mathrm{K} \propto t^{-1}$ in a compressible fluid (\citealt{maclow98}; \citealt*{ostriker99,christensson01,ostriker01}).  In a linear-log plot like Fig. \ref{energy_evolution}, this power-law relationship appears as a decaying exponential.  A significant portion of runs B5, A2, and especially A5 is spent in this phase.  In run B2, the kinetic energy damped sufficiently during the shock-forming phase, allowing gravitational collapse to proceed without a significant decay phase.

	The final phase of these simulations begins when the turbulent energy has damped sufficiently that gravitational contraction can overcome turbulent pressure to force at least one core to collapse.  In Fig. \ref{energy_evolution}, this is marked by a sharp decrease in the total energy as the gravitational energy grows to large negative values.  The Mach 5 runs have a larger kinetic energy for the same amount of mass as the Mach 2 runs, and so can resist gravitational collapse for more flow crossing times.  Similarly, the B runs have more mass than the A runs, and so less decay of turbulent energy is needed before gravitational effects become dominant.

\begin{figure*}
\begin{center}
\includegraphics[width=84mm]{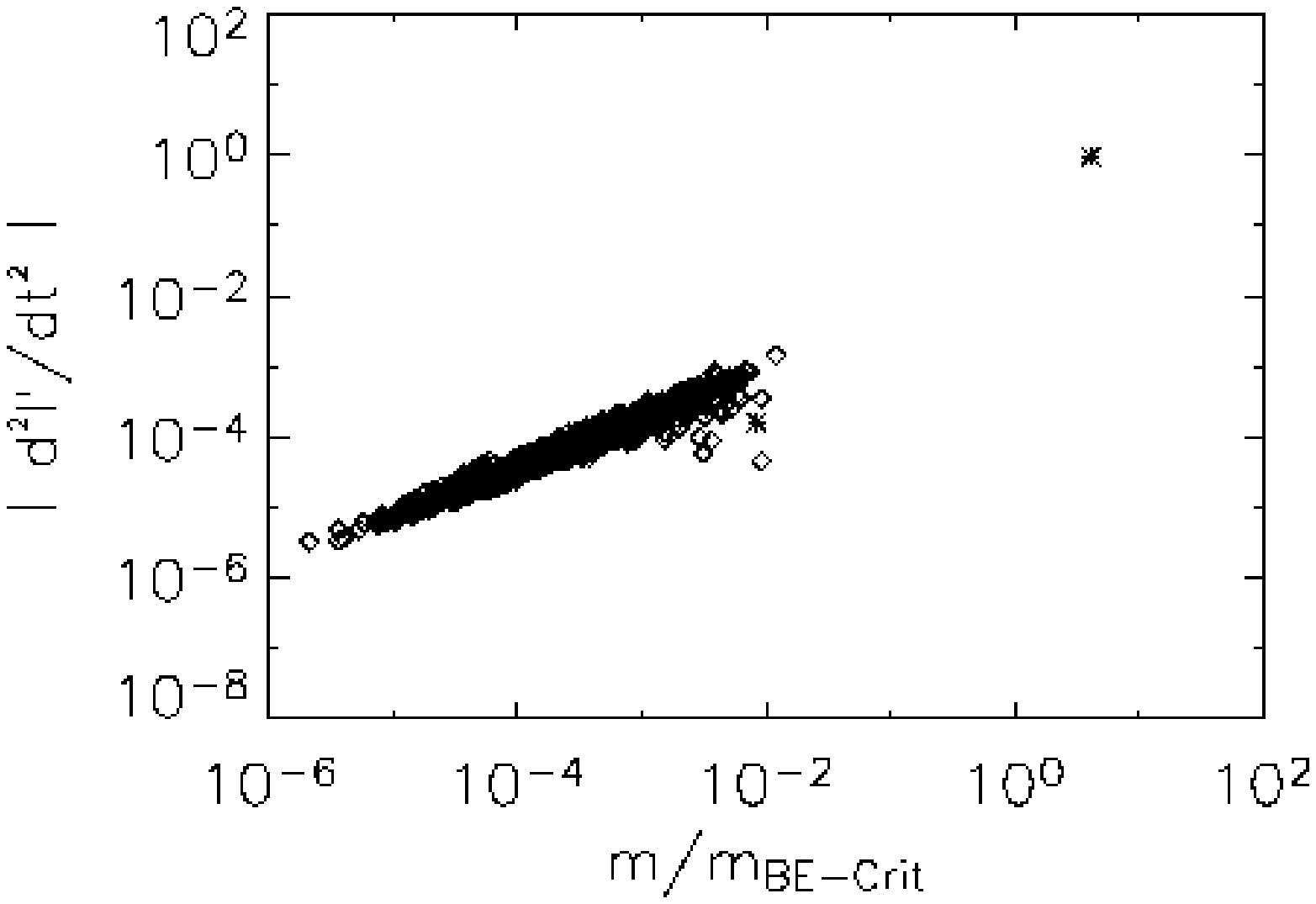}\includegraphics[width=84mm]{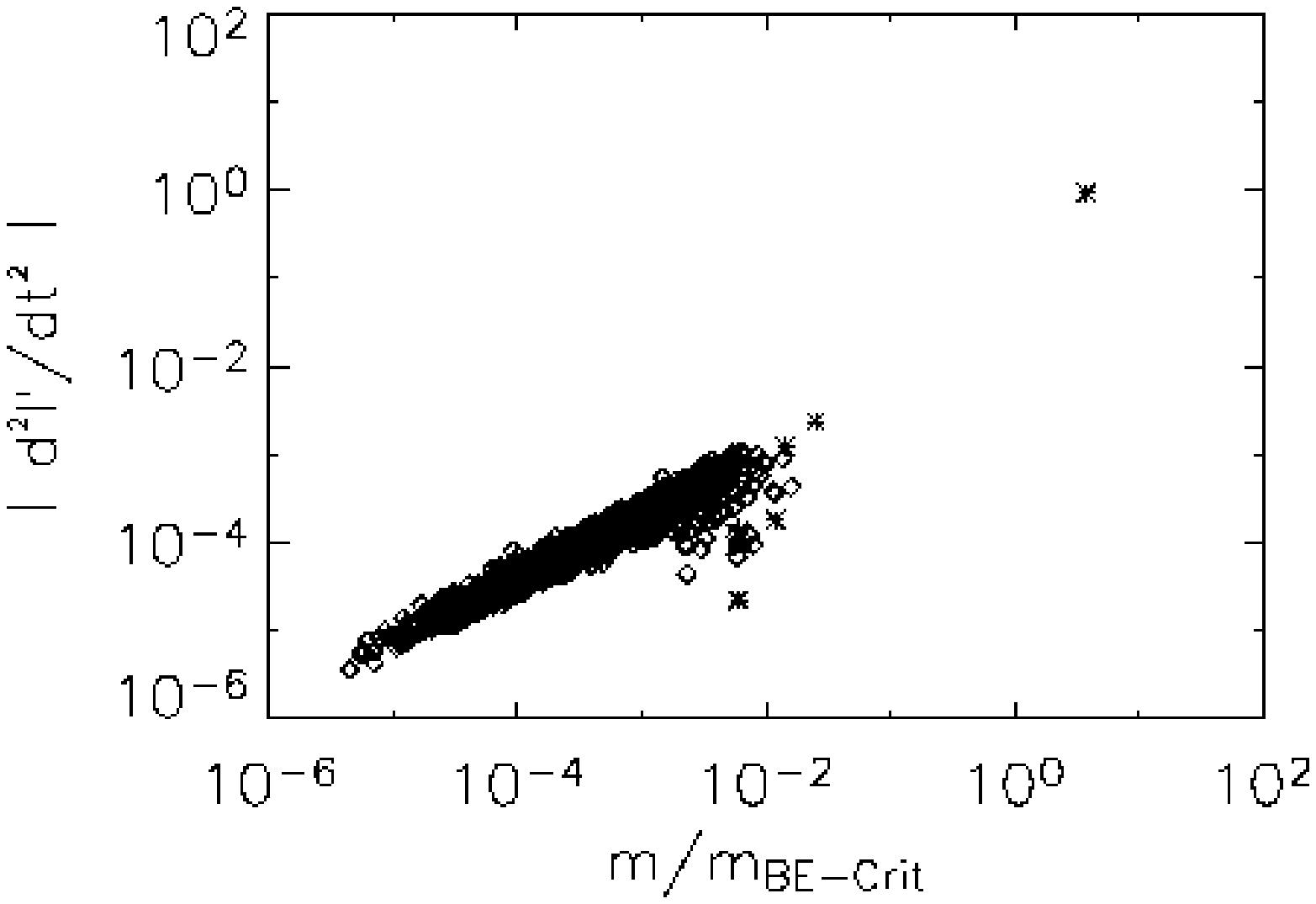}\\
\includegraphics[width=84mm]{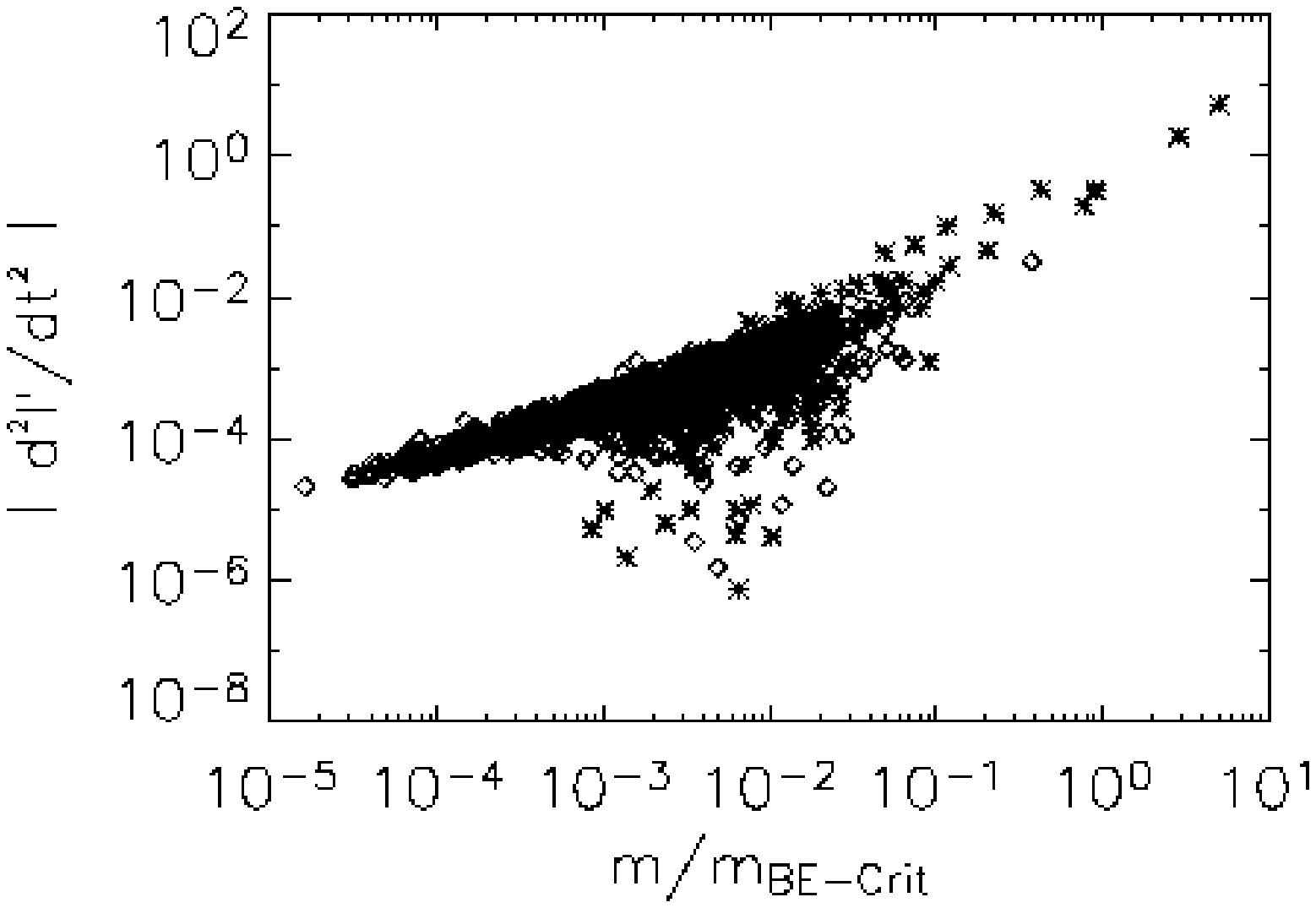}\includegraphics[width=84mm]{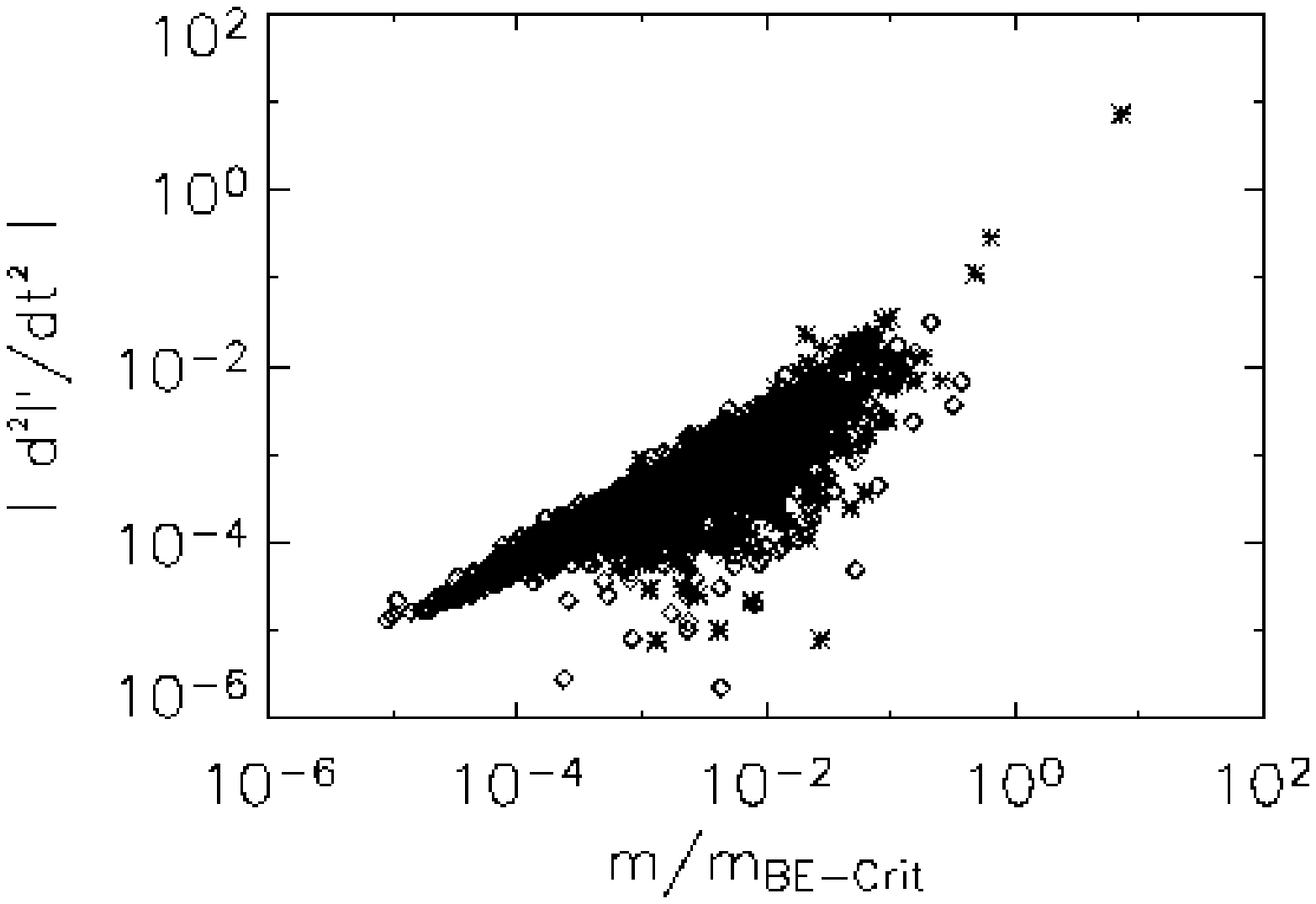}
\caption{Comparison of the stability from the virial equation, versus the stability from the Bonnor-Ebert criterion.  The diamonds represent fluctuations with $\ddot{I}'>0$; the stars represent cores with $\ddot{I}'<0$.  \label{dynamic_measures}}
\end{center}
\end{figure*}

\begin{figure*}
\begin{center}
\includegraphics[width=84mm]{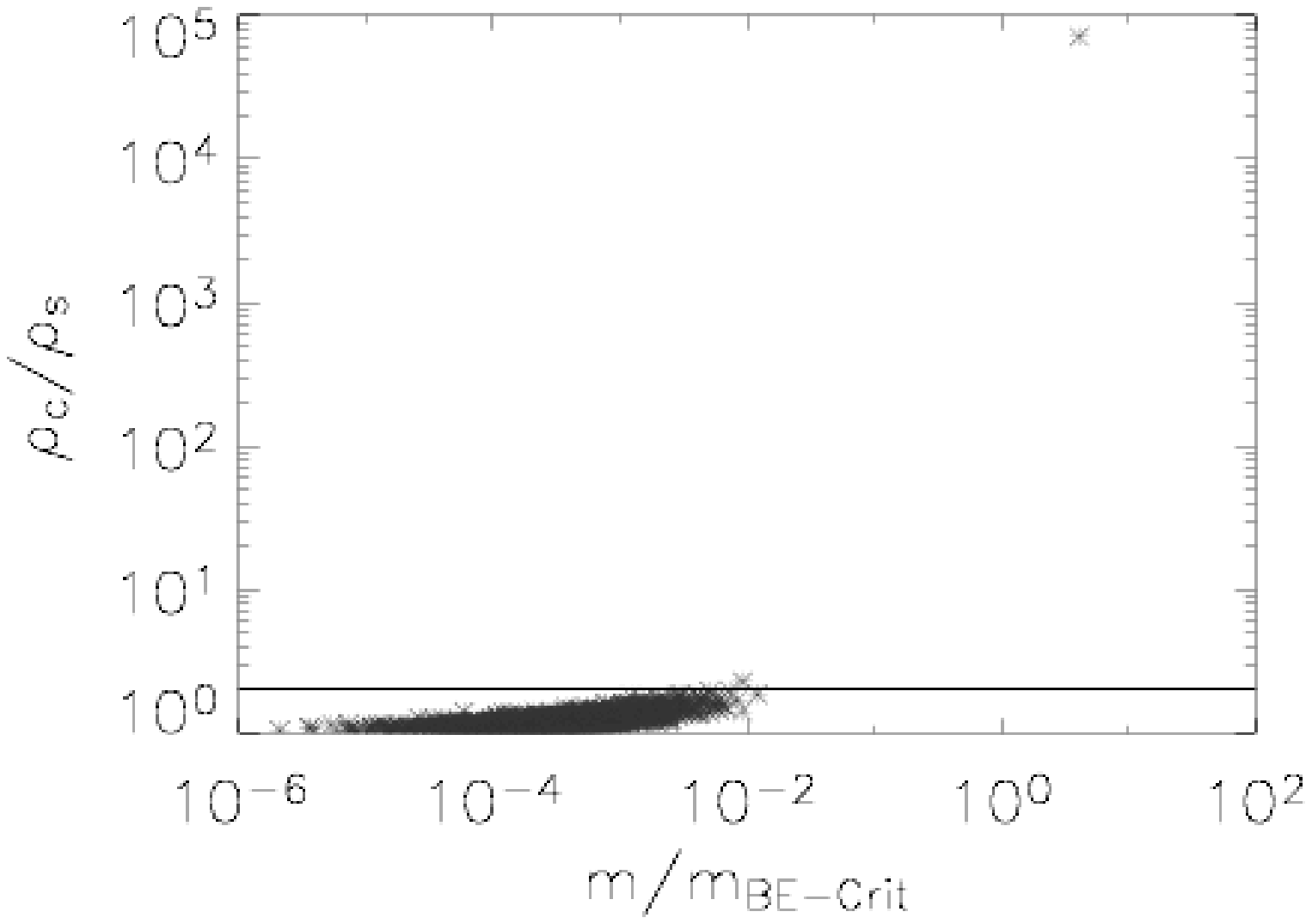}\includegraphics[width=84mm]{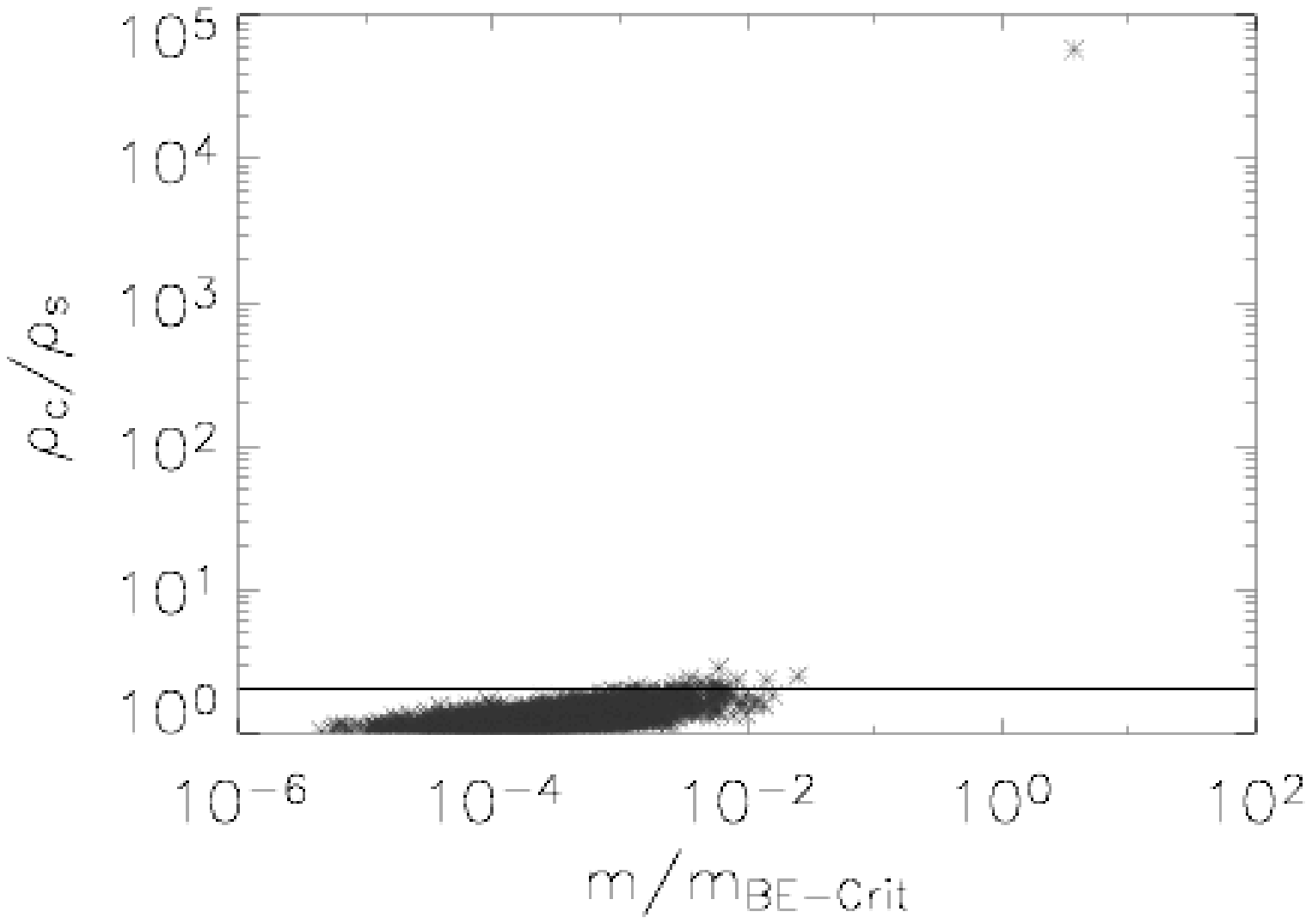}\\
\includegraphics[width=84mm]{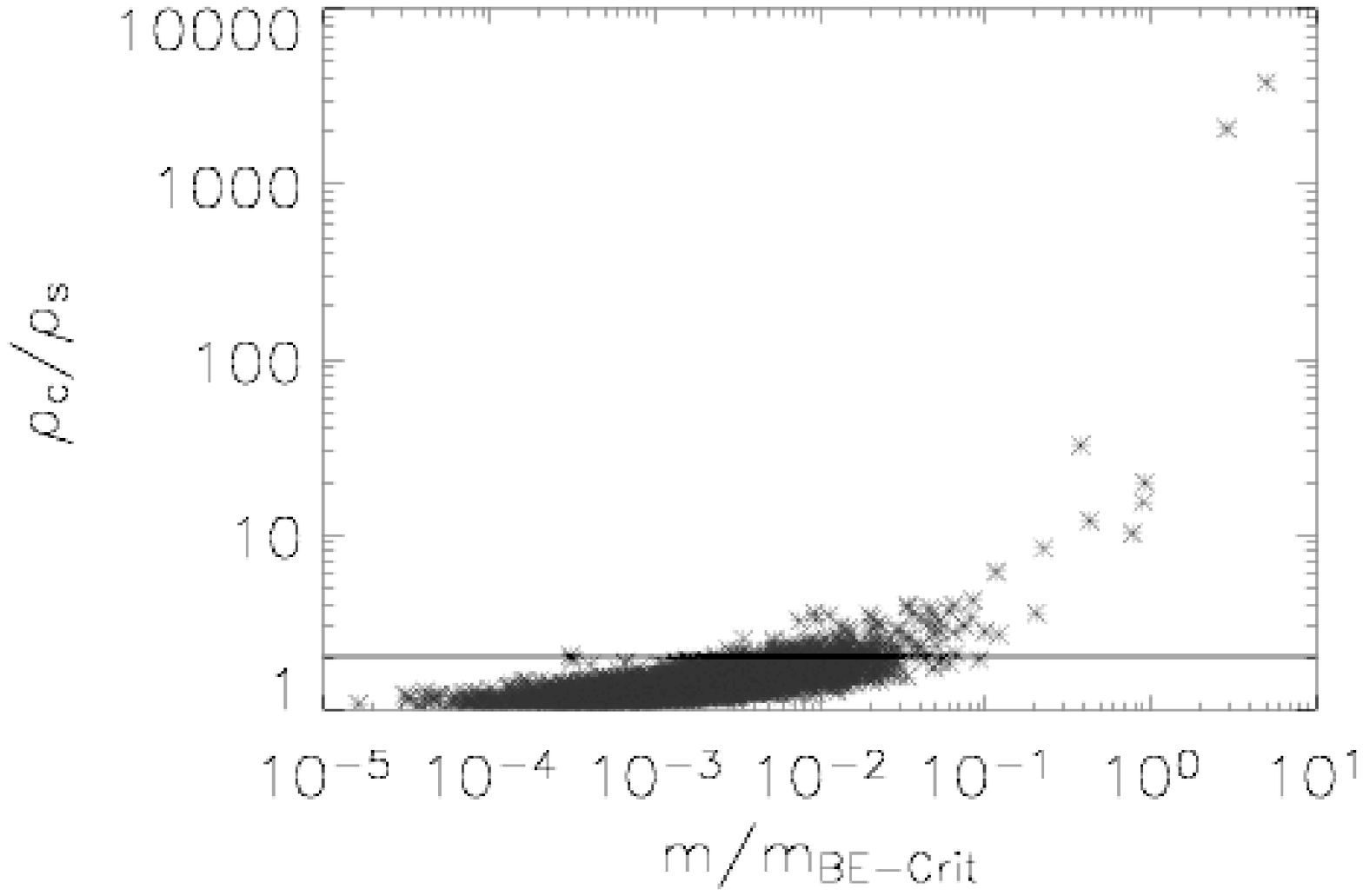}\includegraphics[width=84mm]{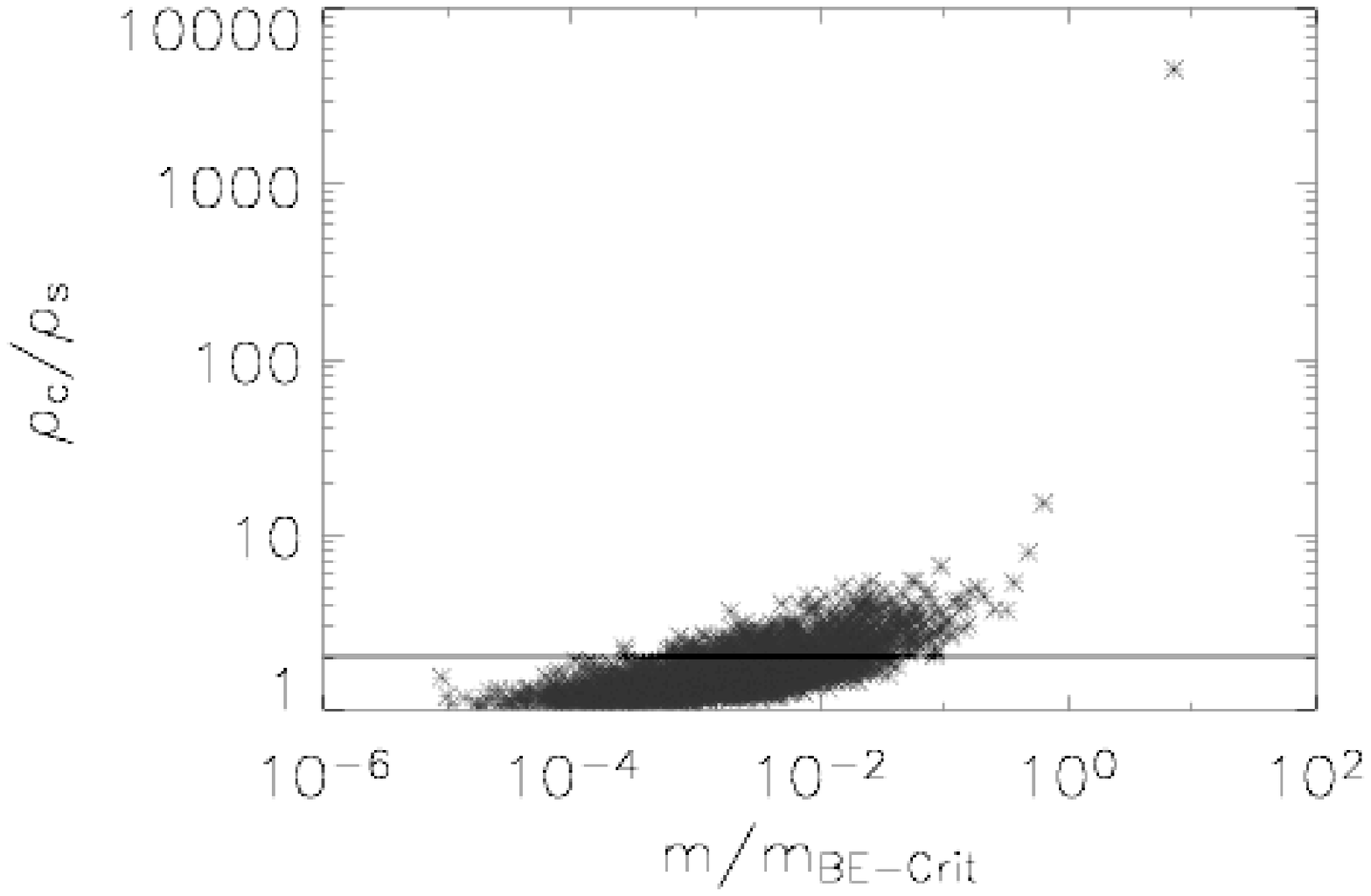}
\caption{The ratio of central density to surface density, as a function of $m/m_\mathrm{BE}$.  The horizontal line in each plot is $\rho_c/\rho_s = 2$.  \label{density_ratio}}
\end{center}
\end{figure*}

	The simulations with 1.1 Jeans masses take many flow crossing times before the turbulent motions damp sufficiently that self-gravity can overcome the combination of turbulent and thermal internal pressure; with turbulence having damped away, a significant amount of the fluid is gathered into a few large cores.  In the simulation with 4.6 Jeans masses, gravitational collapse occurs after only a few flow crossing times, so that the turbulence has not damped as much.  While the first object to experience gravitational runaway collapse causes the simulation timesteps to become very small, thus preventing us from observing the evolution of any other objects, we still find many gravitationally bound cores amongst the wispy filamentary structures in the fluid; we would thus expect to find a number of star-forming cores in such a fluid.

	Insight into the relationship between the time it takes for runaway gravitational collapse to occur can be found in the relationship between the turbulent damping time and the local gravitational free-fall time of an individual fragment.  The damping time of the turbulence in the simulation is related to the flow crossing time, $t_\mathrm{damp} = L/u = L/(c_s M)$, where $L$ is the length of one side of the simulation box, $u = c_s M$ is the root-mean-square velocity of waves in the fluid, and $M$ is the Mach number.  As the gravitational collapse occurs in the enhanced density behind the shocks, the local free-fall time can be calculated from the post-shock density, $t_\mathrm{ff} = (32 G \rho_1/3\pi)^{-1/2}$.  For a system of strong isothermal shocks, $\rho_1 \sim \rho_0 M$, where $\rho_0$ is the average, pre-shock density of the fluid.  Using Equation (\ref{massScaling}), the free-fall time can be expressed in terms of our $n_\mathrm{J}$ parameter, and we can calculate the ratio of the damping time to the free-fall time to be
\begin{eqnarray}
  \frac{t_\mathrm{damp}}{t_\mathrm{ff}} & = & 3.25 \,n_j^{1/3}M^{-1/2}\label{tdamptff}\label{dampff}
\end{eqnarray}
If $t_\mathrm{damp}/t_\mathrm{ff} \ll 1$, we expect turbulence to damp away before collapse can occur to any significant degree, leading to a few main cores that dominate.  For $t_\mathrm{damp}/t_\mathrm{ff} \gg 1$, we expect the fluid to collapse while turbulent motions remain strong, thus resulting in many small cores located within the shocks.  This limit corresponds to the standard turbulent fragmentation scenario (\citealt{klessen00a}; \citealt*{ostriker01}; \citealt{padoan01}).

We plot the relationship in Equation (\ref{tdamptff}), along with the values for our simulations of $(n_\mathrm{J}, M_\mathrm{RMS})$, in Fig. \ref{njvsmE}.  All of our runs have $1.5 \le t_\mathrm{damp}/t_\mathrm{ff} < 4.0$.  The ratios for $t_\mathrm{damp}/t_\mathrm{ff}$ for the runs are consistent with the time it takes for collapse to occur in Fig. \ref{energy_evolution}, in that the simulations with a larger ratio of $t_\mathrm{damp}/t_\mathrm{ff}$ collapse significantly quicker.

\subsection{Virial equation}\label{subsection_virial_stability}

	We can gain a more detailed insight into the dynamics of these fluctuations by analyzing the terms in the Eulerian virial equation \citep{mckee92}:

\begin{eqnarray}
  \frac{1}{2}\ddot{I} + \frac{1}{2}\frac{\dif}{\dif t}\int_S (\rho \boldsymbol{v}r^2)\cdot \dif\boldsymbol{S} \;=\; U+K+W+S \;=\; \frac{1}{2}\ddot{I}'\label{vir_eq}
\end{eqnarray}
where
\begin{eqnarray}
I & = & \int_V \rho r^2 \dif V \\
U & = & 3\int_V P\dif V \\
K & = & \int_V \rho v^2 \dif V \\
W & = & -\int_V \rho\boldsymbol{r}\cdot\nabla\Phi\dif V \\
S & = & -\int_S [P\boldsymbol{r}+\boldsymbol{r}\cdot(\rho\boldsymbol{vv})]\cdot \dif\boldsymbol{S}
\end{eqnarray}
where $U$ is the thermal energy of the fluctuation, $K$ is the internal kinetic energy (including both collapse motion and internal turbulence), and $W$ is the gravitational potential energy.  Each dot over a symbol represents a time derivative.  The terms $U$ and $K$ act to support the fluctuation, while $W$ drives gravitational collapse.  The two terms in $S$ represent the thermal pressure and turbulent pressure on the surface of the cloud.  The second term on the left-hand side of Equation (\ref{vir_eq}) is a time derivative of the moment-of-inertia flux through the surface of the fluctuation.  We are unable to directly calculate this quantity, even though it could be non-negligible and of either sign.  We will treat it as if it were negligible, such that we will call $\ddot{I}'=0$ virial equilibrium.  It should be remembered that this will be merely an estimate of the true virial equilibrium of the fluctuations.

The internal energy and kinetic energy terms must always be positive.  For an isolated fluctuation, the gravitational term will be negative (counteracting the effects of thermal pressure), but the fluctuations in our simulations are not in isolation --- they can be disrupted by tidal effects from nearby fluctuations.  On average, about a quarter of the fluctuations found in each of our simulations have a gravitational term which acts to increase $\ddot{I}$.  The surface pressure term will usually be negative (but in theory could be positive); for all of the fluctuations in each of our simulations, the surface pressure acts as a confining force.

\begin{figure*}
\begin{center}
\includegraphics[width=84mm]{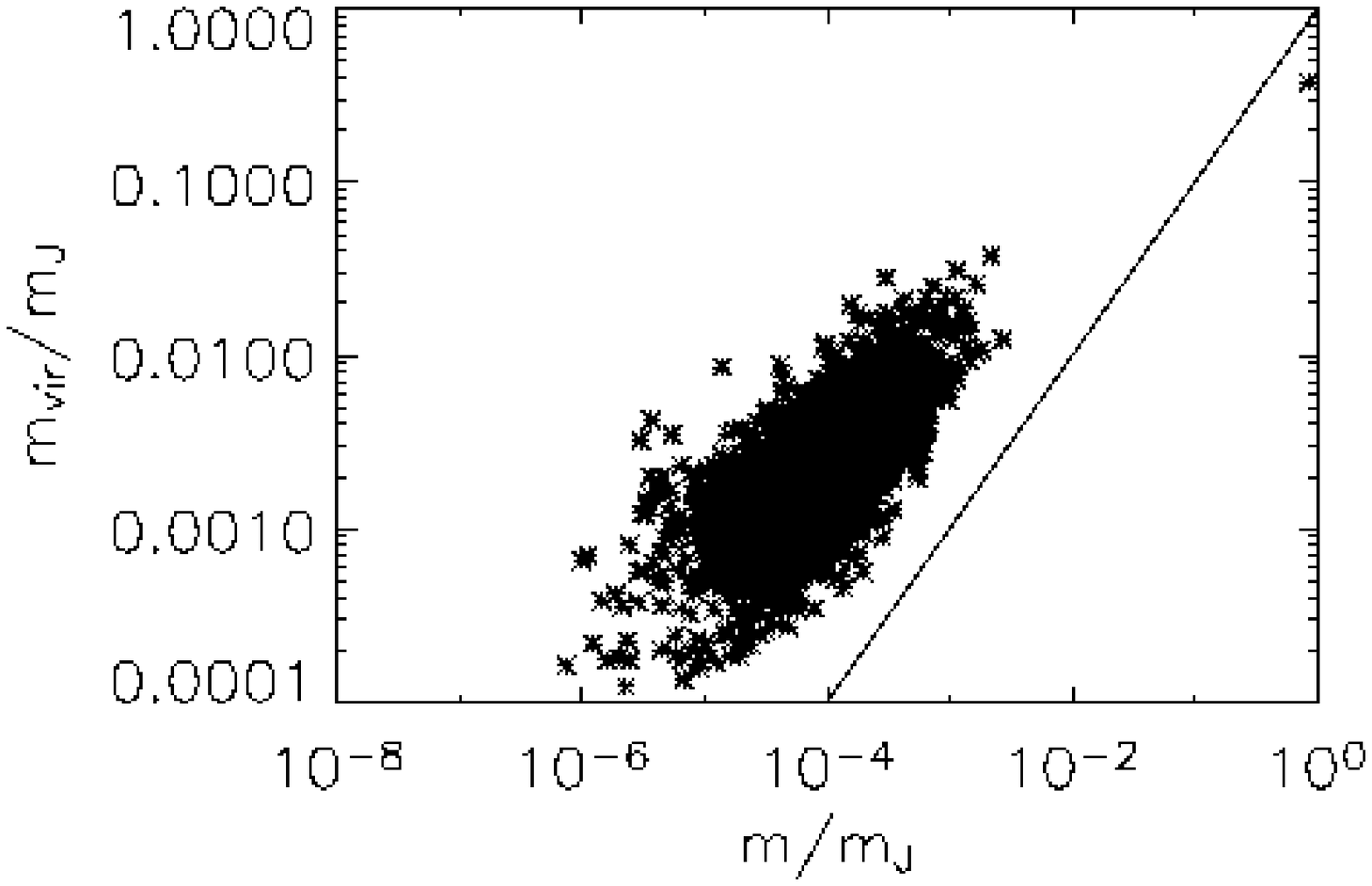}\includegraphics[width=84mm]{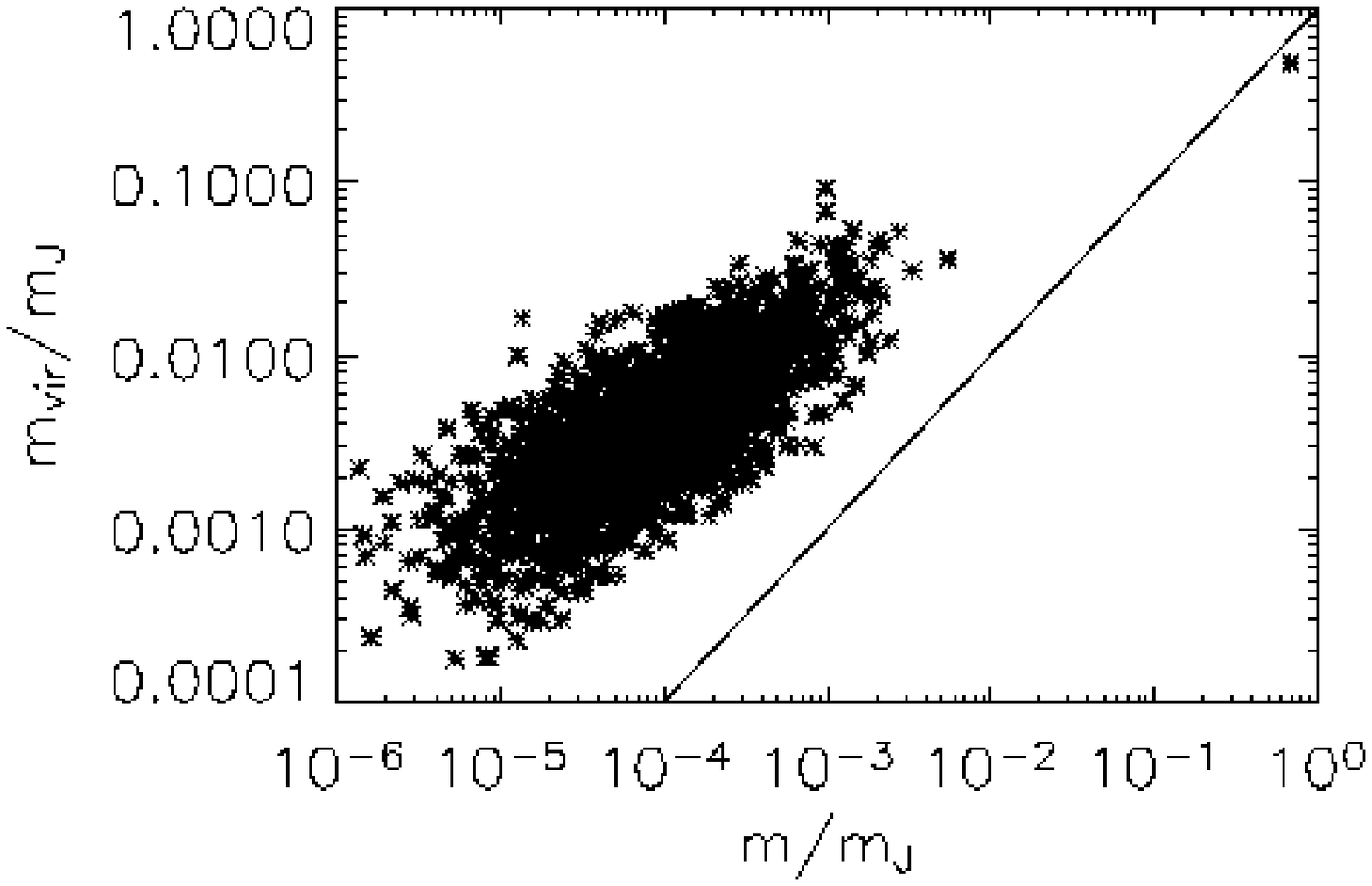}\\
\includegraphics[width=84mm]{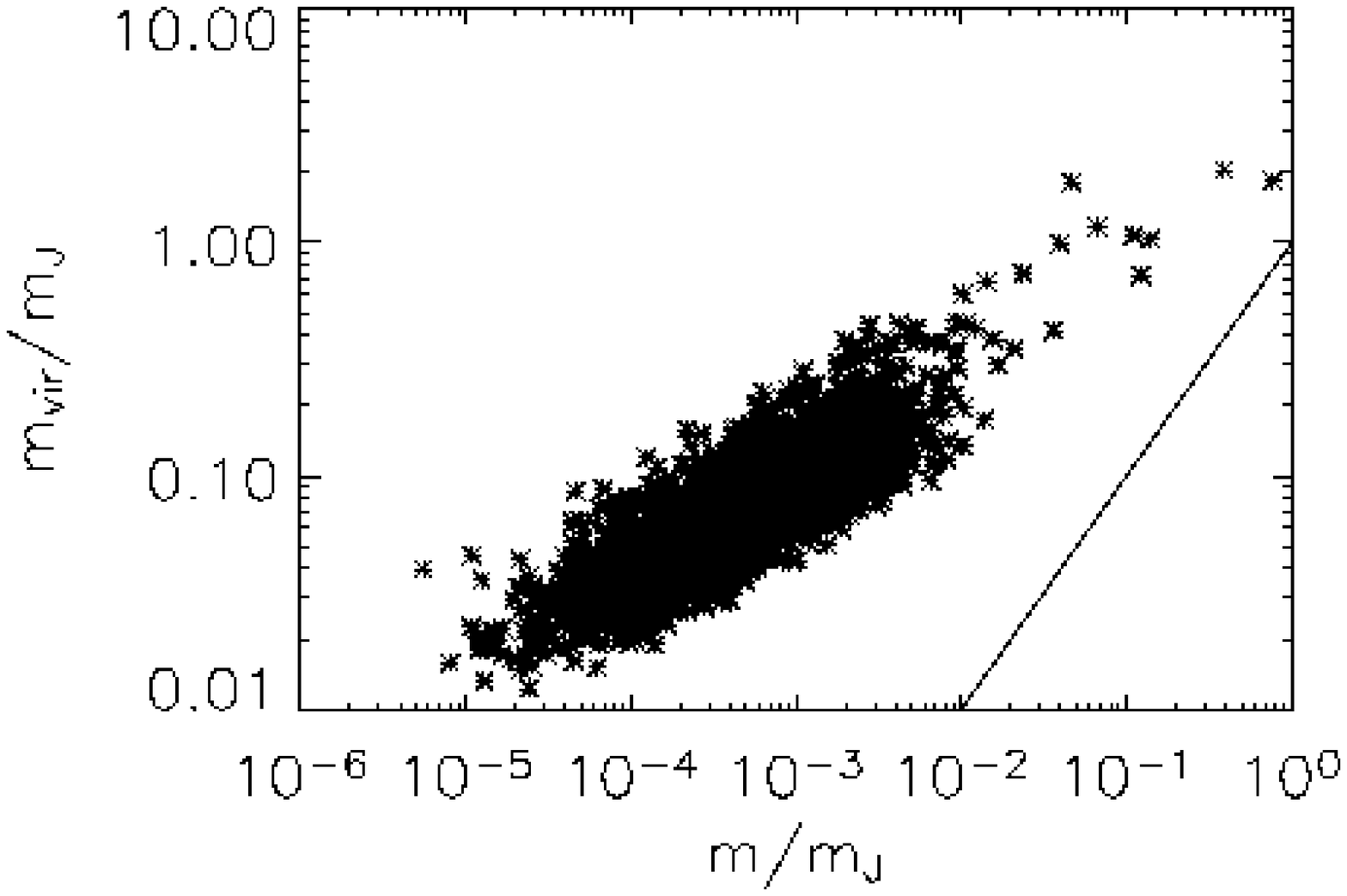}\includegraphics[width=84mm]{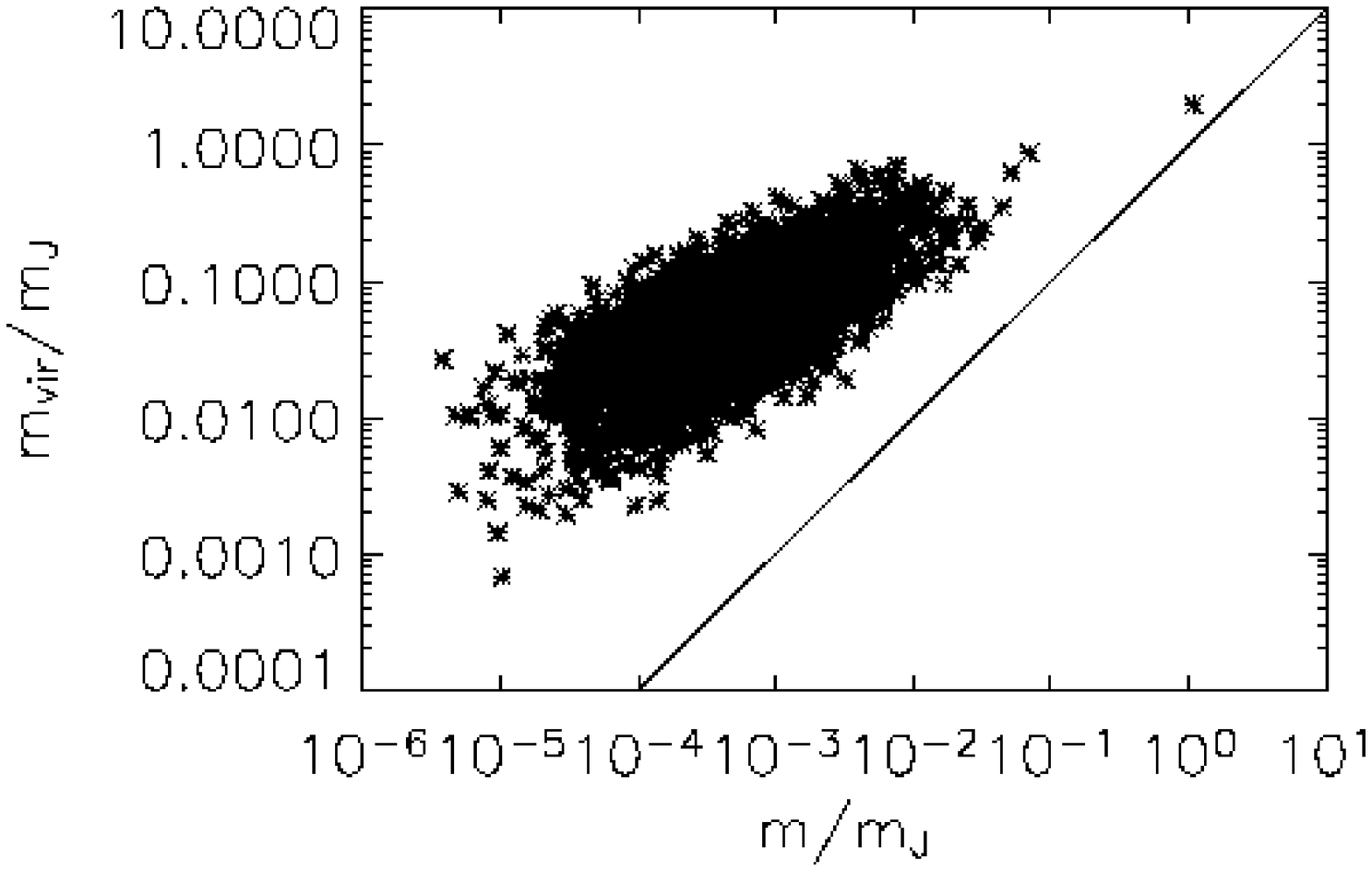}
\caption{The relationship between the mass that would be calculated by applying the virial theorem and the actual mass of the fluctuation.  The solid line is the equality $m_\mathrm{vir} = m$.  \label{virialmass}}
\end{center}
\end{figure*}

We compare the surface terms in the virial equation to the gravitational term in Fig. \ref{internal_dynamics}.  For both axes, we normalize the virial terms to the internal kinetic and thermal terms of the virial equation.  The solid line marks the locus of virial equilibrium on the graph.  The fluctuations that are the most out of equilibrium (as measure by the largest values of $|\ddot{I}'|$) cluster at the most negative values of $W/(K+U)$ (as a strongly self-gravitating core is expected to be) and the least negative values of $S/(K+U)$.

Most of the fluctuations in each of the runs have $\ddot{I}' > 0$ (for the plots in Fig. \ref{internal_dynamics}, this is equivalent to the fluctuation lying above the line $\ddot{I}'=0$), indicating that they are likely to disperse.  These fluctuations have relatively little gravitational energy, but the surface pressure can be significant.  In the A runs, only a tiny number of fluctuations are out of virial equilibrium with $\ddot{I}' < 0$, such that they will collapse.  Even these collapse-unstable fluctuations are not far out of equilibrium, lying close to the virial equilibrium line.  We see a different behaviour in the B runs; there are many fluctuations that are virial-collapse-unstable, with some fluctuations that are far out of equilibrium.  This is not surprising, as the A runs have had time for turbulence to damp away and for the system to reach a quasi-equilibrium state, but we were forced by our Jeans resolution criterion to stop the evolution of the B runs while significant turbulent motions remain.

We use the virial equation to divide our set of density fluctuations into two subsets.  The first is what we will call \textbf{bound cores} .  These cores have $\ddot{I}<0$, indicating that they are bound by pressure and/or gravity.  Every density fluctuation that is not a bound core will be called an \textbf{unbound fluctuation}.

\subsection{Bonnor-Ebert stability}\label{subsection_be_stability}

	For each fluctuation we can calculate the mass of a critical Bonnor-Ebert sphere from the velocity dispersion and surface pressure \citep{ebert55}:
\begin{eqnarray}
m_\mathrm{BE} = \sqrt{\frac{17.6}{4\pi G^3 P_S}}\,\,c_s^4\label{becrit}
\end{eqnarray}
where $P_S$ is the surface pressure, $c_s$ is the sound speed, and $G$ is the gravitational constant.  $m_\mathrm{BE}$ represents the maximum mass the fluctuation could have, given its temperature and surface pressure, before it becomes unstable to collapse.  If the mass of an individual fluctuation is greater than $m_\mathrm{BE}$, it suggests that that fluctuation might be in a state of collapse.  Conversely, if the mass of a fluctuation is less than $m_\mathrm{BE}$, the fluctuation is either stable or not gravitationally bound.

Out of the few thousand objects our algorithm identifies as fluctuations, only the most massive one or two fluctuations are Bonnor-Ebert-unstable.  The surface pressures of the other fluctuations are low enough that the Bonnor-Ebert critical mass is larger than the actual fluctuation mass.

We can compare the stability criterion established by Bonnor and Ebert to the stability criterion from the virial equation.  We plot the absolute value of $\ddot{I}'$ against the ratio of the mass of the fluctuation to the Bonnor-Ebert critical mass in Fig. \ref{dynamic_measures}; the fluctuations marked by stars are the ones that are likely to collapse from virial arguments.  We can see that the fluctuations that are Bonnor-Ebert-unstable are also the most virial-unstable, although many fluctuations which are stable according to the Bonnor-Ebert criterion are unstable according to the virial equation.  We also note that the fluctuations with $\ddot{I}'<0$ (marked by stars) follow a distribution with a steeper slope than the fluctuations with $\ddot{I}'>0$ (marked by diamonds).

	Our fluctuation-finding method is sensitive to any local density fluctuation, but not all density maxima in our simulations would be detected in surveys for cores.  In observed maps of molecular clouds, cores are detected by virtue of having a substantial density contrast between centre and edge.  We examine the relationship between this density contrast and $m/m_\mathrm{BE}$ in Fig. \ref{density_ratio}.  A substantial fraction of fluctuations in the B runs have a significant density contrast between surface and center, but only a handful of fluctuations in the A runs have $\rho_c/\rho_s > 2$.  The region $\rho_c/\rho_s \ge 2$ is the three-dimensional analogue of what the \citet*{jijina99} data set would consider a core.  

\subsection{Virial masses}

	It is useful to compare the true mass of the fluctuations found in our simulations to the mass that would be predicted by standard observation techniques.  A common method used with molecular line observations is to assume a core is in virial equilibrium (such that $2K+W = 0$; this doesn't consider surface effects), and calculate the appropriate mass given the size and velocity dispersion of the core.  If one assumes that the cores are uniform spheres (as if often done, for example \citet{goodman93}; \citet*{jijina99}), then the virial mass is $m_\mathrm{vir} = \frac{5}{3}\frac{\sigma^2 R}{G}$, where $\sigma$ is the turbulent velocity dispersion.  We plot the relationship between this predicted virial mass and the true mass of our fluctuations in Fig. \ref{virialmass}.  
	For all four runs, this measure of the virial mass generally overestimates the mass of the fluctuations by two orders of magnitude; the average ratio of $m_\mathrm{vir}/m$ is $40.6$ for run A2, $66.4$ for run A5, $211.9$ for run B2, and $178.2$ for run B5.  Furthermore, we find that the relationship between $m_\mathrm{vir}$ and $m$ is approximately a straight line on a log-log plot, but with a slope less than one ($0.53$ for runs A2 and A5, $0.37$ for run B2, and $0.45$ for run B5).

	Observational comparisons between the mass estimated by virial techniques and masses estimated from other methods also find that the assumption of virial equilibrium leads to masses larger than that measured by other methods.  A study by \citet{vandertak00} found that $m_\mathrm{vir} \approx 2.77 m_\mathrm{other}$, and that the slope of the $\log(m_\mathrm{vir})-\log(m_\mathrm{other})$ plot to be slightly less than 1.0.  Our results suggest a much larger discrepancy between true mass and virial-estimated mass.

	The discrepancy between $m_\mathrm{vir}$ and $m$ is likely due to three effects.  First, many of the fluctuations in our simulations are not in equilibrium; turbulent motions play a large role in the dynamics of these fluctuations.  This is especially true in the B runs, which have more Jeans masses on the computational grid.  Second, the simple virial mass estimator neglects the effects of the surface terms, which as we have already shown are dynamically significant, especially for low-mass fluctuations.  Finally, for cores that are centrally condensed, the gravitational energy will be greater than that of a uniform core.  This will reduce the virial mass estimate, as less mass is needed for there to be virial equilibrium between kinetic and gravitational energies.

\section{PHYSICAL PROPERTIES OF SIMULATED CORES}\label{structure}
\subsection{Global images and line maps}

\begin{figure*}
\begin{center}
\includegraphics[width=168mm]{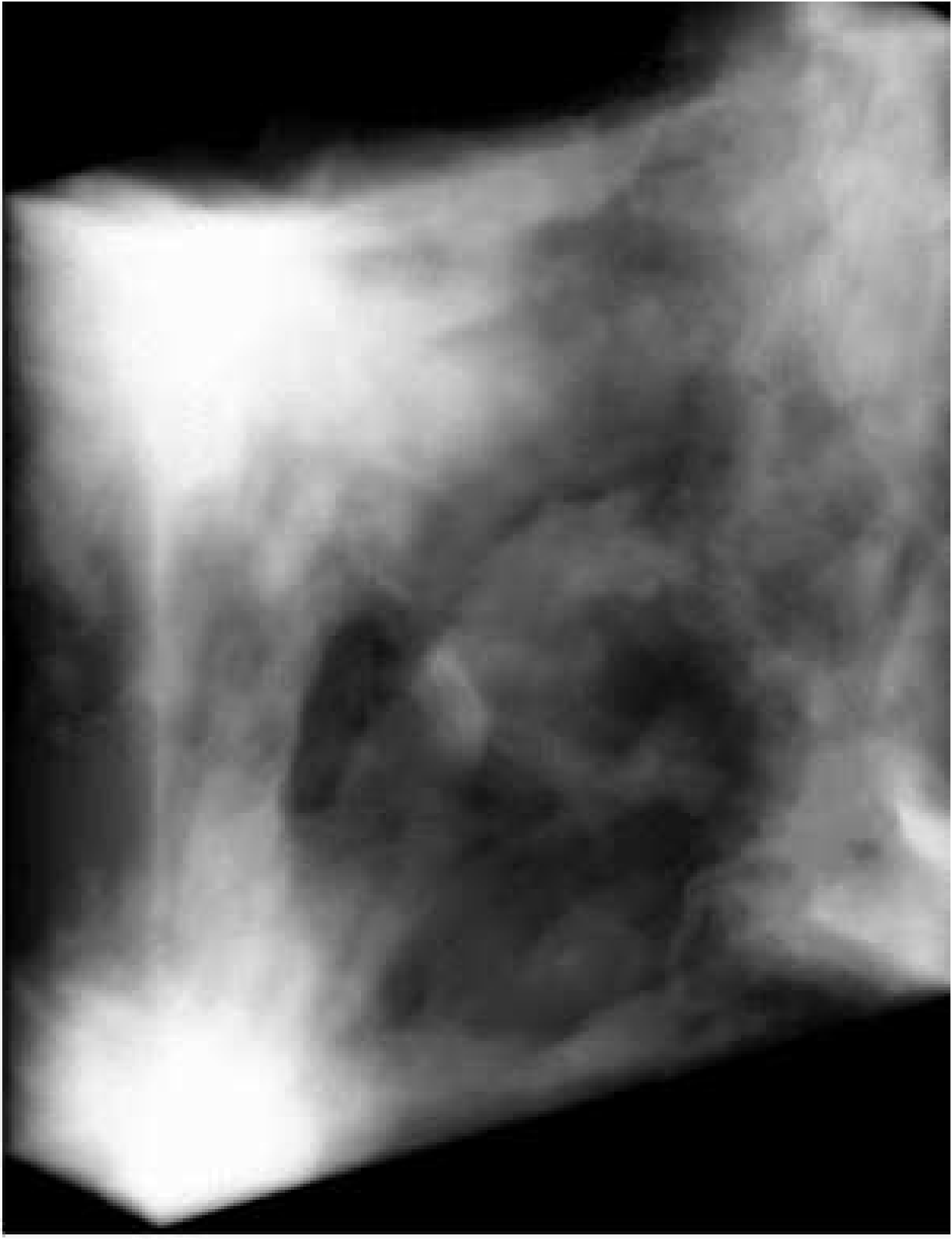}
\caption{Volume rendering of the density in a 256x256x32 slab from one of our simulation runs, B5.\label{volrend}}
\end{center}
\end{figure*}

\begin{figure*}
\begin{center}
\includegraphics[width=84mm]{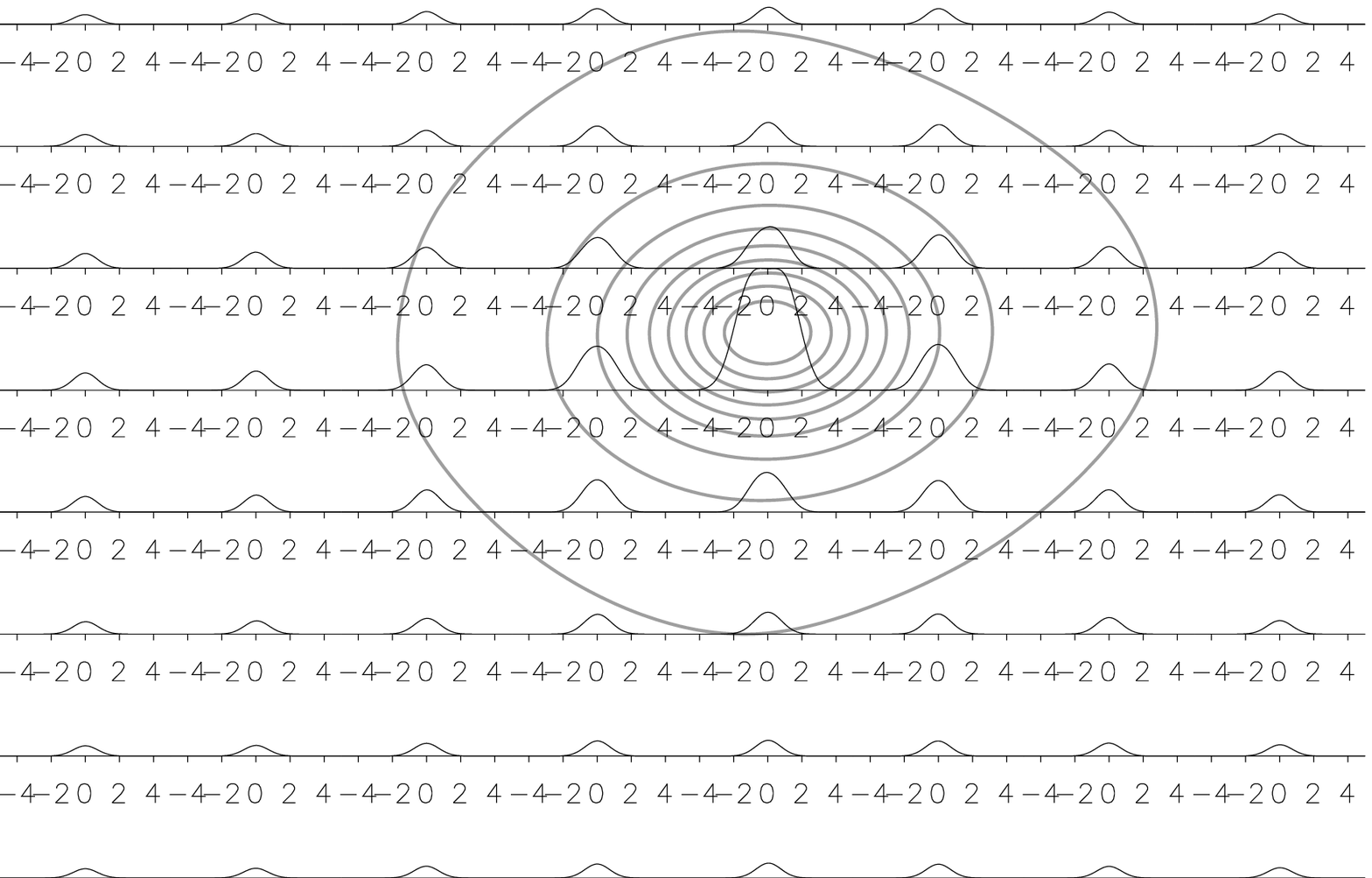}\hspace{5mm}\includegraphics[width=84mm]{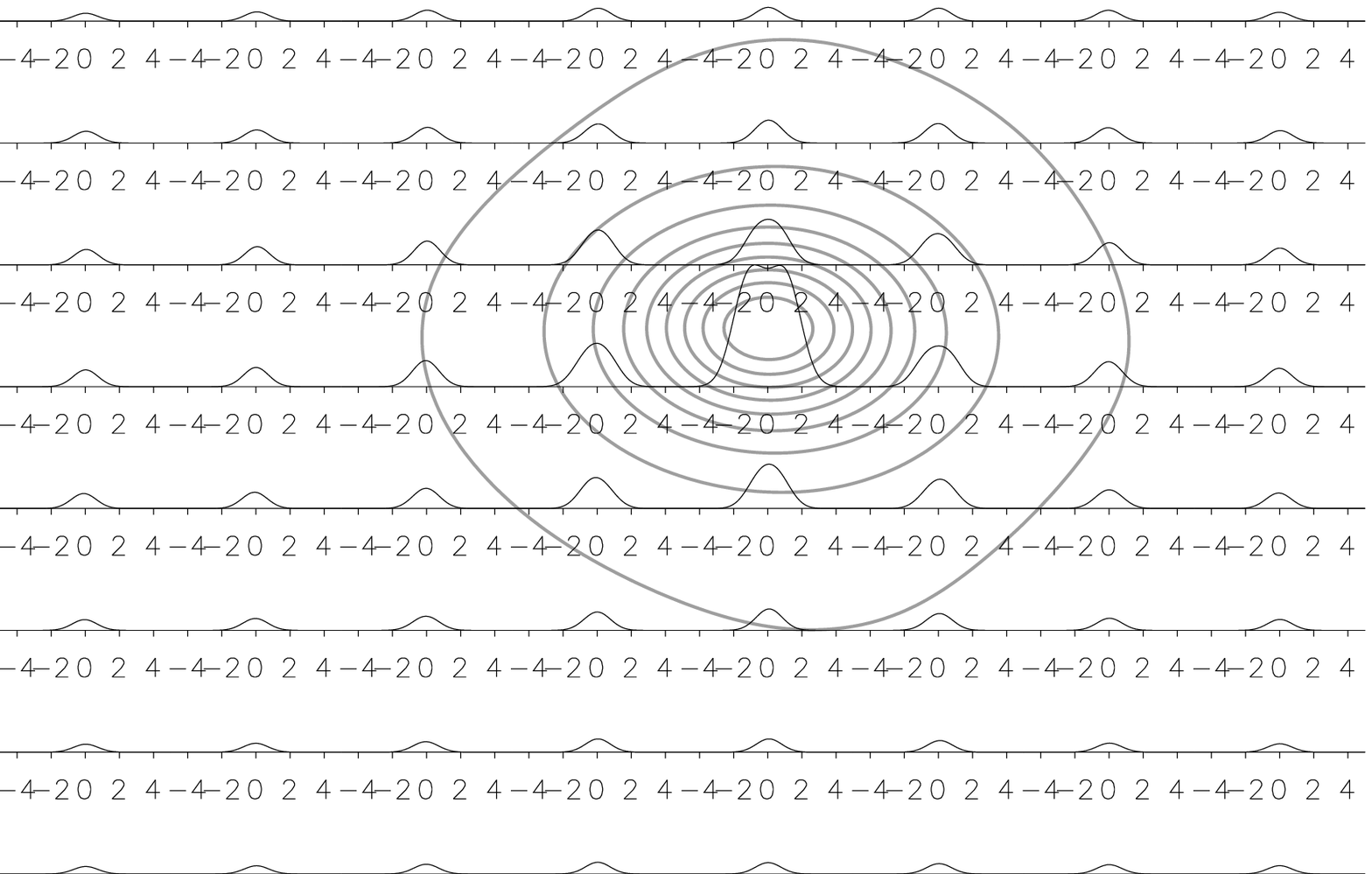}\\\vspace{10mm}
\includegraphics[width=84mm]{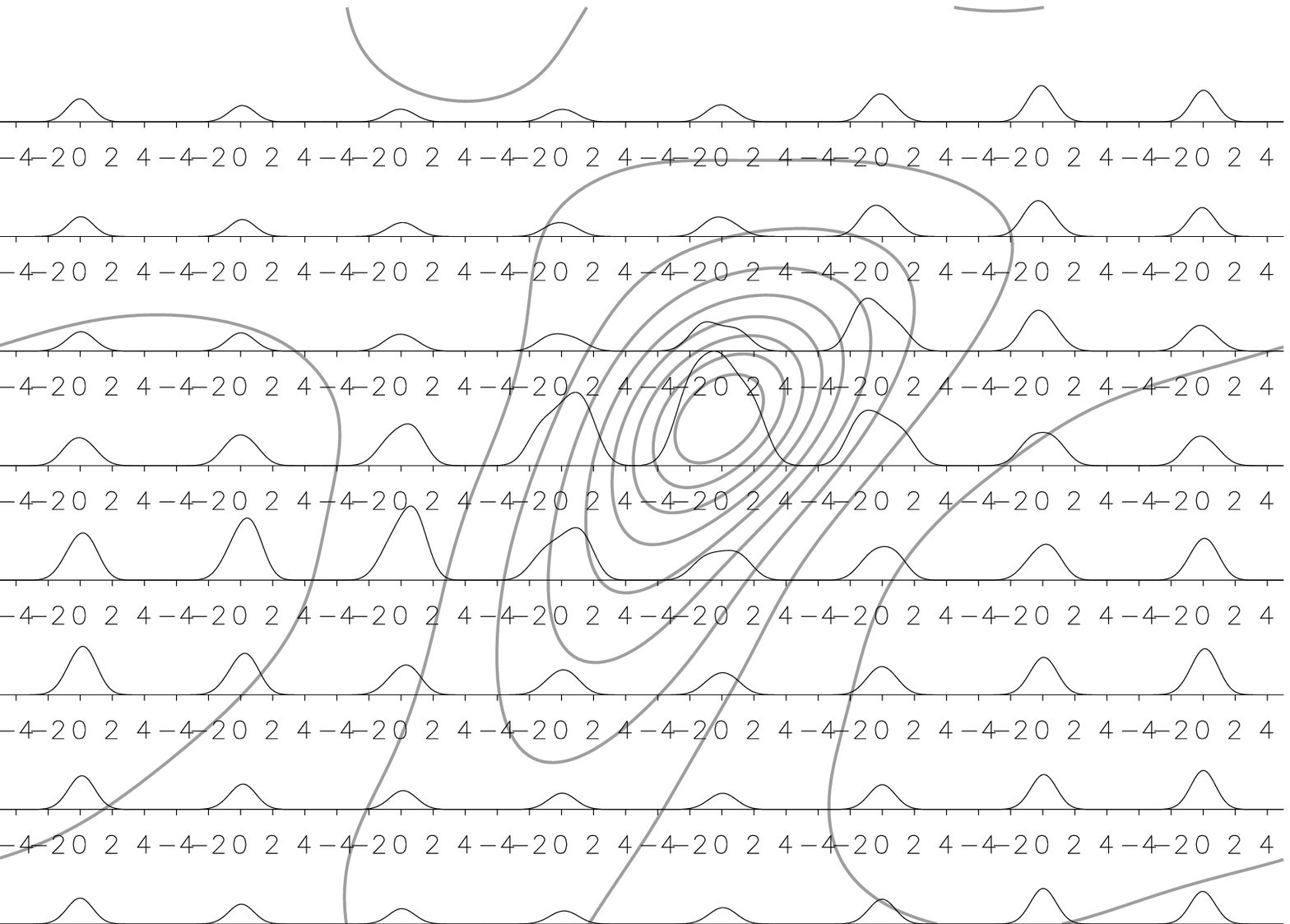}\hspace{5mm}\includegraphics[width=84mm]{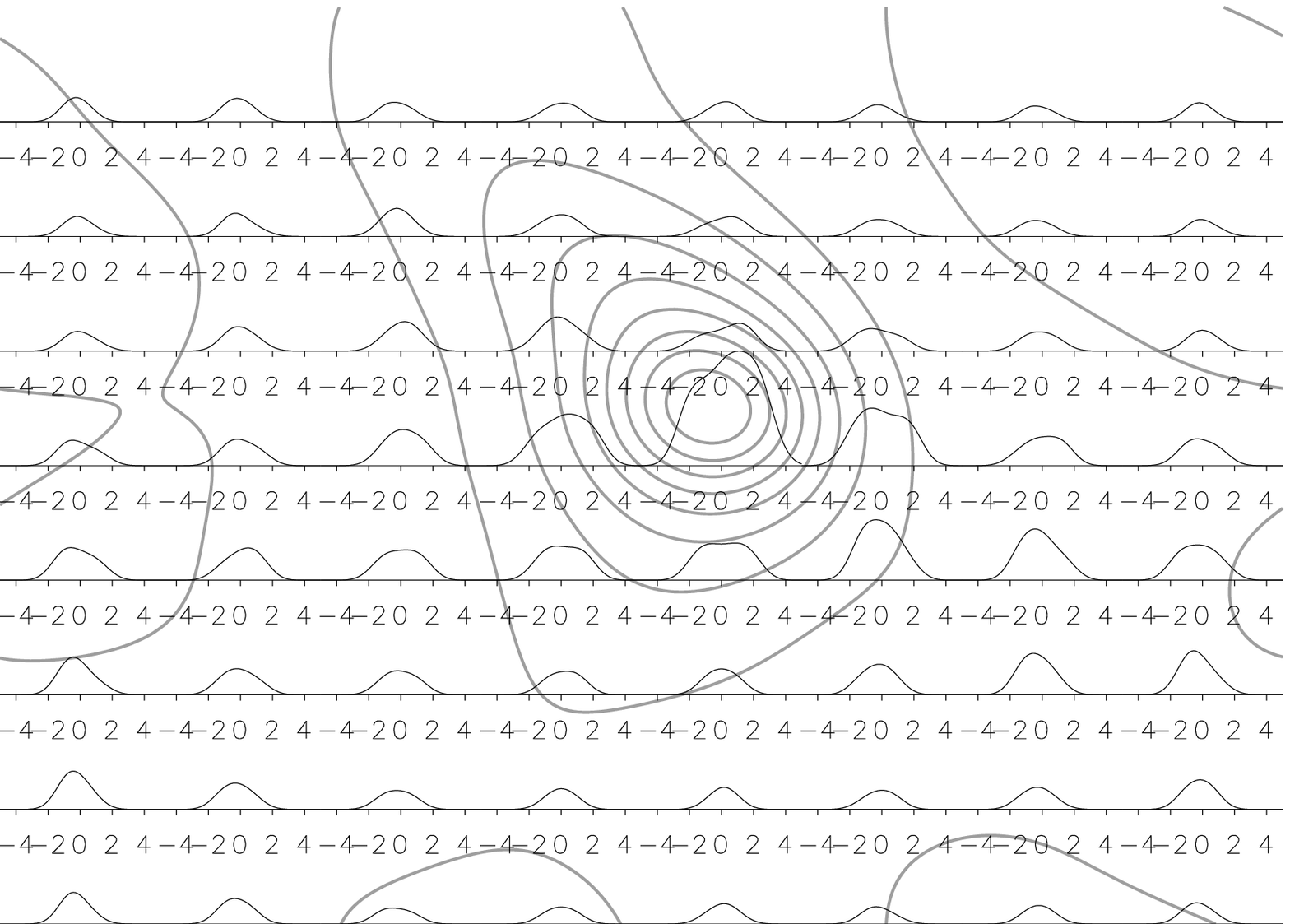}
\caption{Line profiles and contour maps of the column density of our simulations.  The abscissa on each of the line profiles is in units of the Mach number.  The line profiles in each 32x32 pixel block are binned together; the column density map has been convolved with a Gaussian of FWHM of 32 pixels as well, to simulate a typical observation.  Note that the data has been shifted so that the peak in the column density is located at the centre of the image.  The contours are at (10,20,30,40,50,60,70,80,90) per cent of the maximum column density value.\label{cdimages}}
\end{center}
\end{figure*}

\begin{figure*}
\begin{center}
\includegraphics[width=84mm]{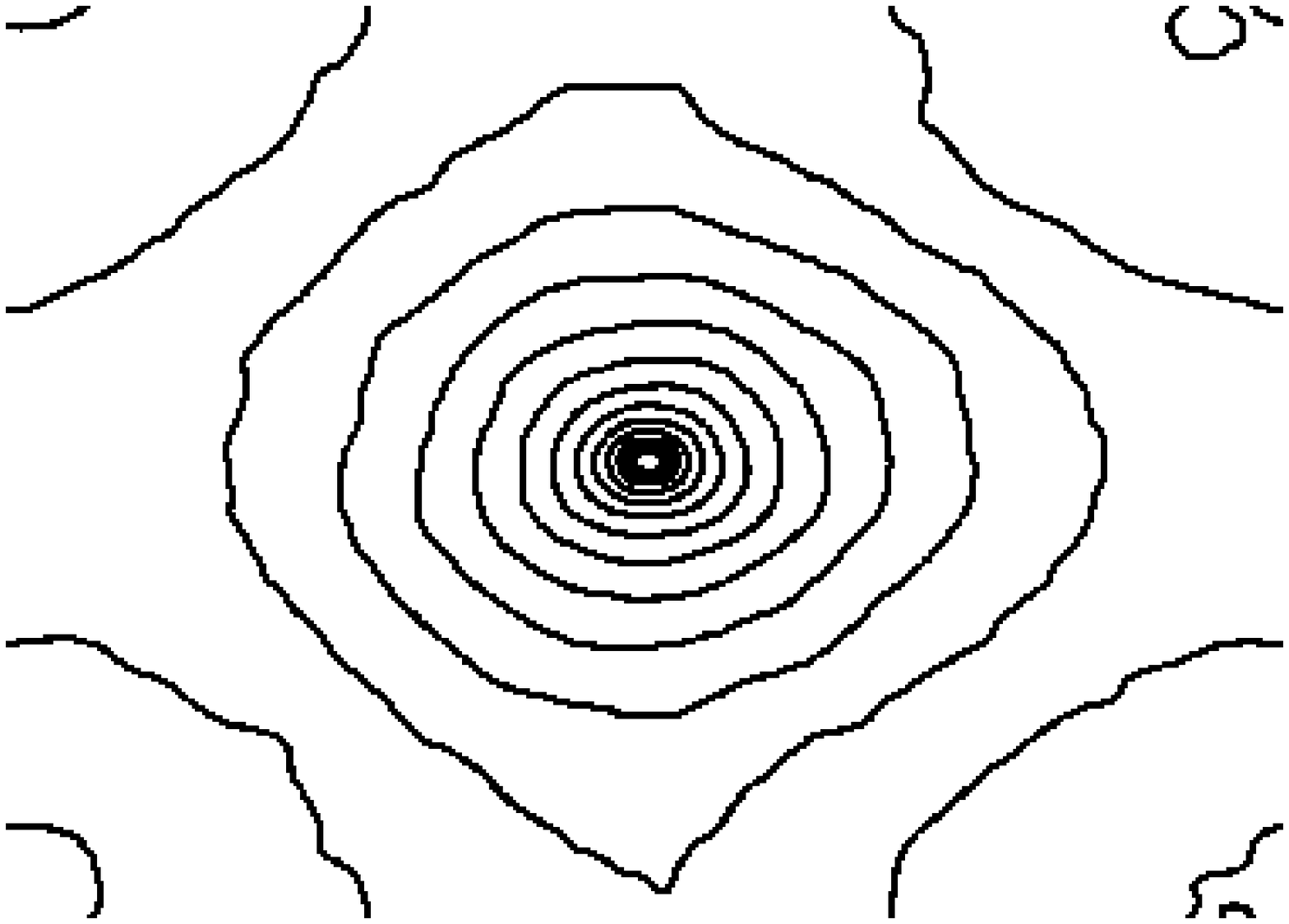}\hspace{5mm}\includegraphics[width=84mm]{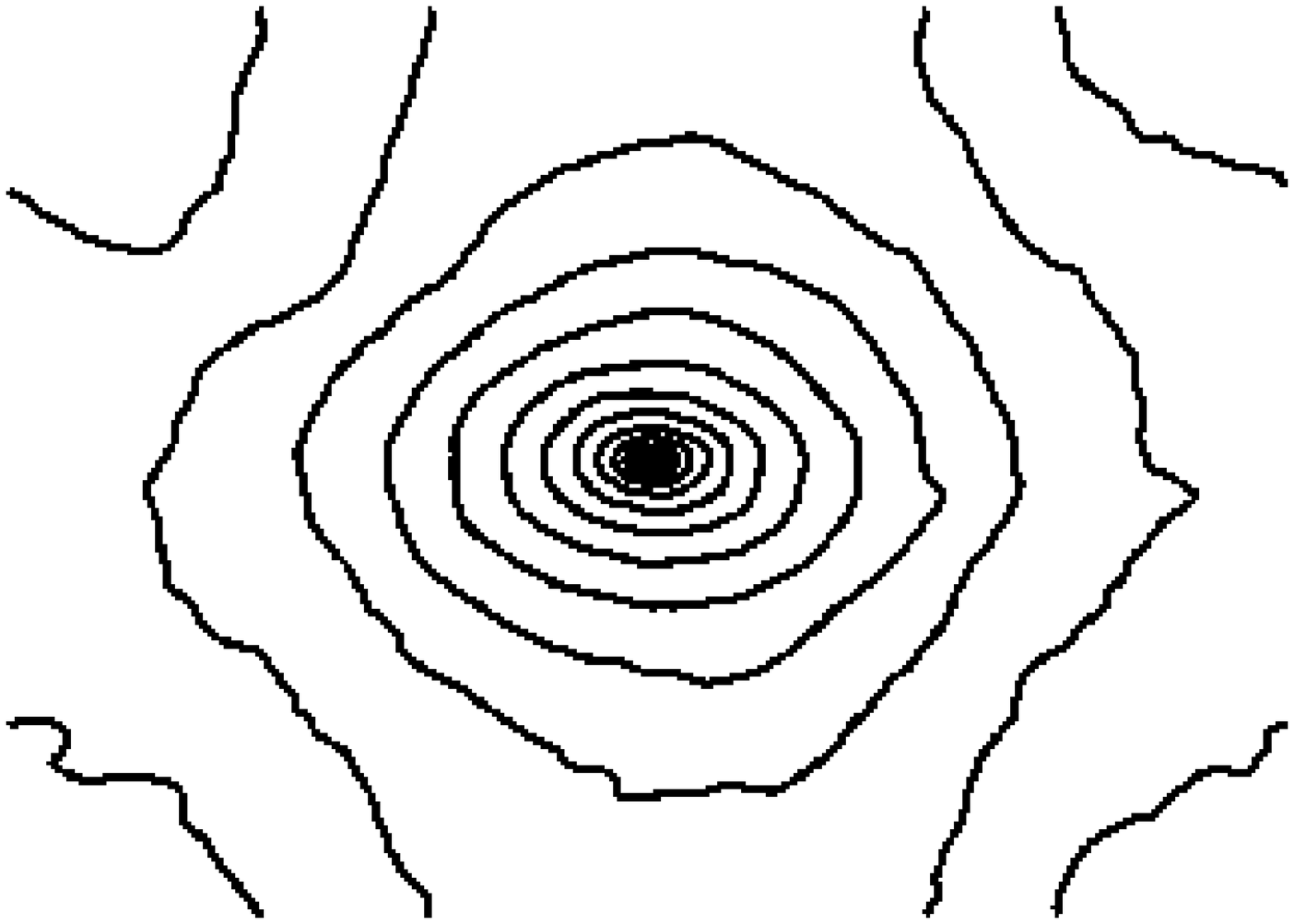}\\\vspace{10mm}
\includegraphics[width=84mm]{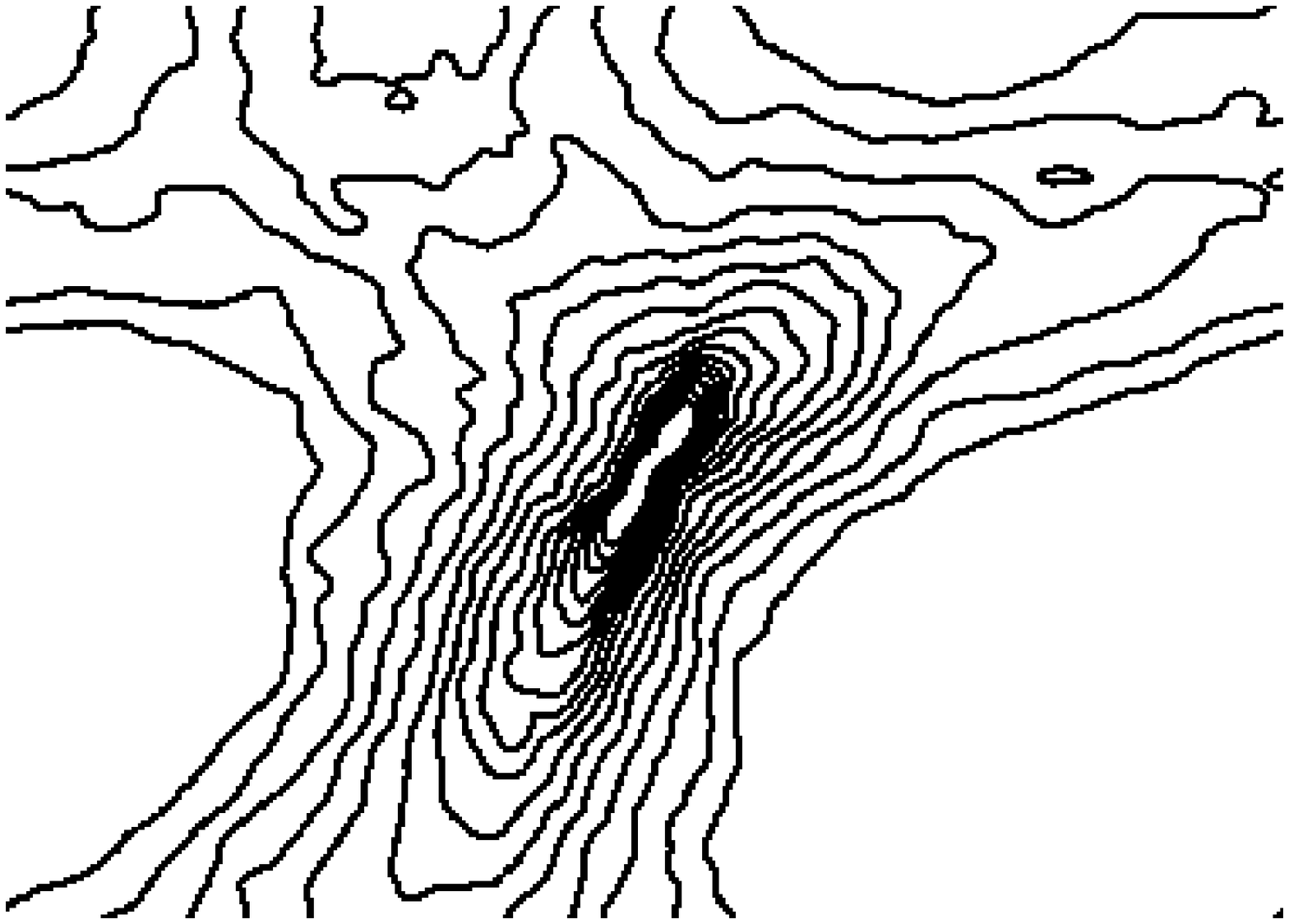}\hspace{5mm}\includegraphics[width=84mm]{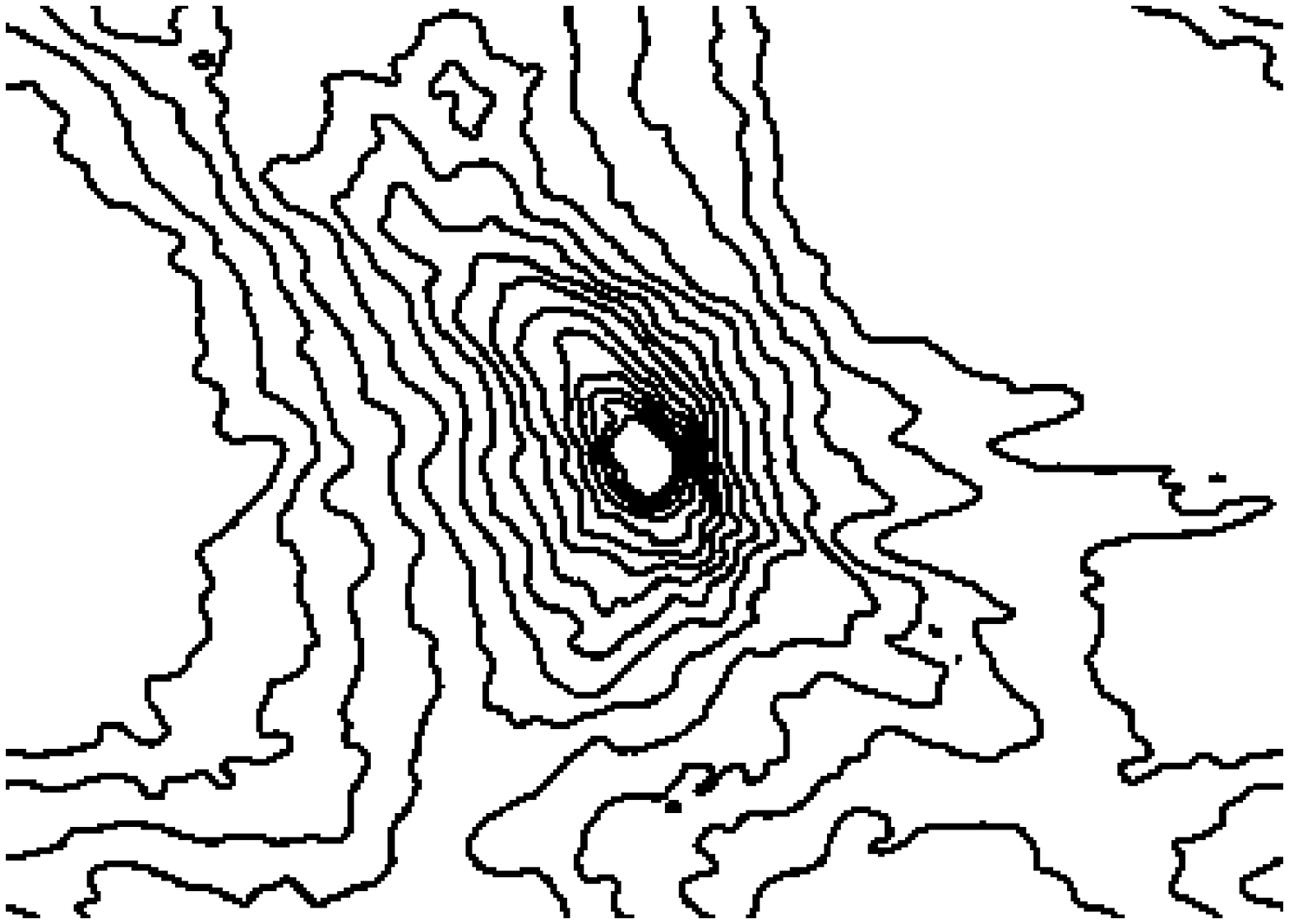}
\caption{Unconvolved contour maps of the column density of our simulations.  As in Fig. \ref{cdimages}, we have shifted the data so that the peak of the column density distribution is in the centre of the image.  The contour levels are logarithmically spaced, starting at a column density of $2.4 \times 10^{23} \;\mathrm{cm}^{-2}$ for the A runs (when scaled to the 0.1 pc size in Table \ref{table_sim_list}) and a column density of $1.9\times 10^{23}\;\mathrm{cm}^{-2}$ for the B runs (when scaled to the 0.32 pc size in Table \ref{table_sim_list}), and increasing a factor of 1.5 with each successive contour level.\label{cdimages2}}
\end{center}
\end{figure*}

\begin{figure*}
\begin{center}
\includegraphics[width=84mm]{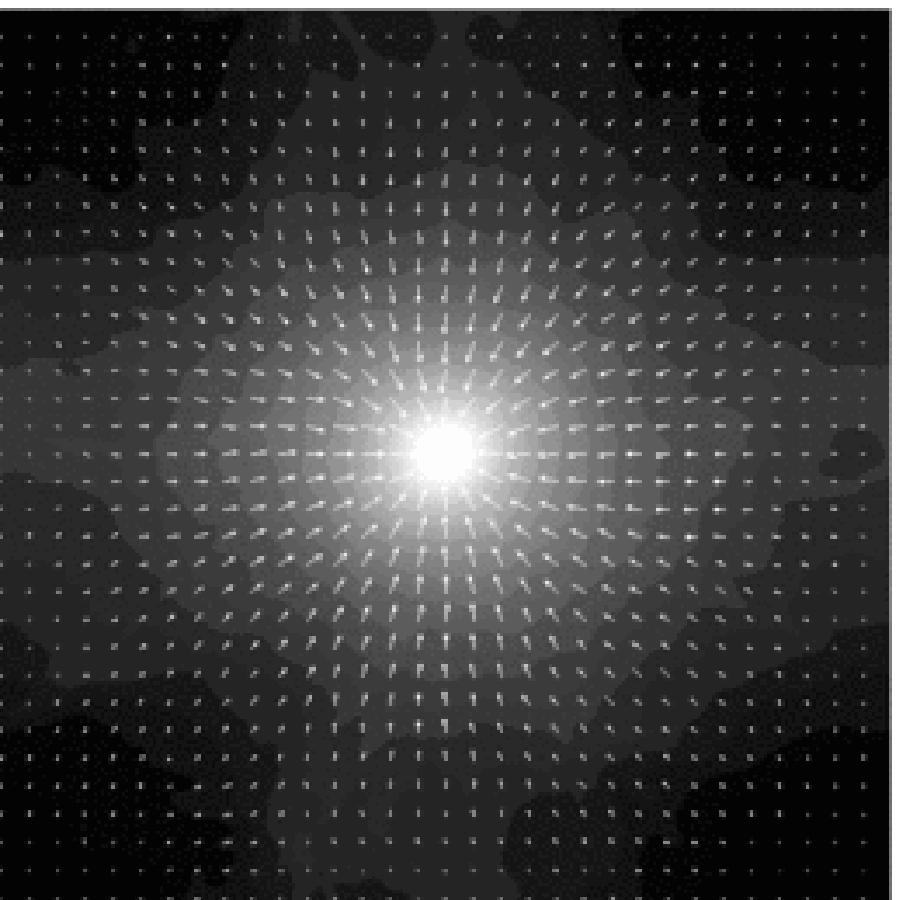}\hspace{5mm}\includegraphics[width=84mm]{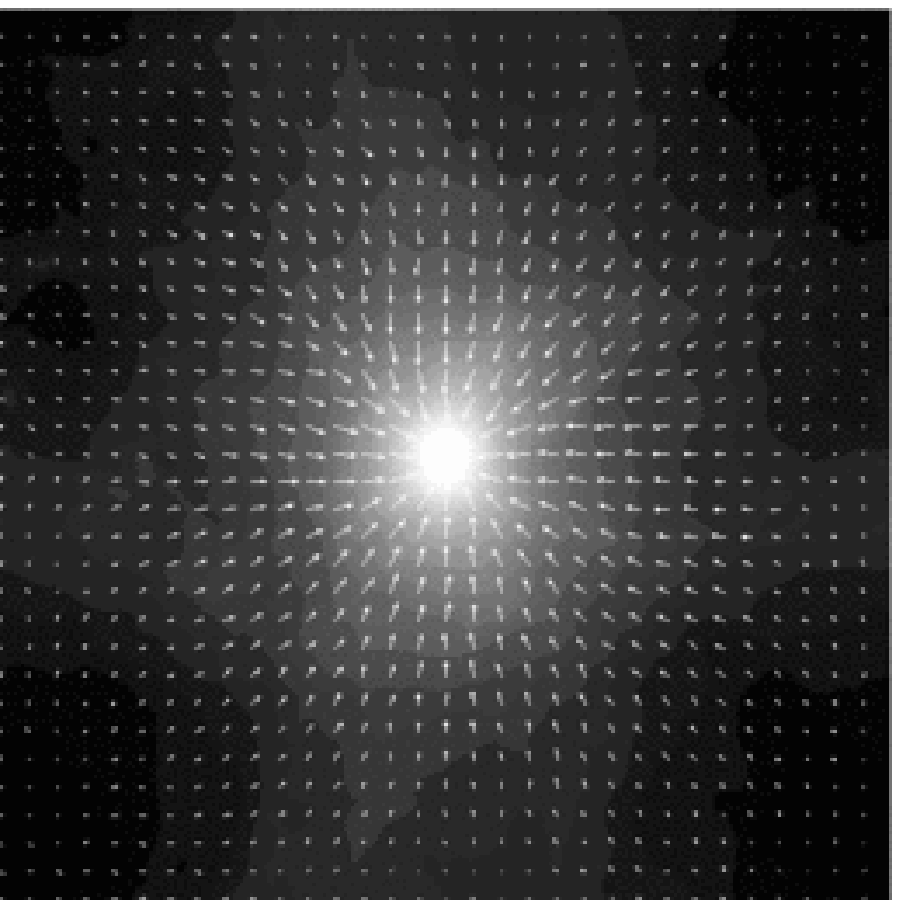}\\\vspace{10mm}
\includegraphics[width=84mm]{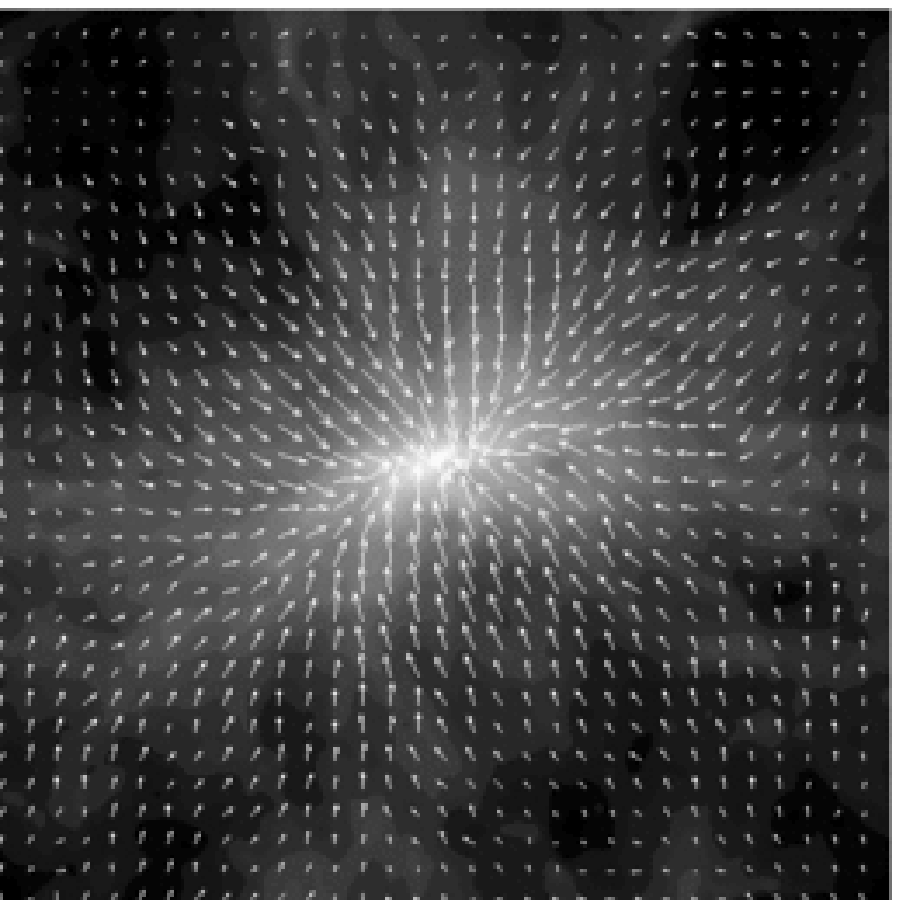}\hspace{5mm}\includegraphics[width=84mm]{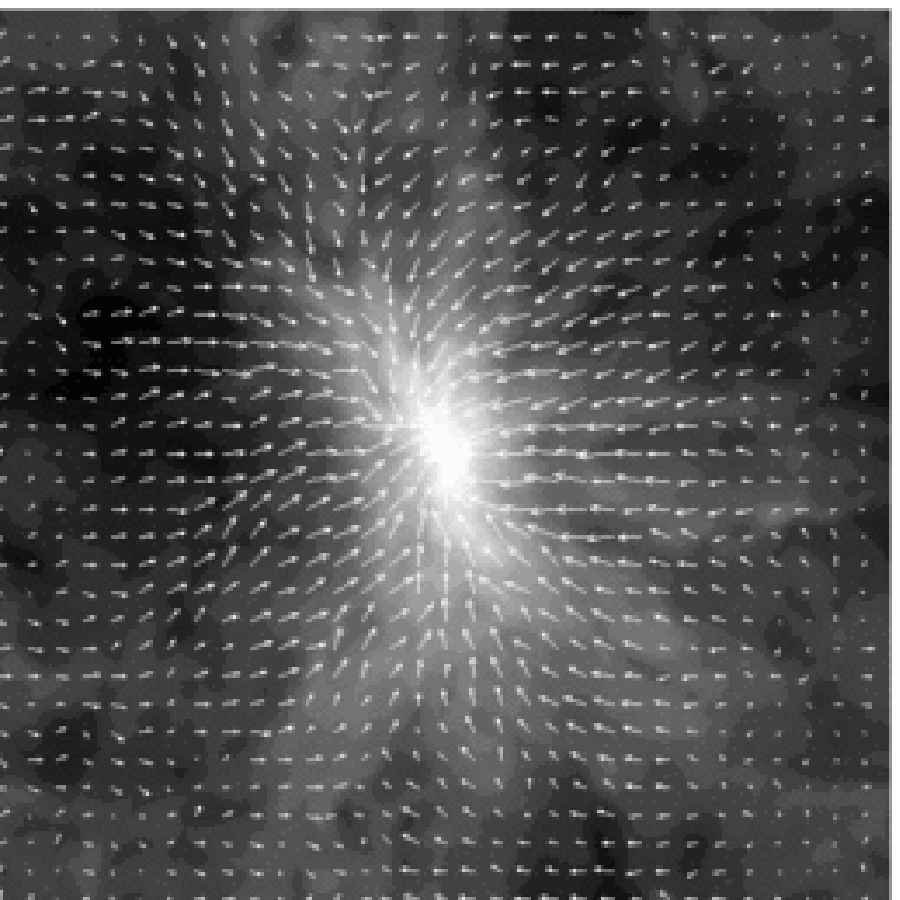}
\caption{A slice through the centre of the most massive core in each of the simulation runs.  The greyscale contours show the density distribution; the arrows represent the velocity flow in the plane of the image.  The length of the velocity vectors are in proportion to the speed of the fluid flow, normalized such that a vector with a length equal to the distance between the tails of the vectors has a speed of Mach 2.\label{figslice}}
\end{center}
\end{figure*}

\begin{figure*}
\begin{center}
\includegraphics[width=84mm]{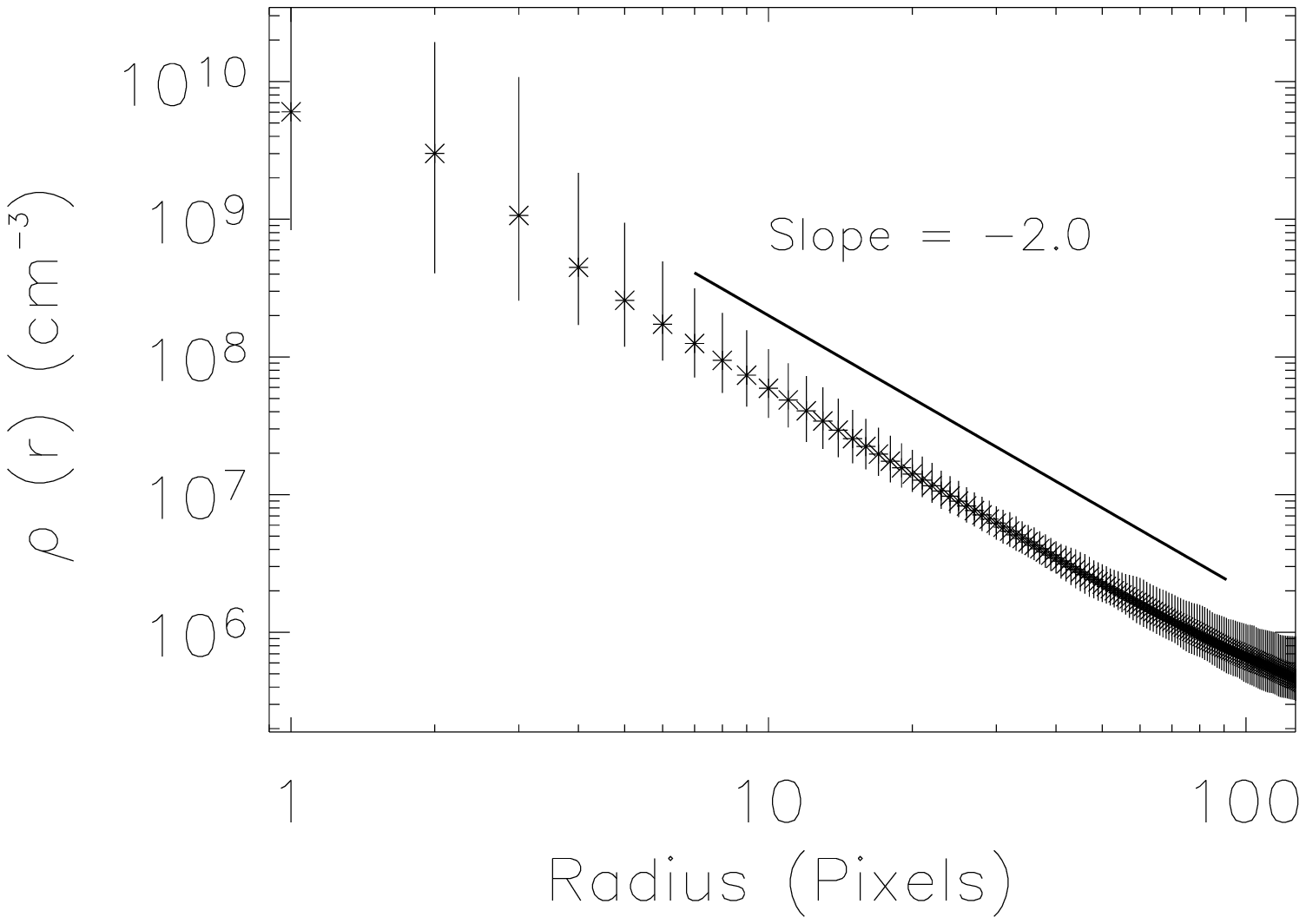}\includegraphics[width=84mm]{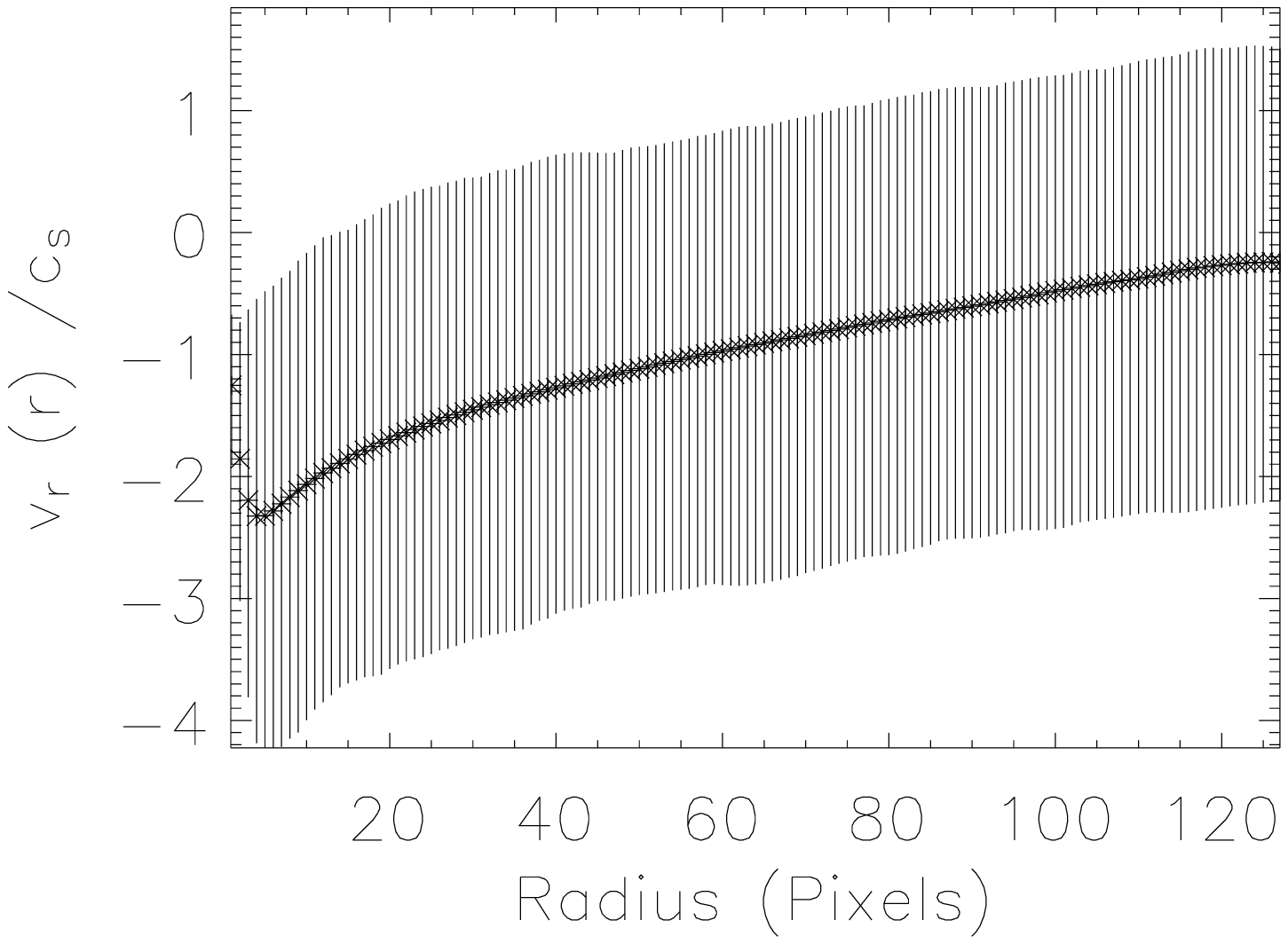} \\
\includegraphics[width=84mm]{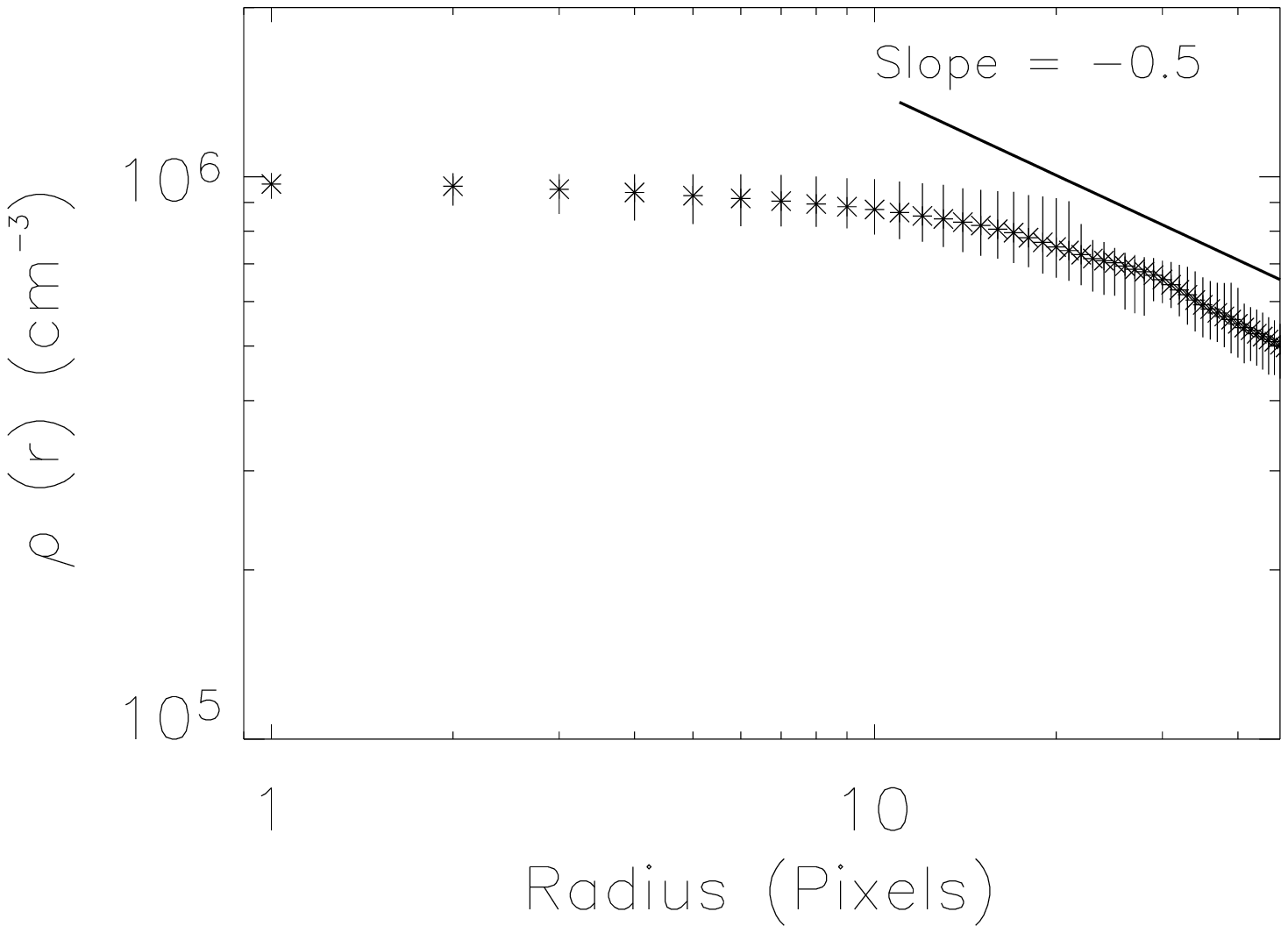}\includegraphics[width=84mm]{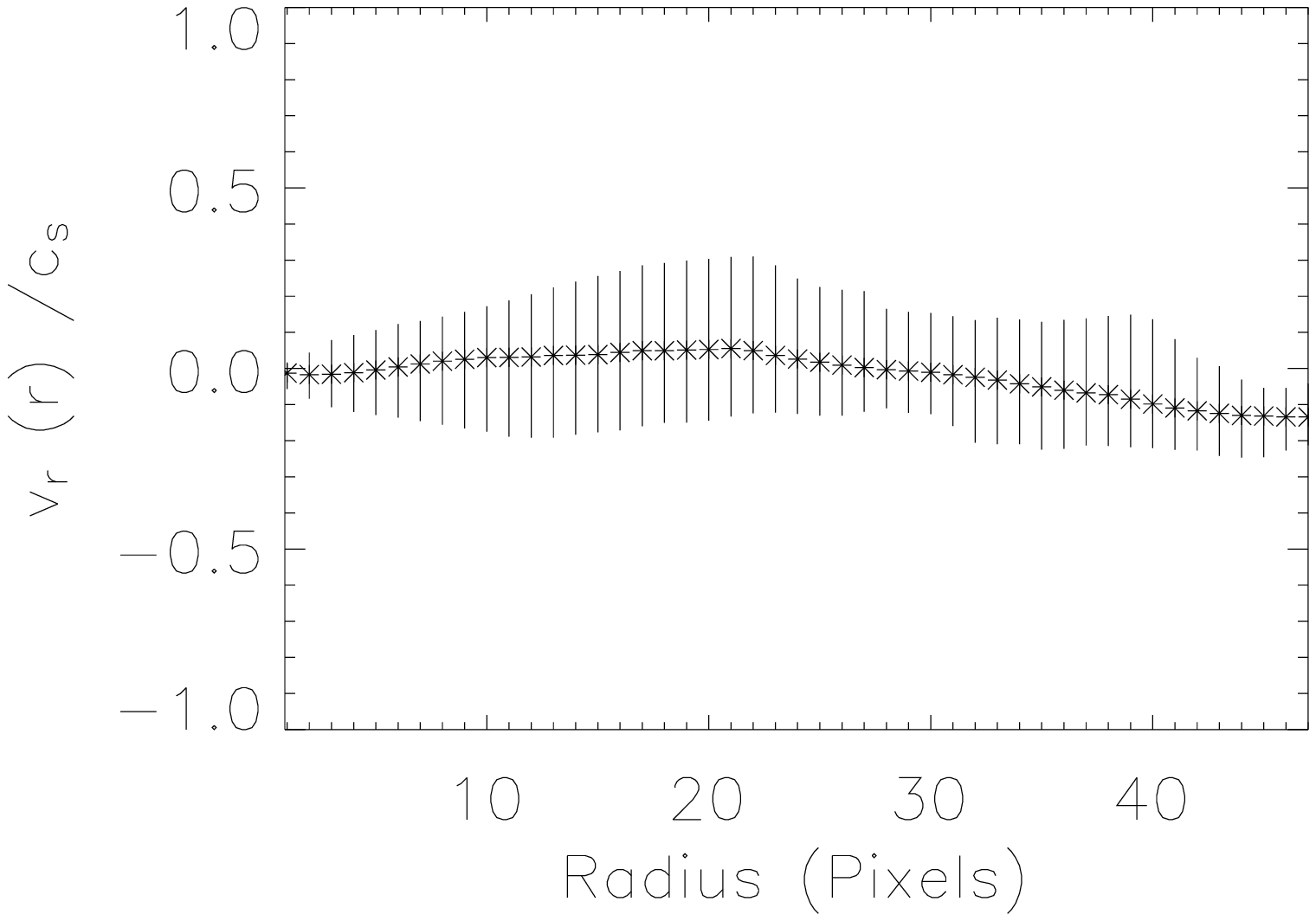} \\
\includegraphics[width=84mm]{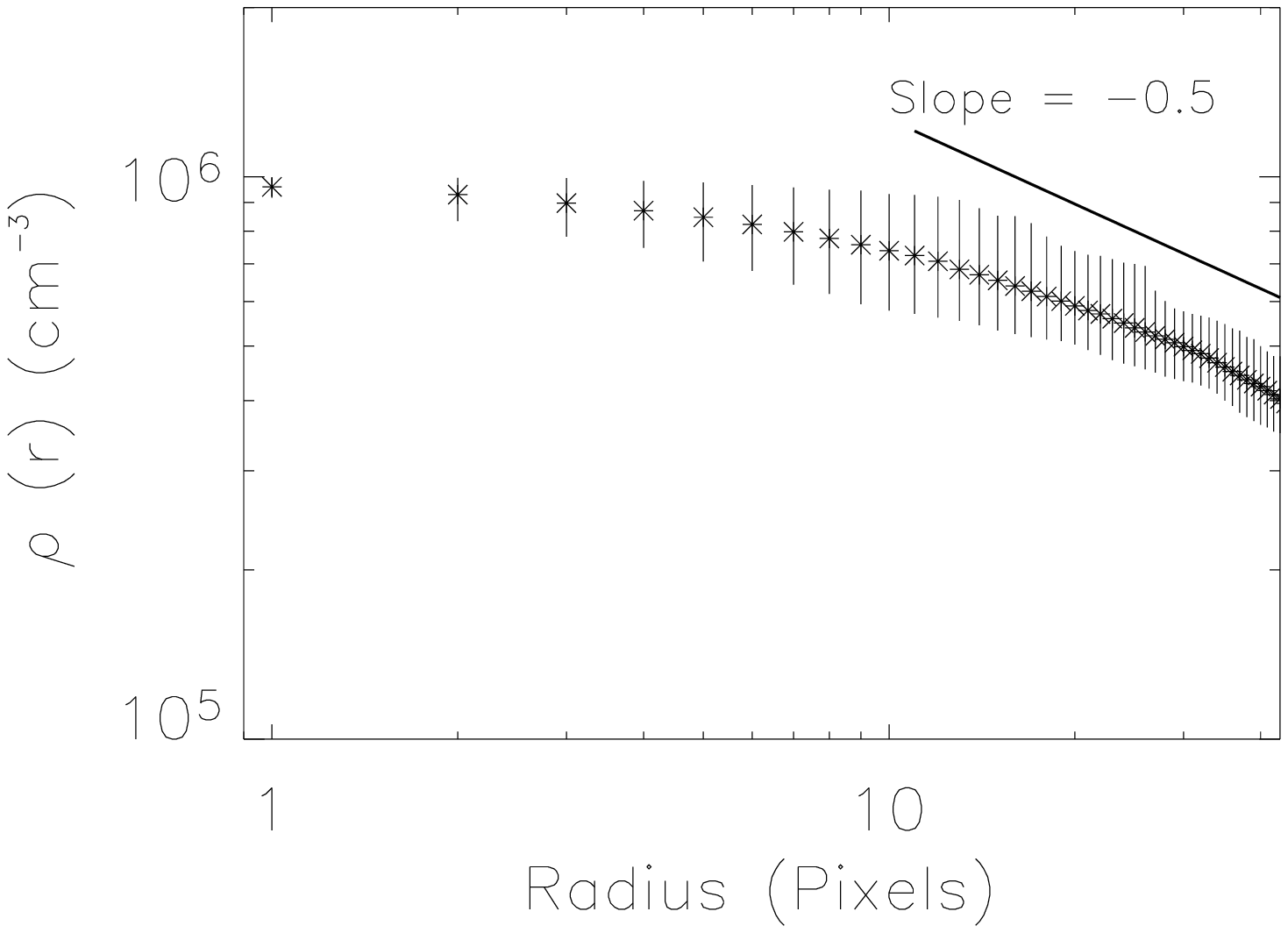}\includegraphics[width=84mm]{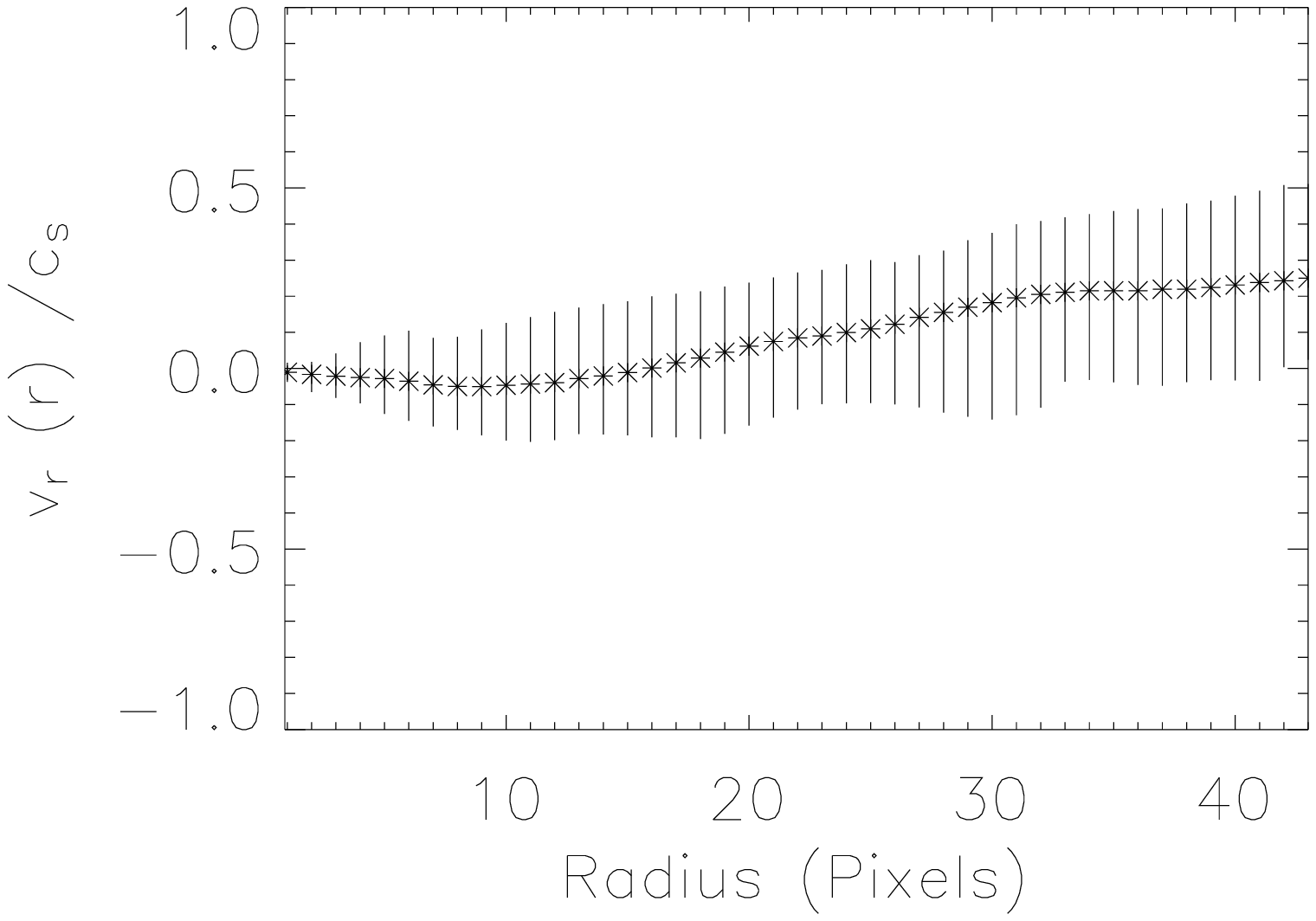}
\caption{Density and infall velocity plots for run A2.  The most massive core is on the top, followed by the second and third most massive.  The first core is bound according to the criterion in Section \ref{subsection_virial_stability}; the second and third cores are both unbound.  The abscissa is in units of pixels; for a box size of $0.1$ pc (from Table \ref{table_sim_list}), one pixel corresponds to $81$ AU.\label{dvrprof_A2}}
\end{center}
\end{figure*}

\begin{figure*}
\begin{center}
\includegraphics[width=84mm]{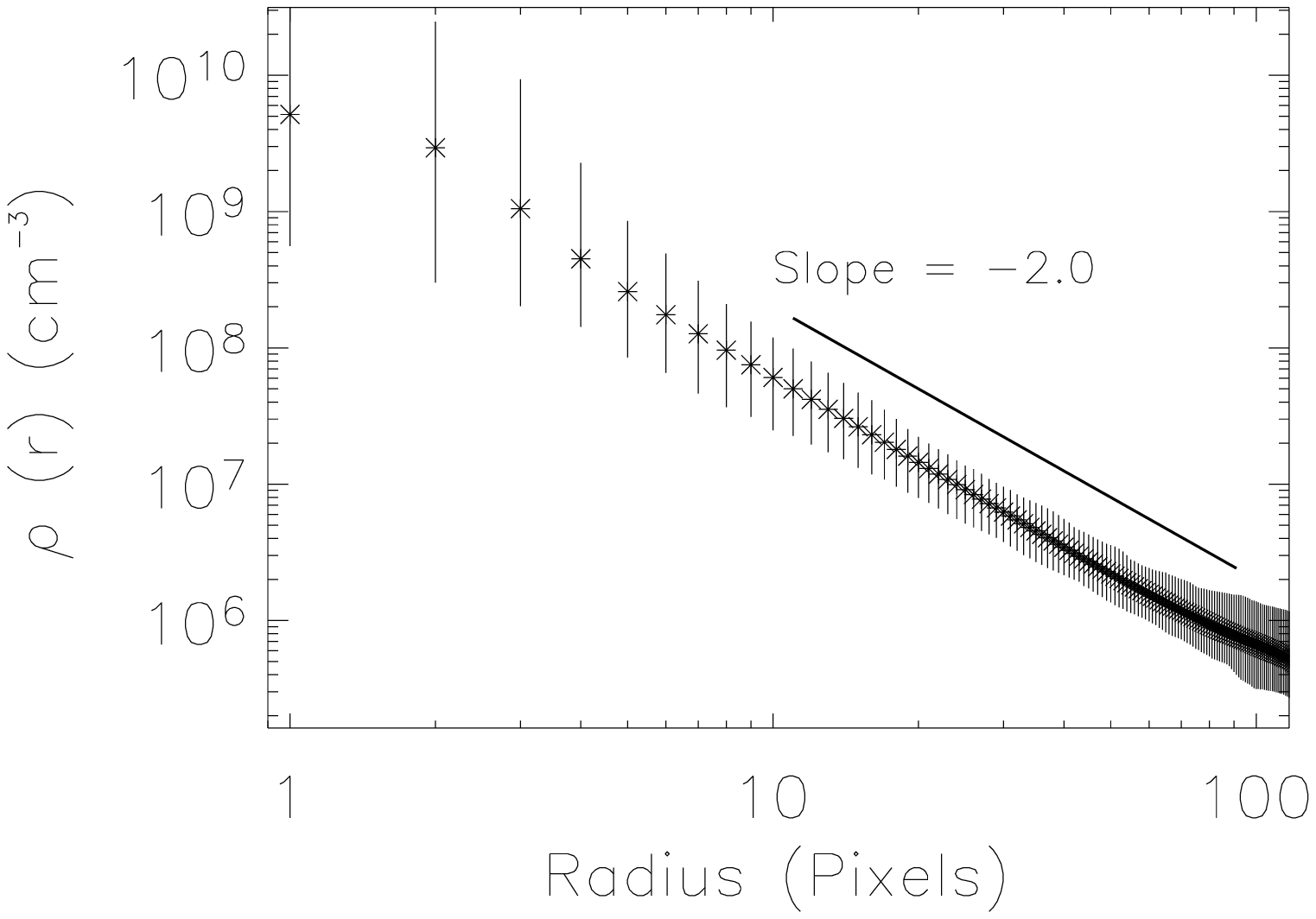}\includegraphics[width=84mm]{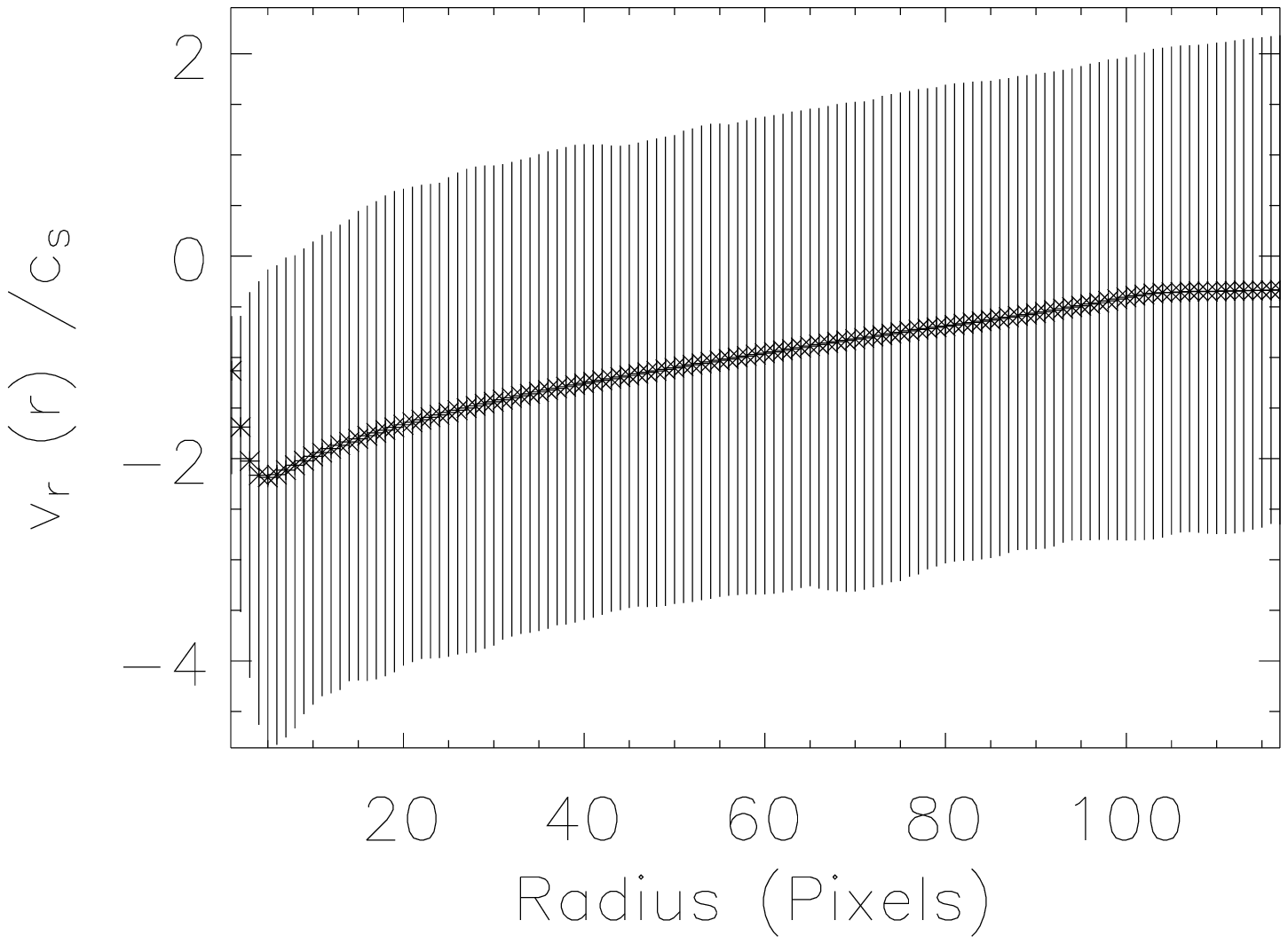} \\
\includegraphics[width=84mm]{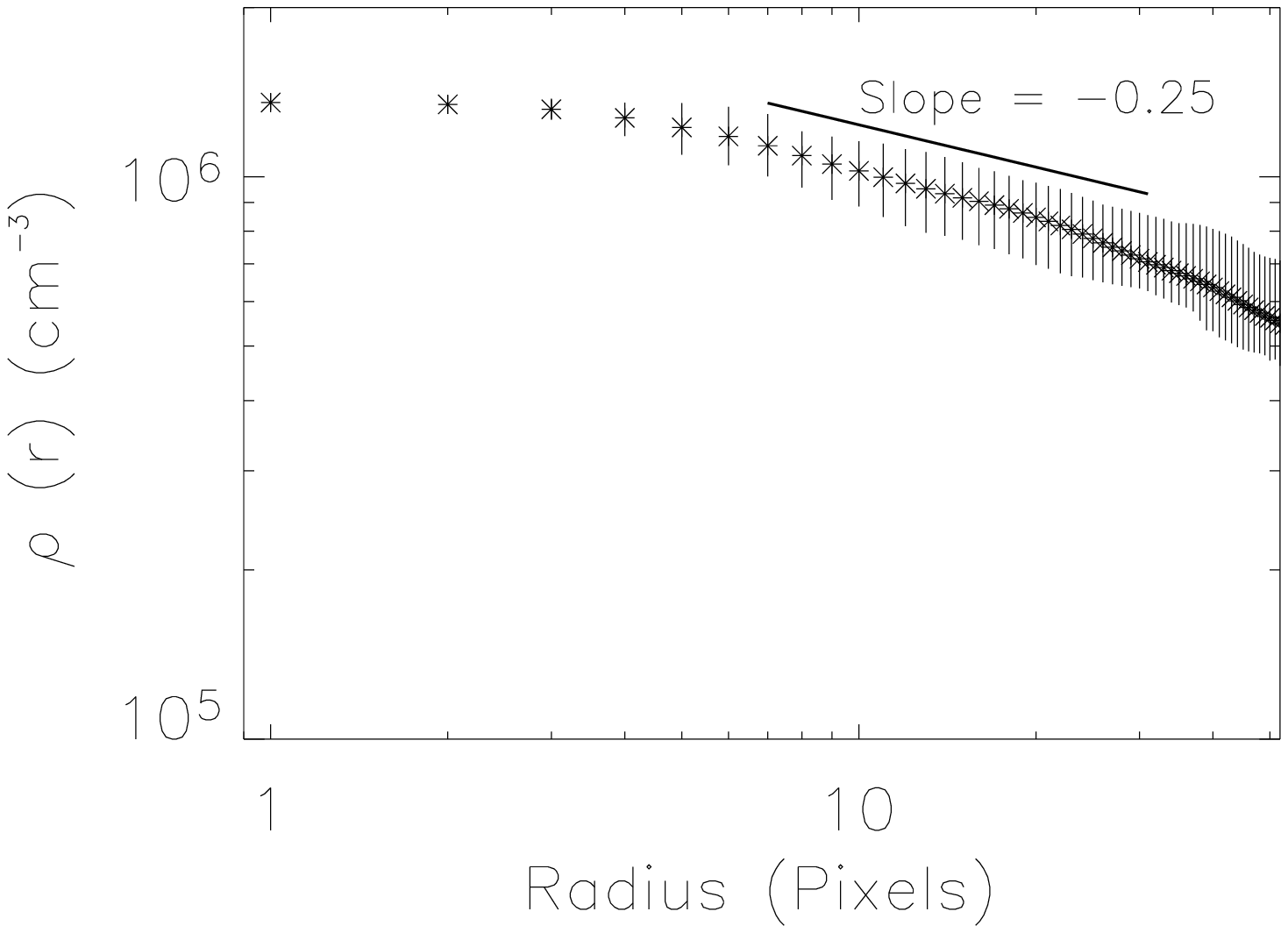}\includegraphics[width=84mm]{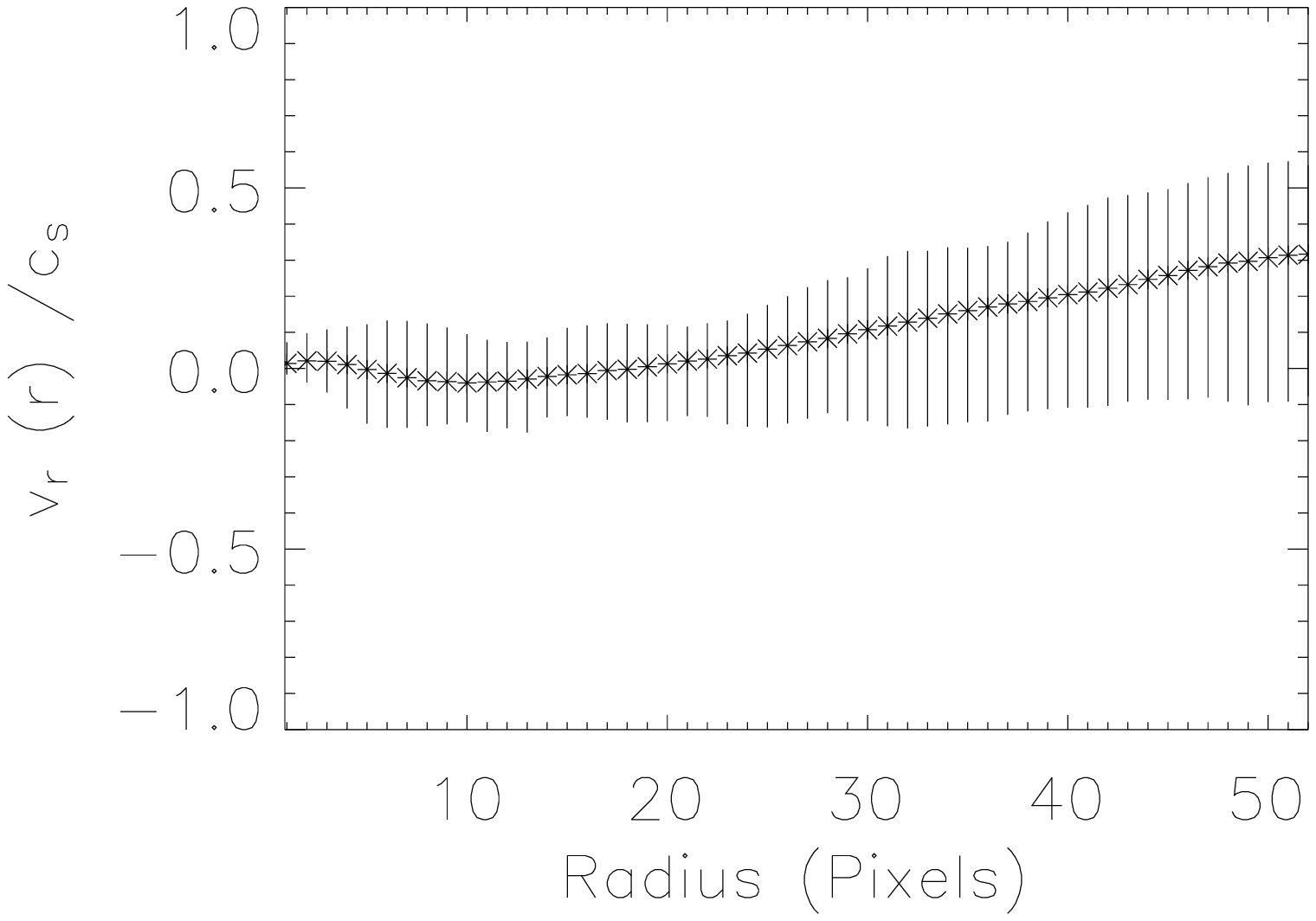} \\
\includegraphics[width=84mm]{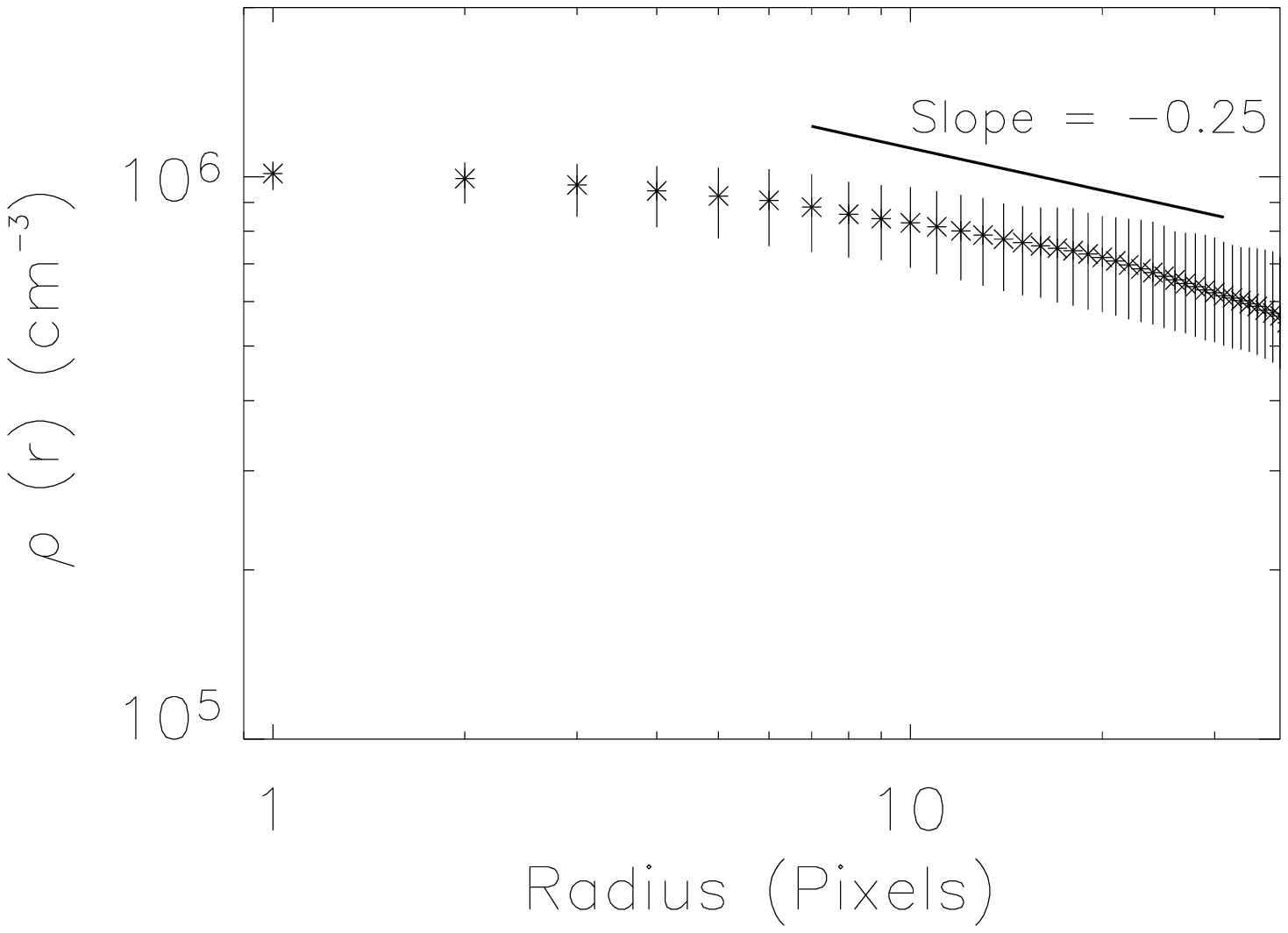}\includegraphics[width=84mm]{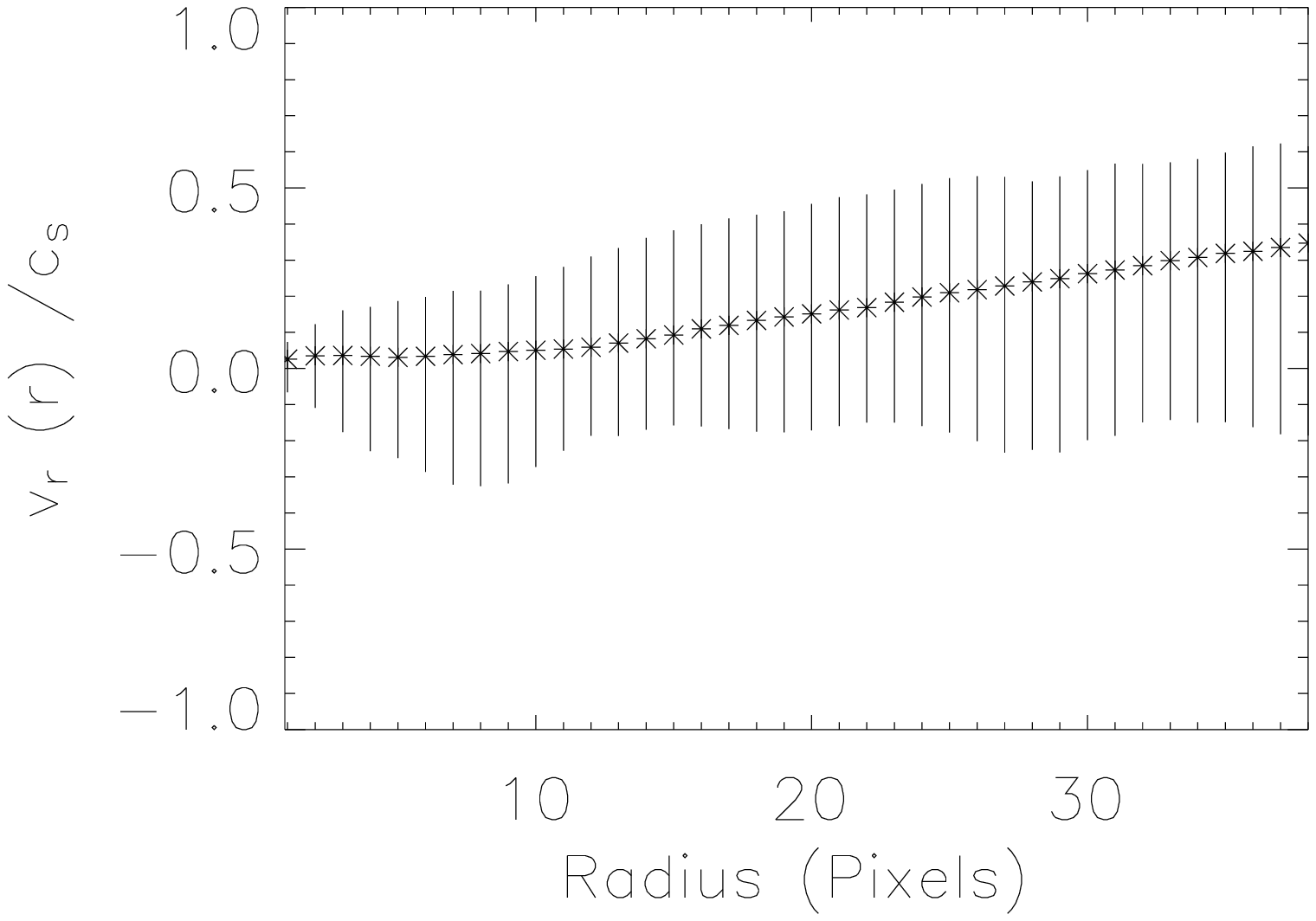}
\caption{Density and infall velocity plots for run A5.  The most massive core is on the top, followed by the second and third most massive. The first and second cores are bound according to the criterion in Section \ref{subsection_virial_stability}; the third core is unbound.  The abscissa is in units of pixels; for a box size of $0.1$ pc (from Table \ref{table_sim_list}), one pixel corresponds to $81$ AU.\label{dvrprof_A5}}
\end{center}
\end{figure*}

\begin{figure*}
\begin{center}
\includegraphics[width=84mm]{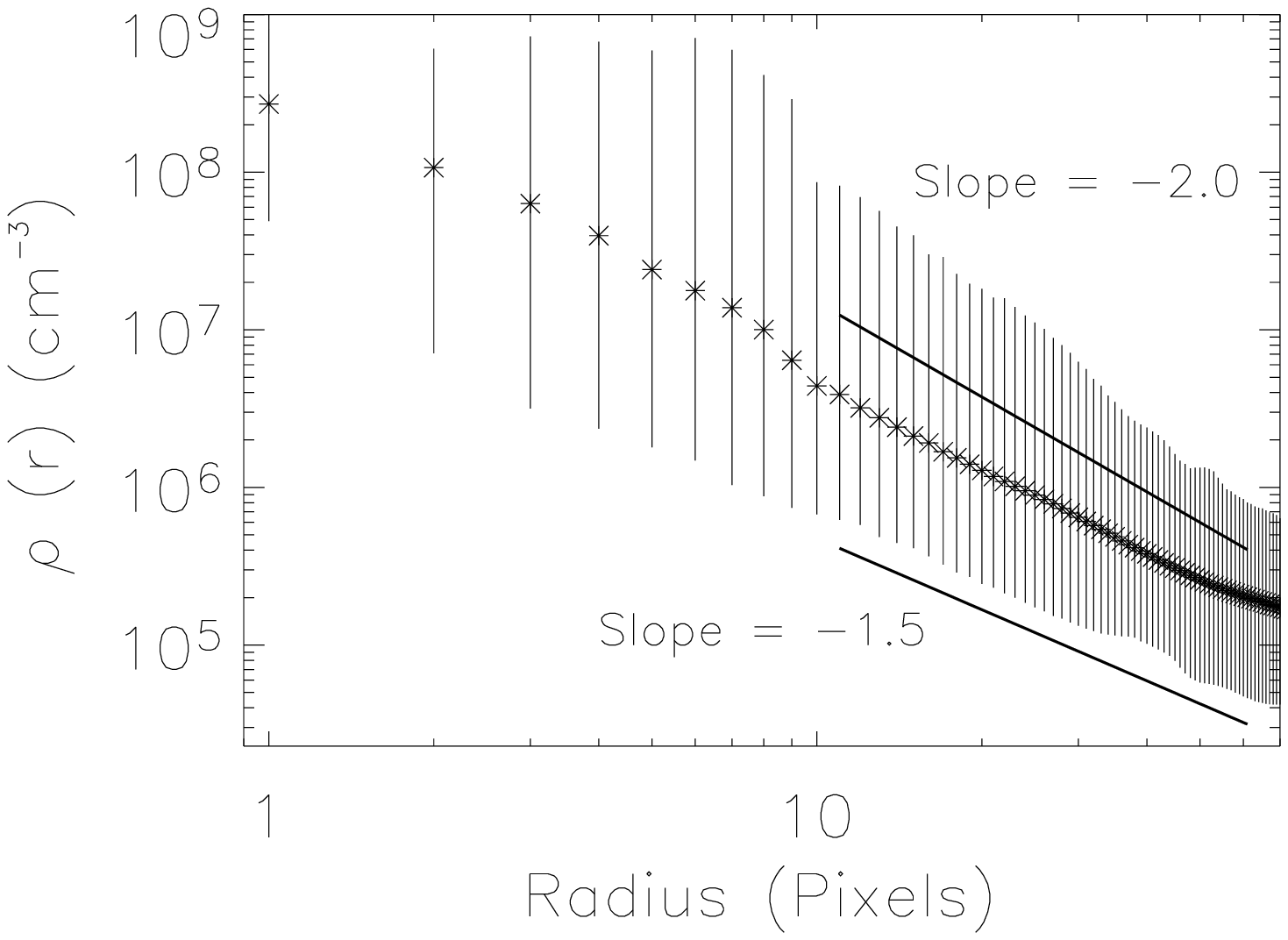}\includegraphics[width=84mm]{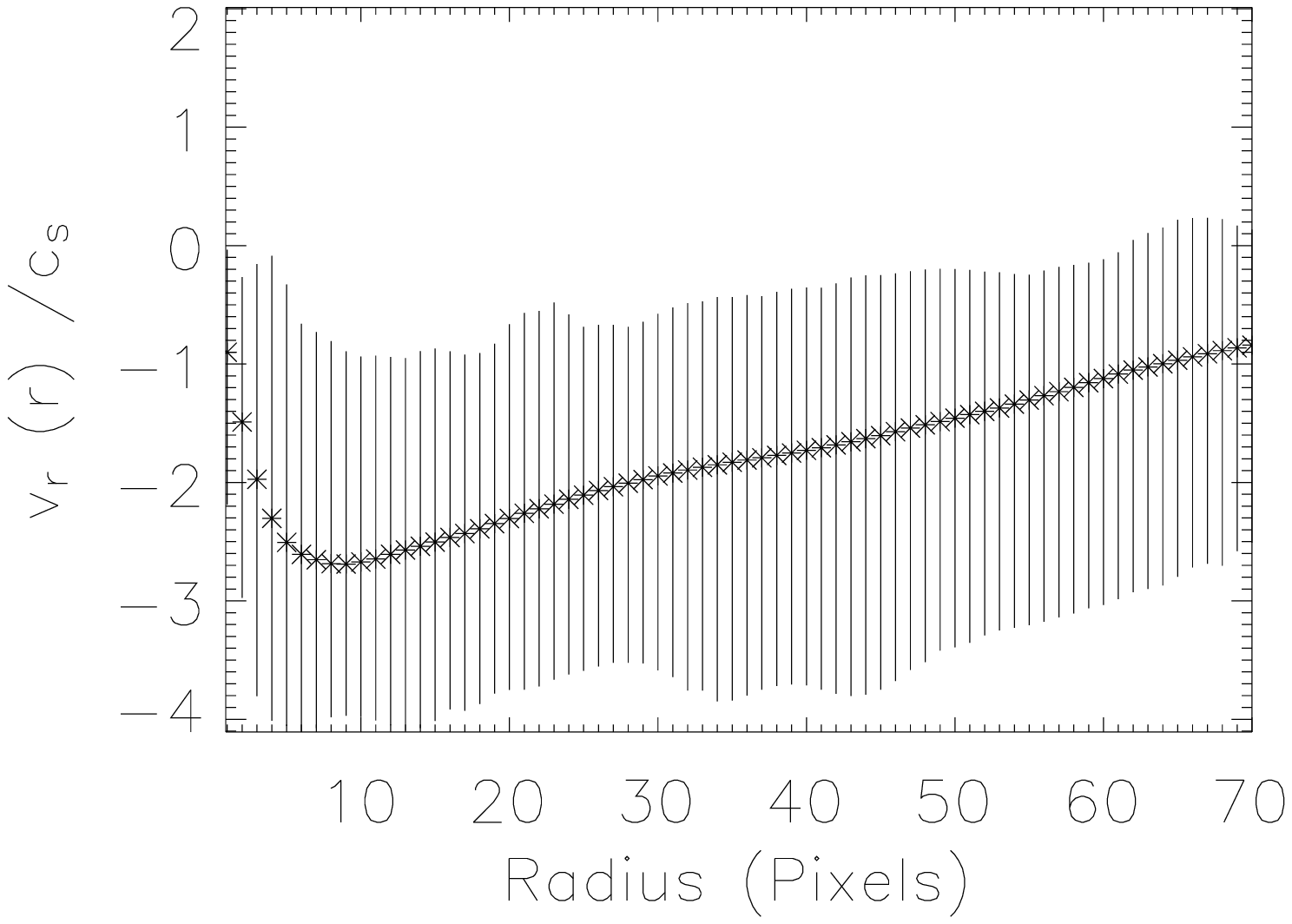} \\
\includegraphics[width=84mm]{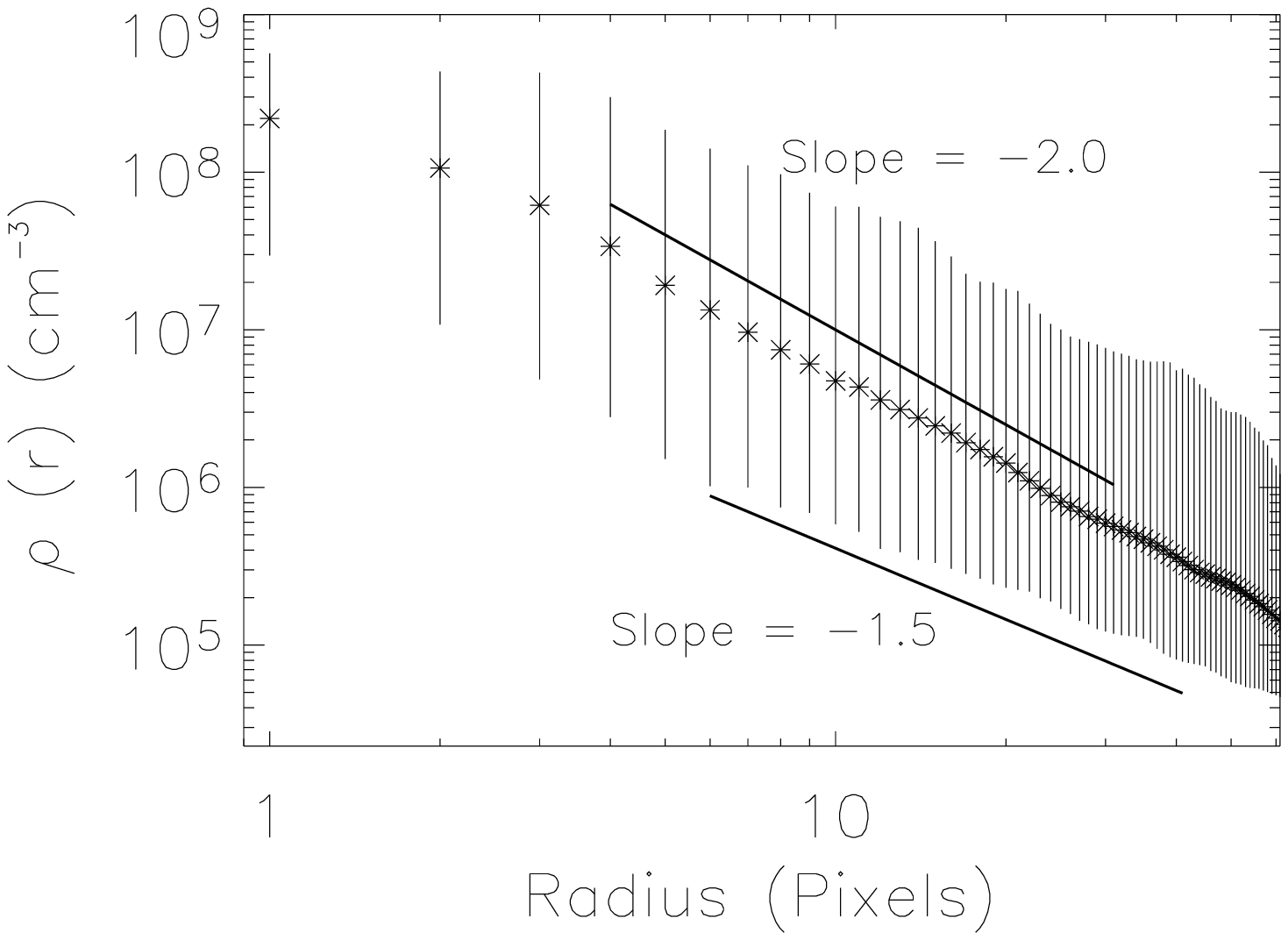}\includegraphics[width=84mm]{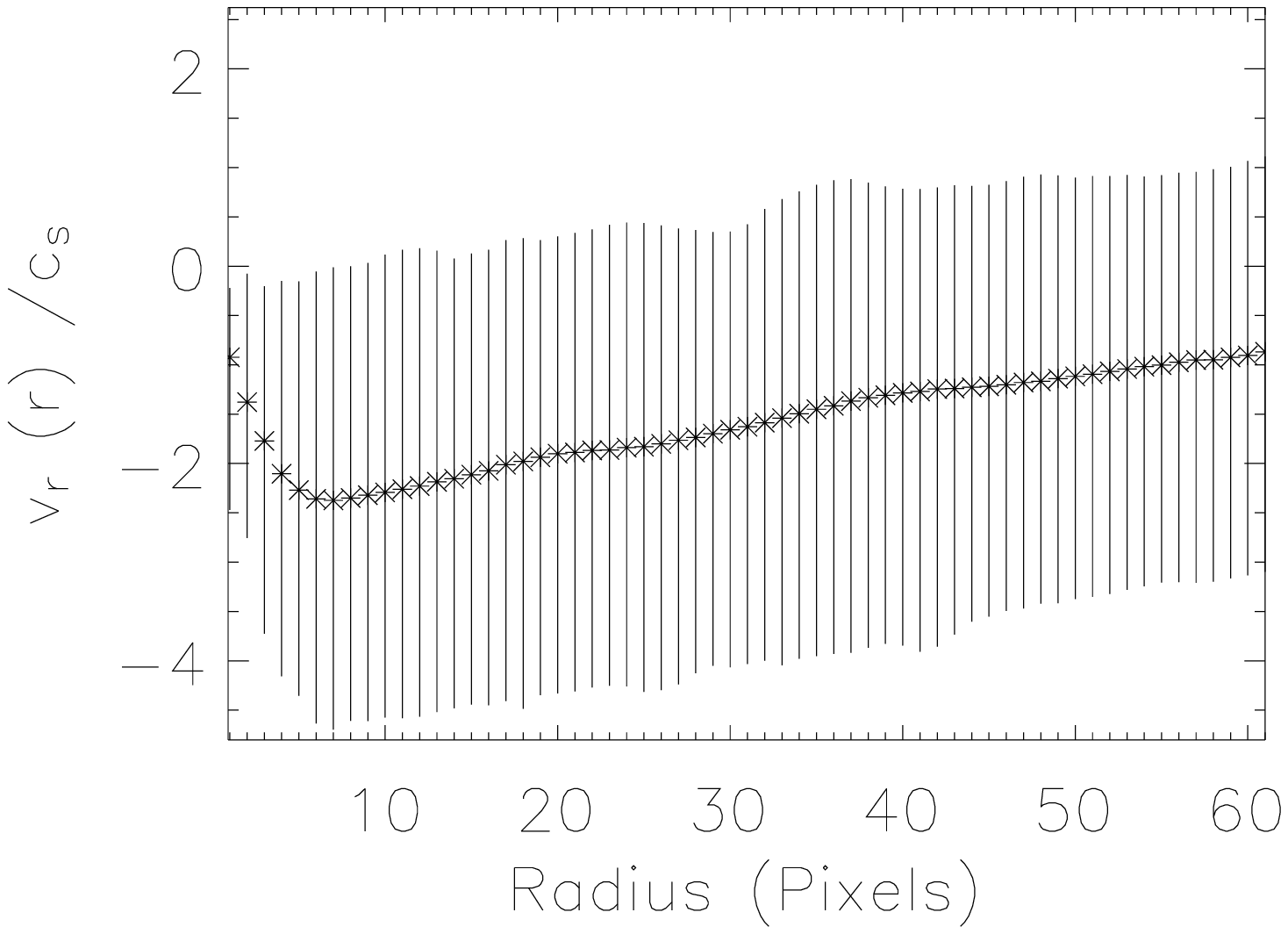} \\
\includegraphics[width=84mm]{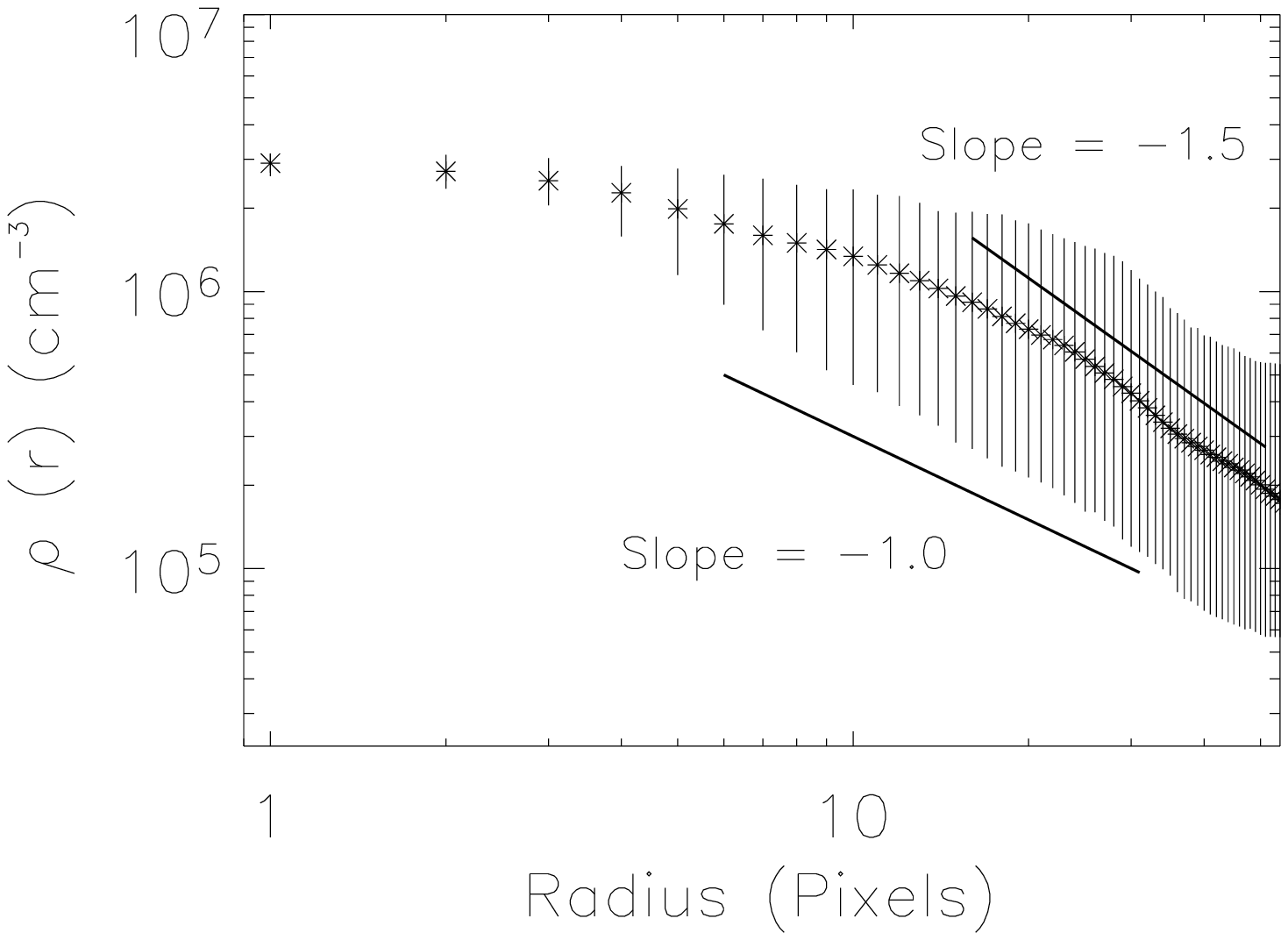}\includegraphics[width=84mm]{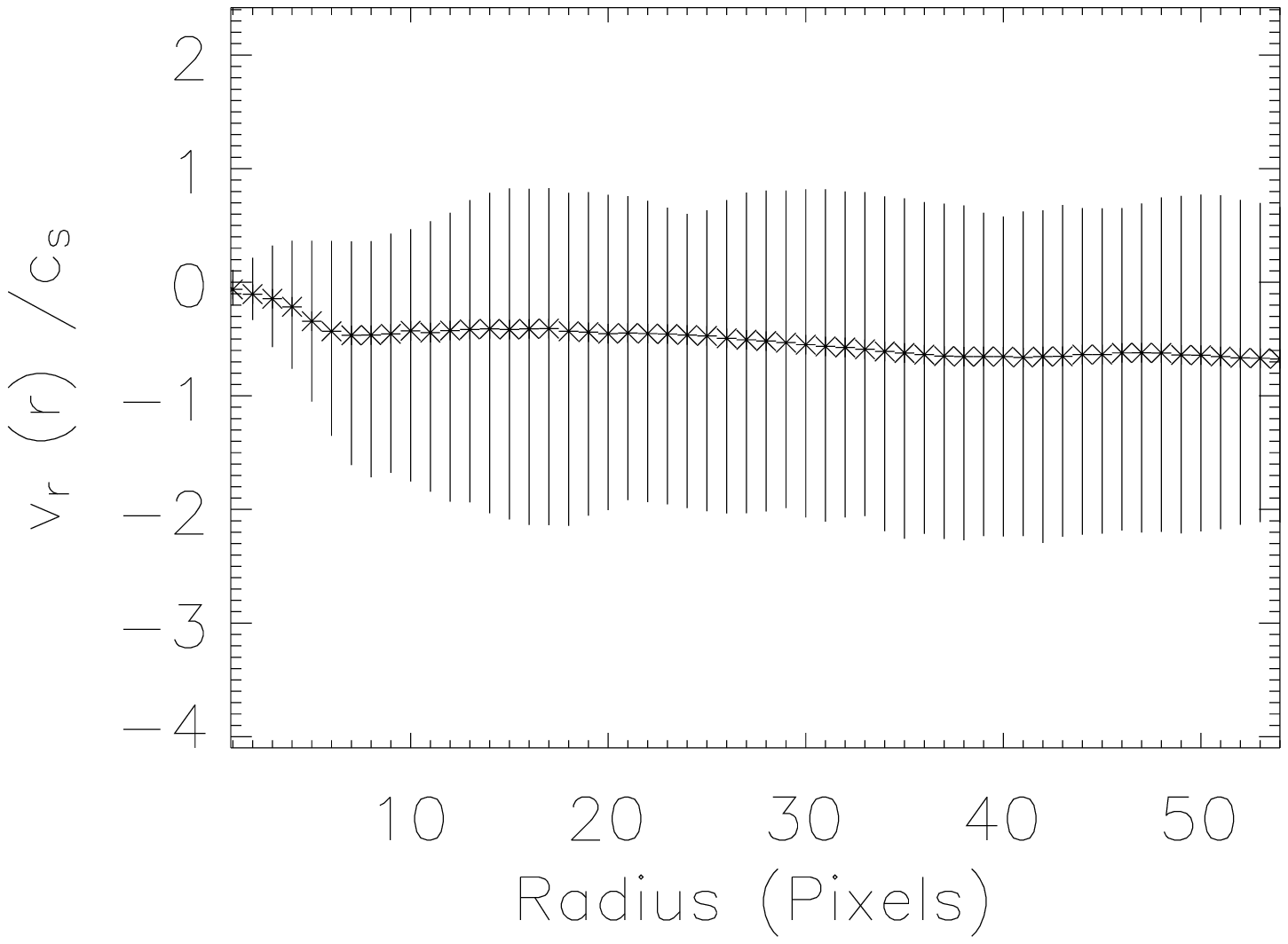}
\caption{Density and infall velocity plots for run B2.  The most massive core is on the top, followed by the second and third most massive; all three cores are bound according to the criterion in Section \ref{subsection_virial_stability}. The abscissa is in units of pixels; for a box size of $0.32$ pc (from Table \ref{table_sim_list}), one pixel corresponds to $258$ AU.\label{dvrprof_B2}}
\end{center}
\end{figure*}

\begin{figure*}
\begin{center}
\includegraphics[width=84mm]{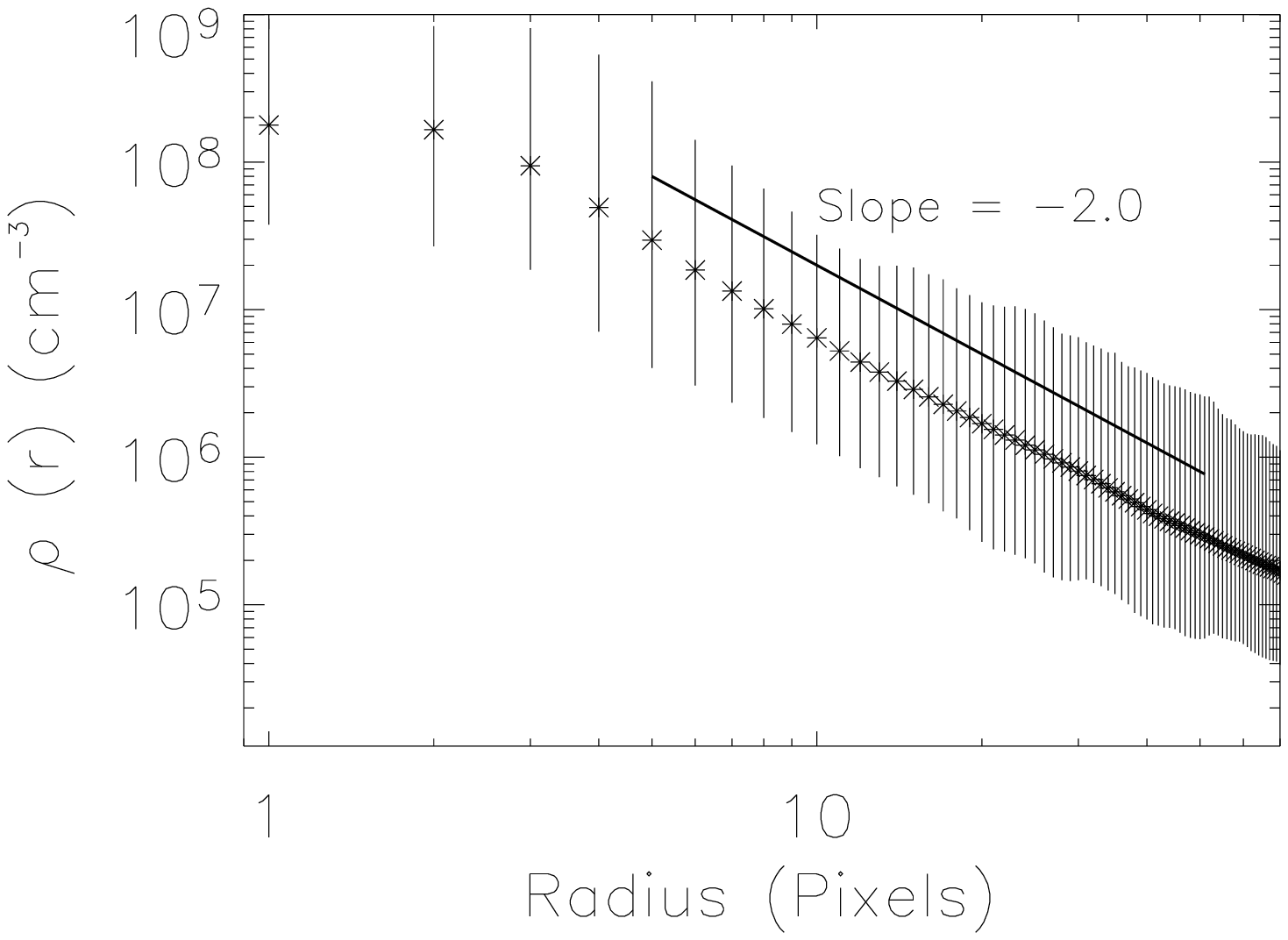}\includegraphics[width=84mm]{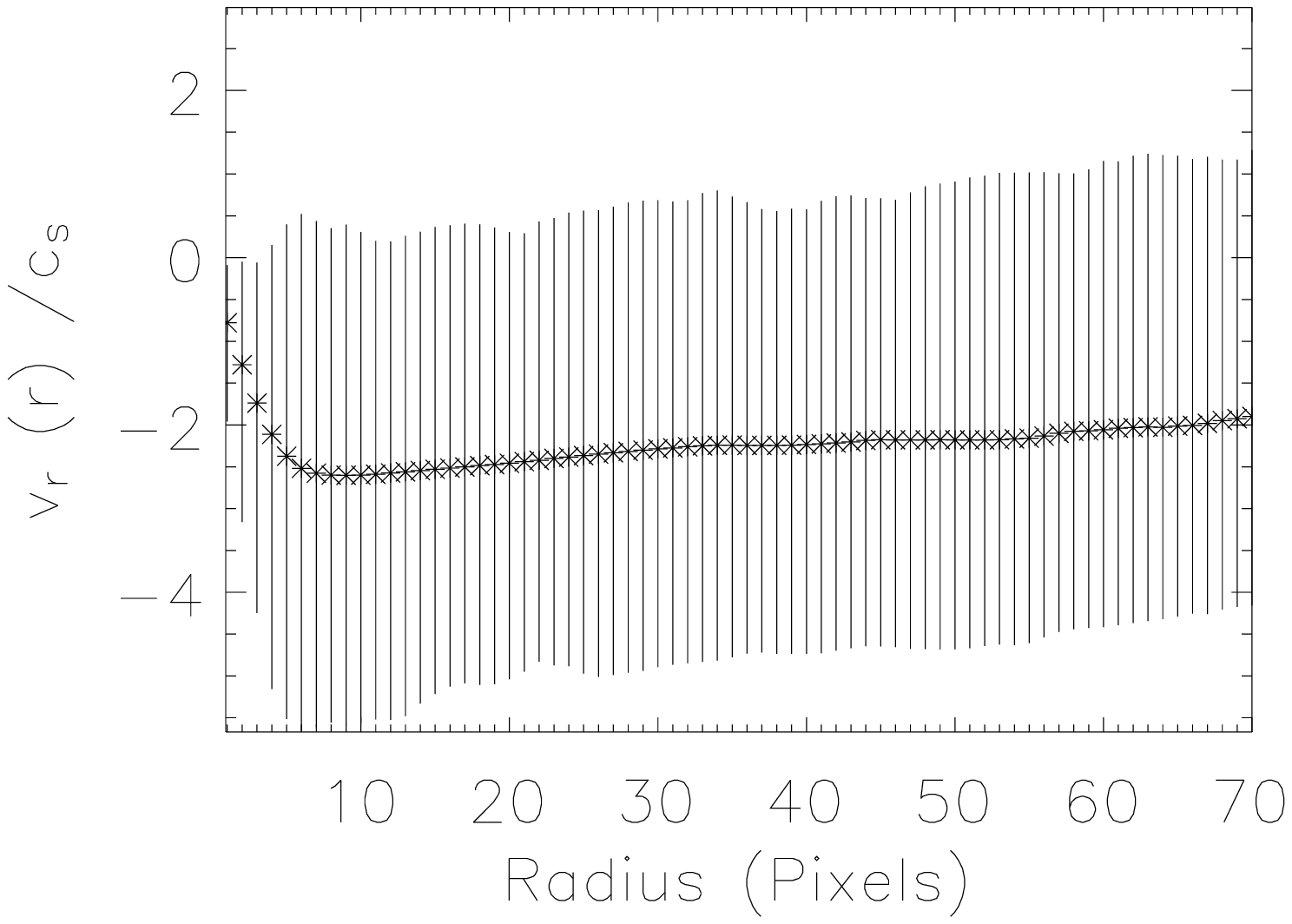} \\
\includegraphics[width=84mm]{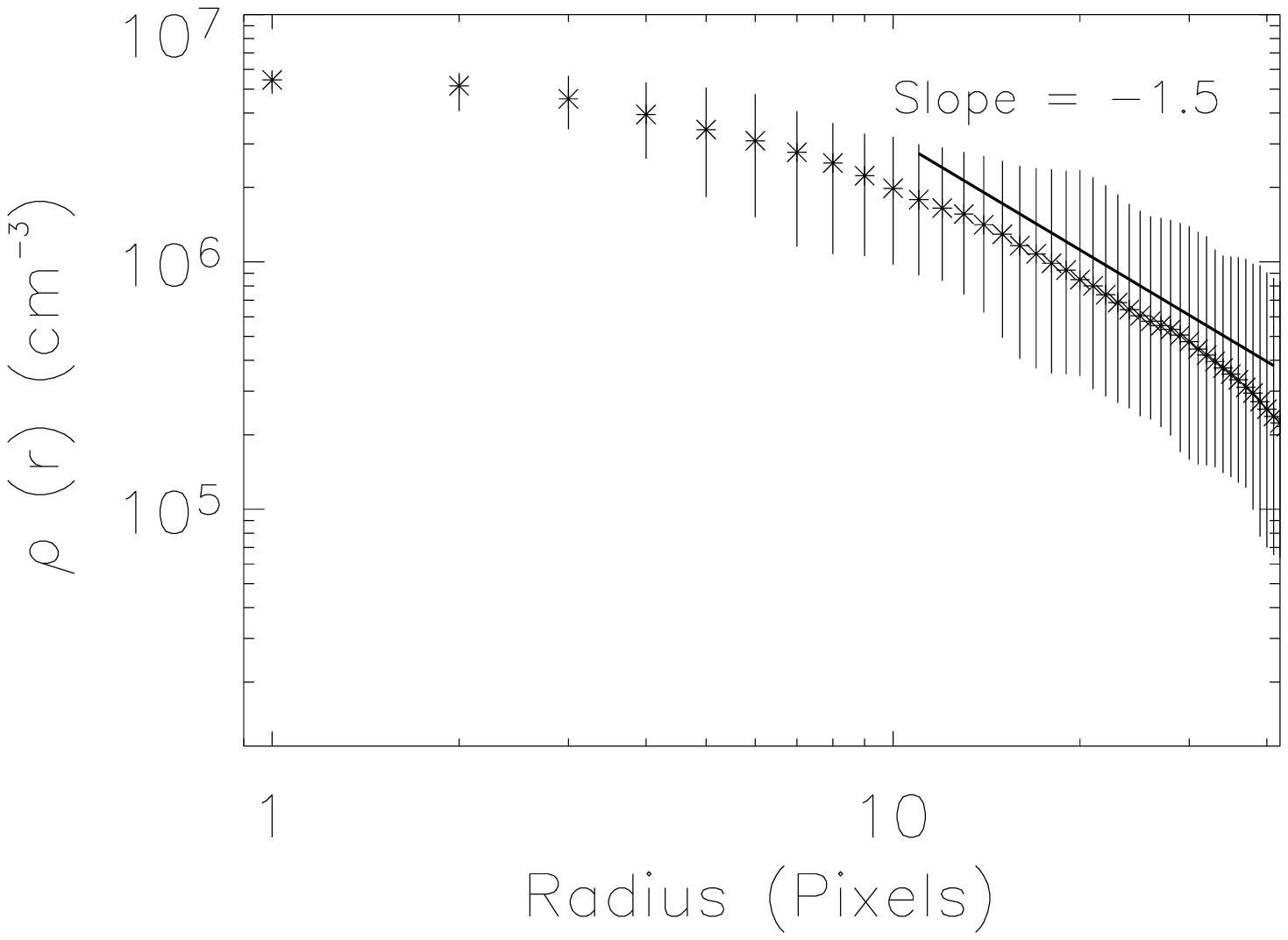}\includegraphics[width=84mm]{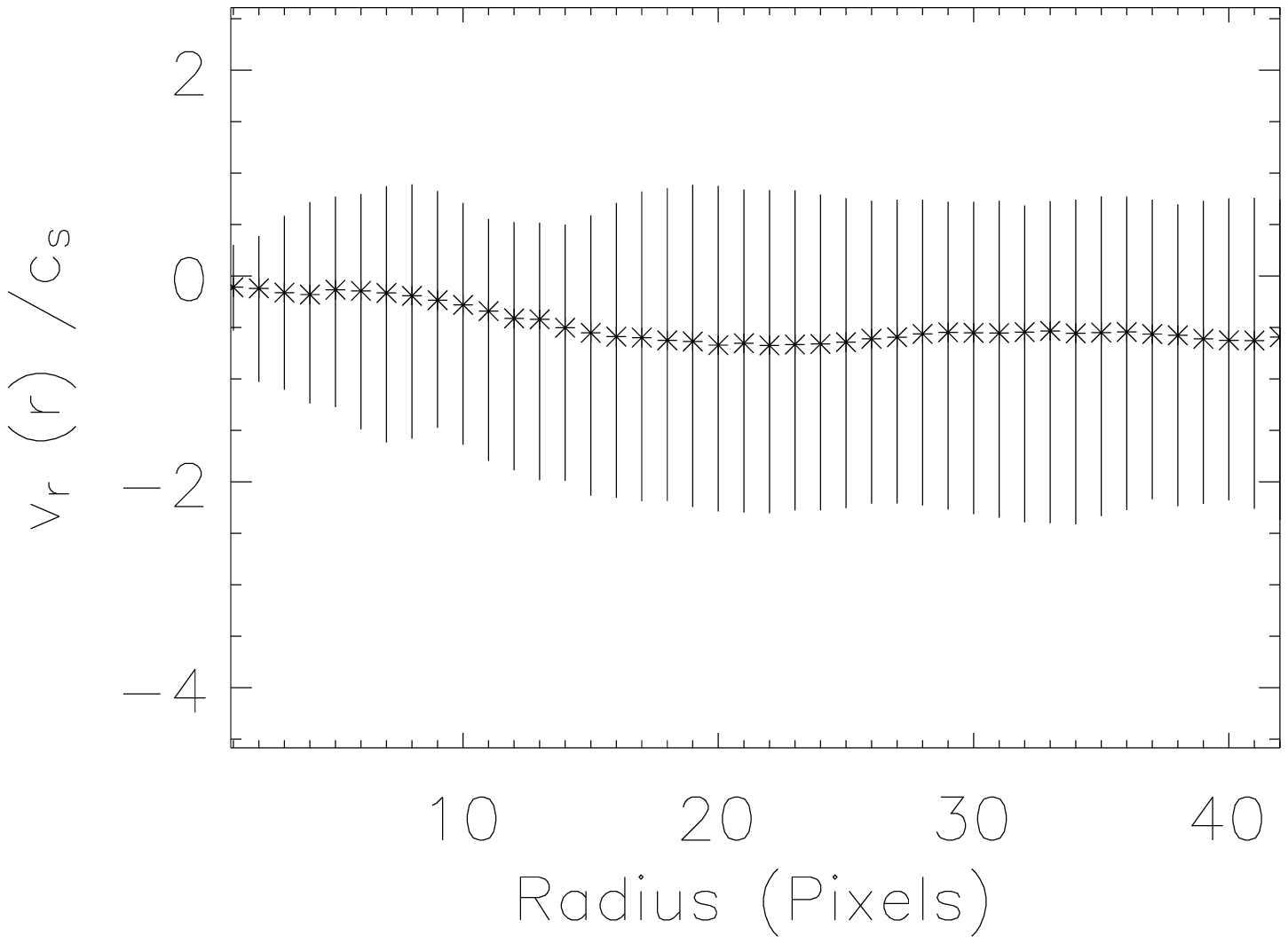} \\
\includegraphics[width=84mm]{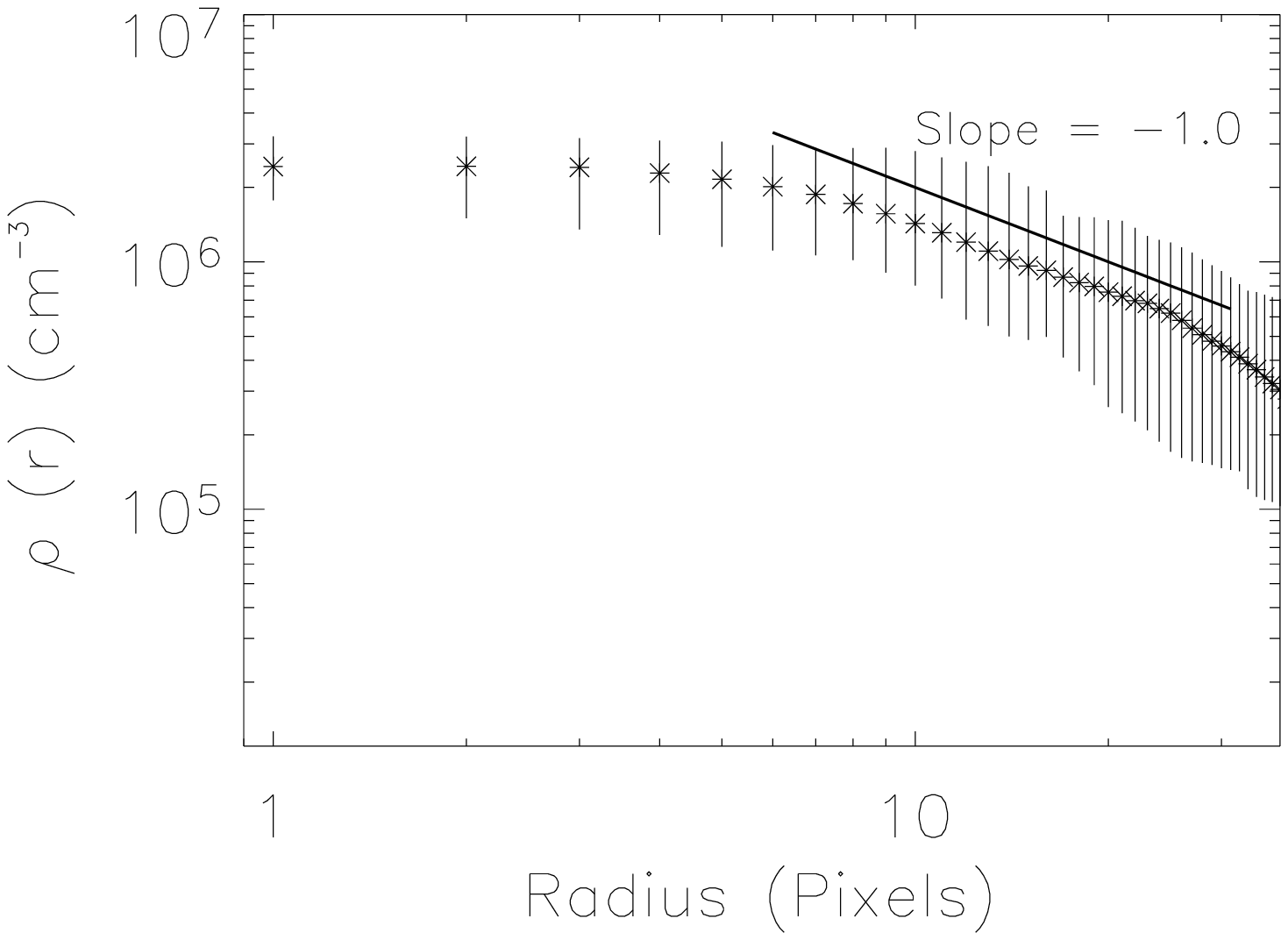}\includegraphics[width=84mm]{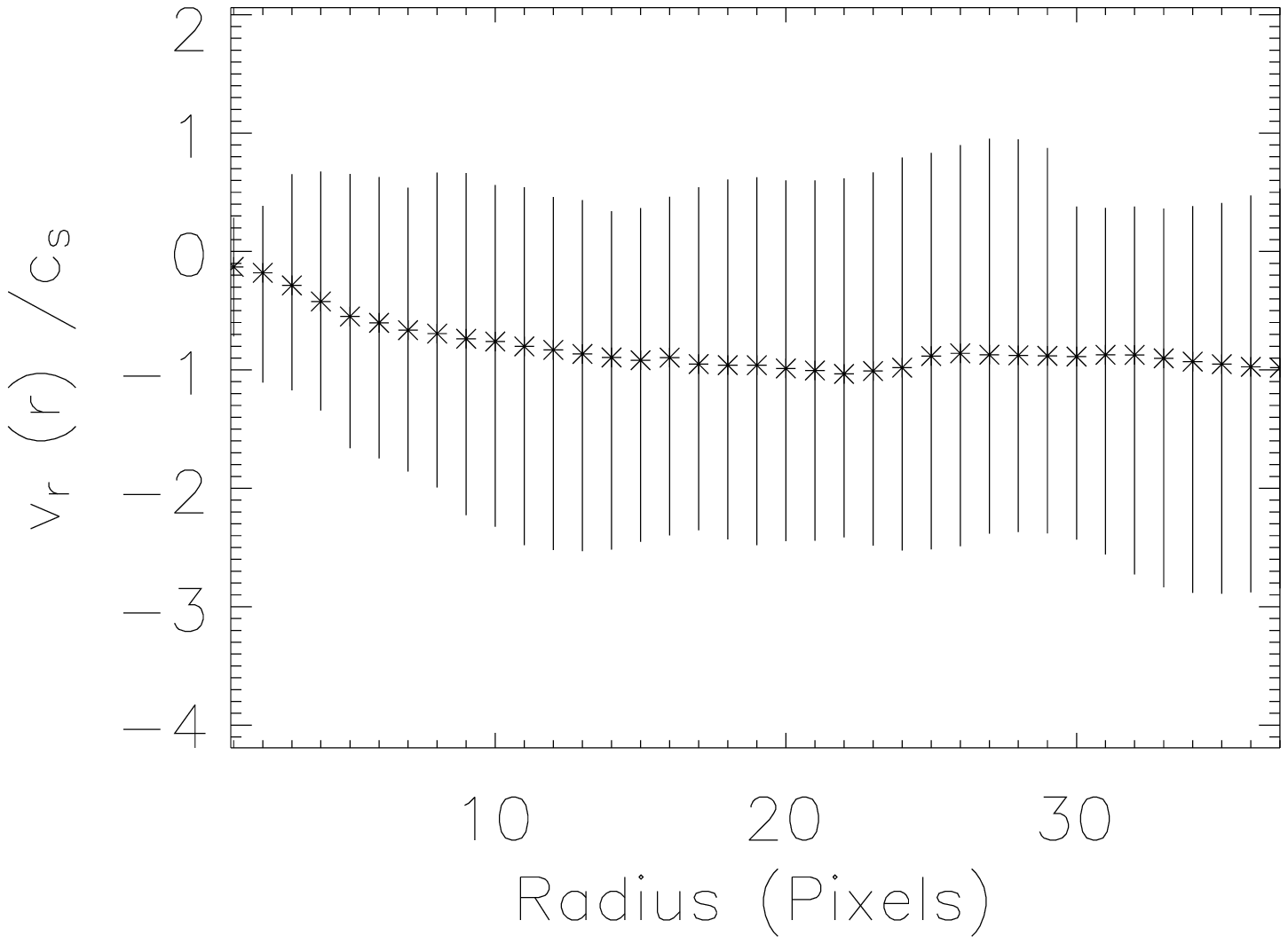}
\caption{Density and infall velocity plots for run B5.  The most massive core is on the top, followed by the second and third most massive; all three cores are bound according to the criterion in Section \ref{subsection_virial_stability}. The abscissa is in units of pixels; for a box size of $0.32$ pc (from Table \ref{table_sim_list}), one pixel corresponds to $258$ AU.\label{dvrprof_B5}}
\end{center}
\end{figure*}

\begin{figure*}
\begin{center}
\includegraphics[width=84mm]{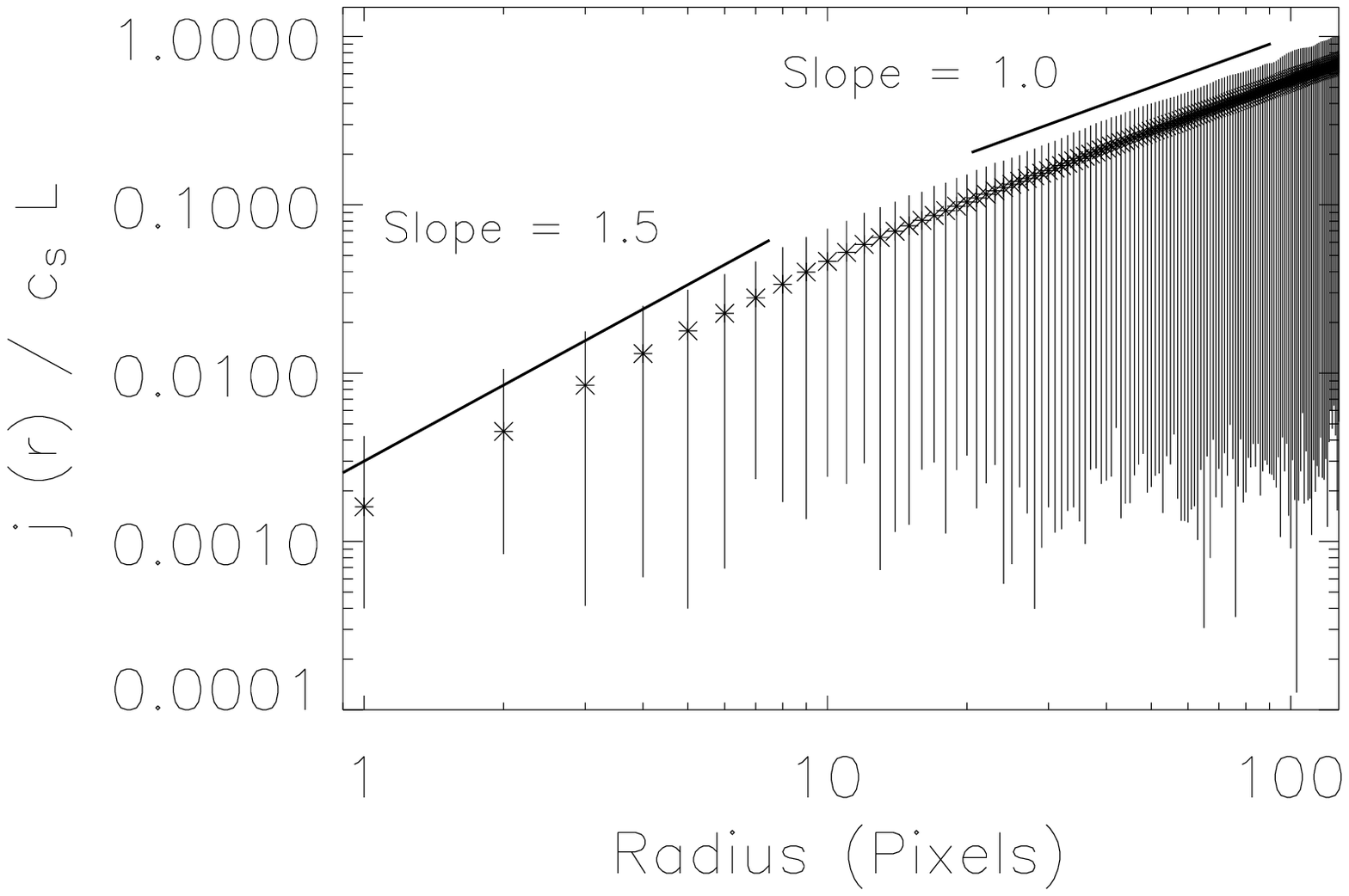}\includegraphics[width=84mm]{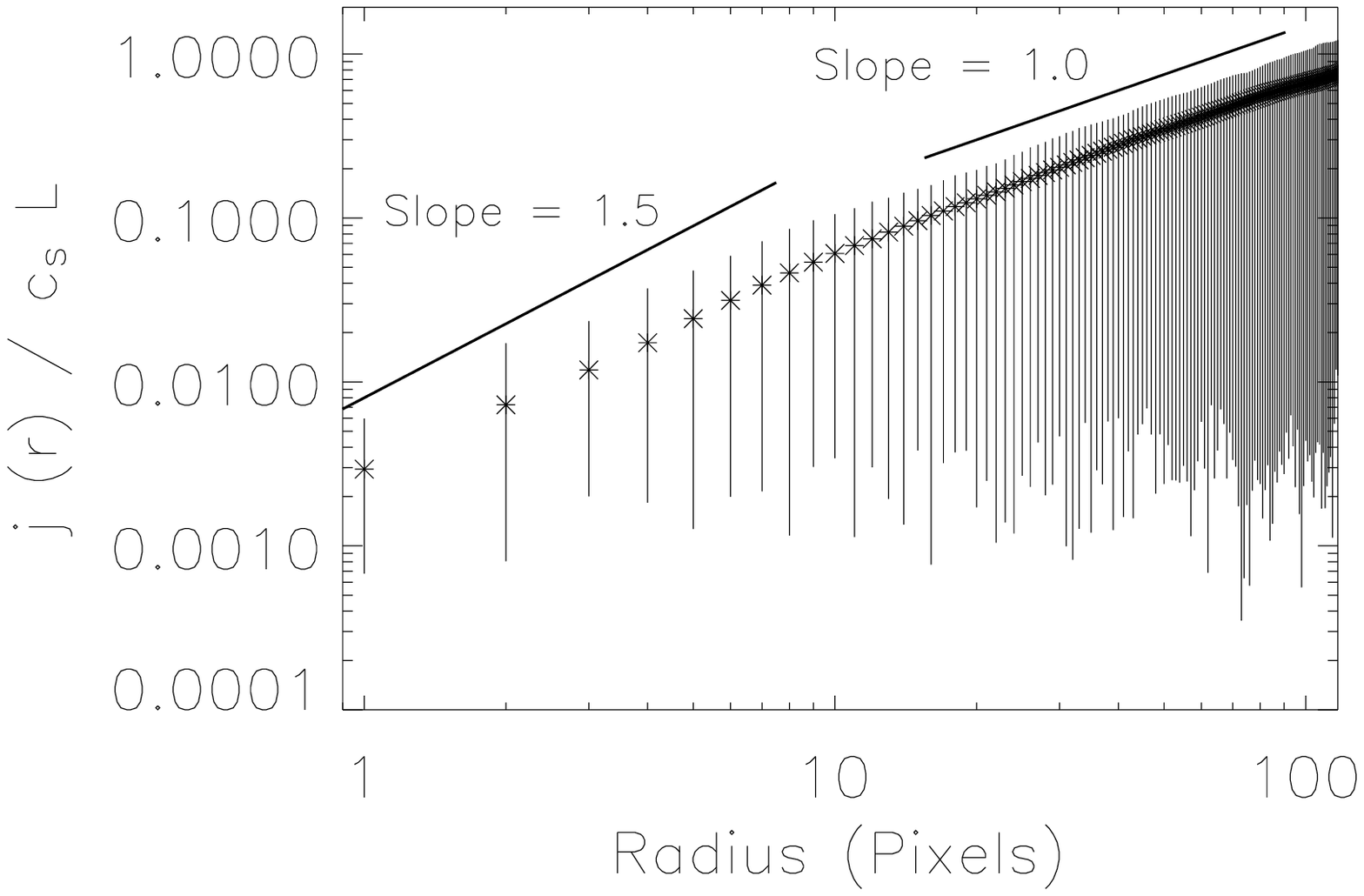} \\
\includegraphics[width=84mm]{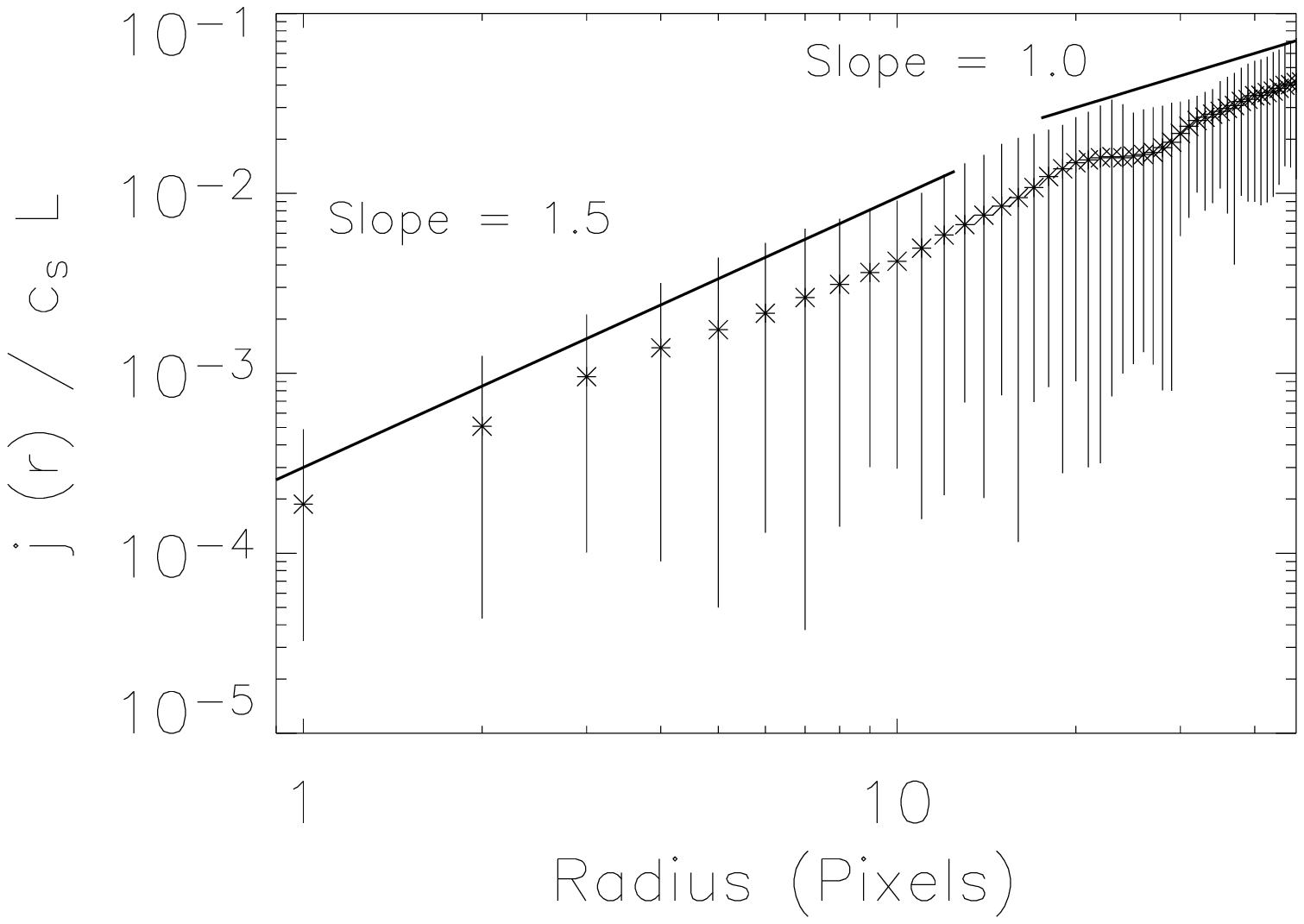}\includegraphics[width=84mm]{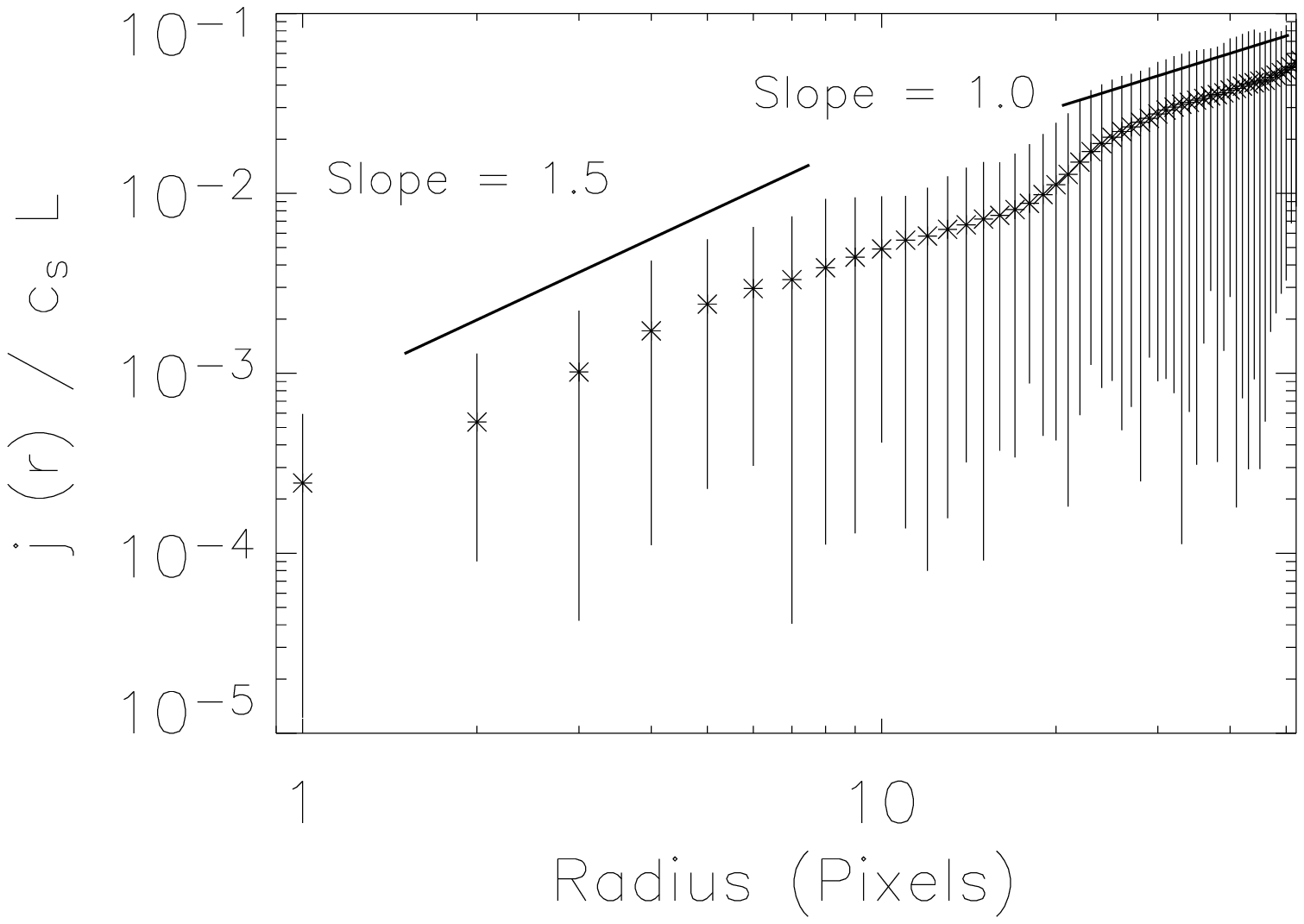} \\
\includegraphics[width=84mm]{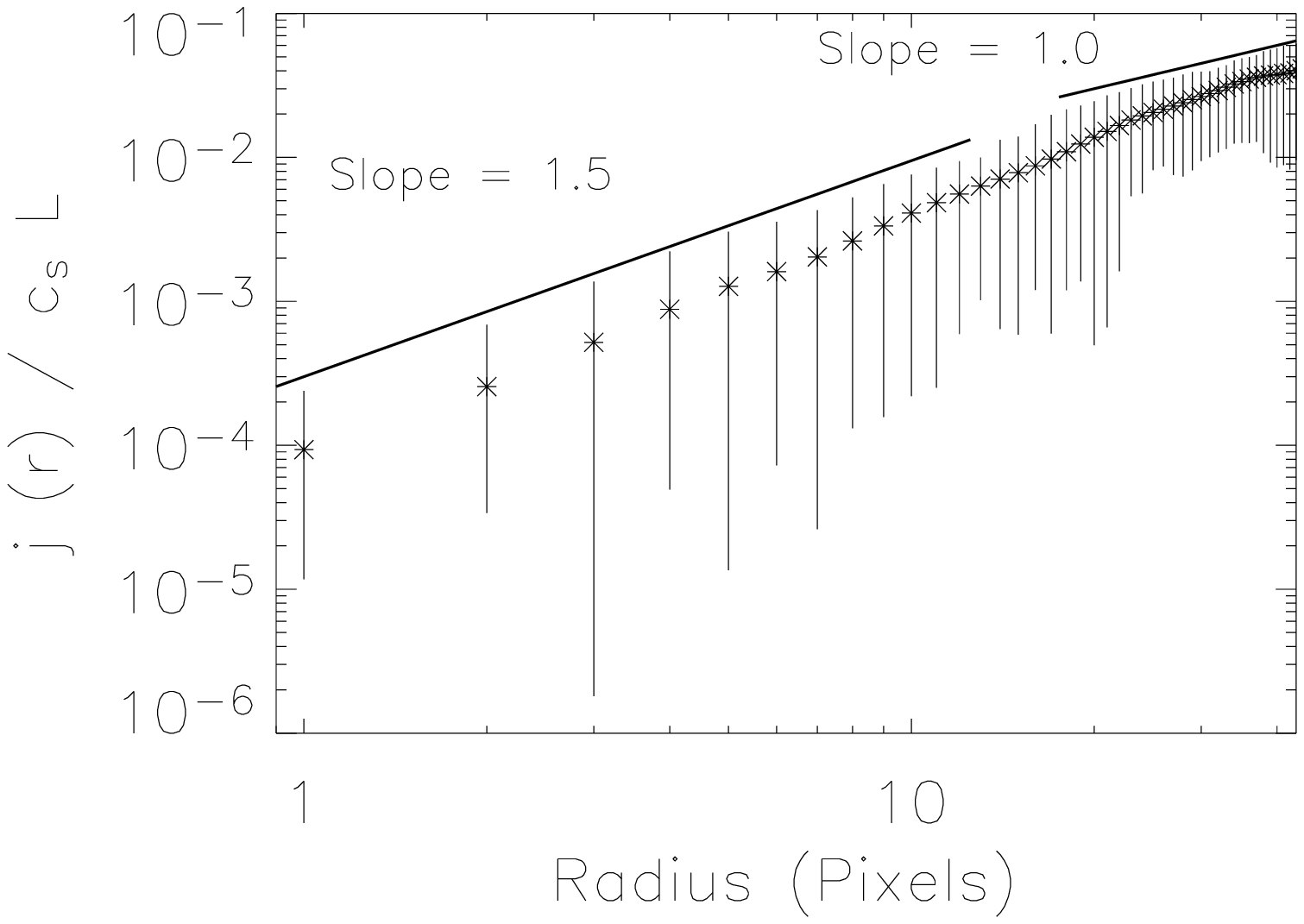}\includegraphics[width=84mm]{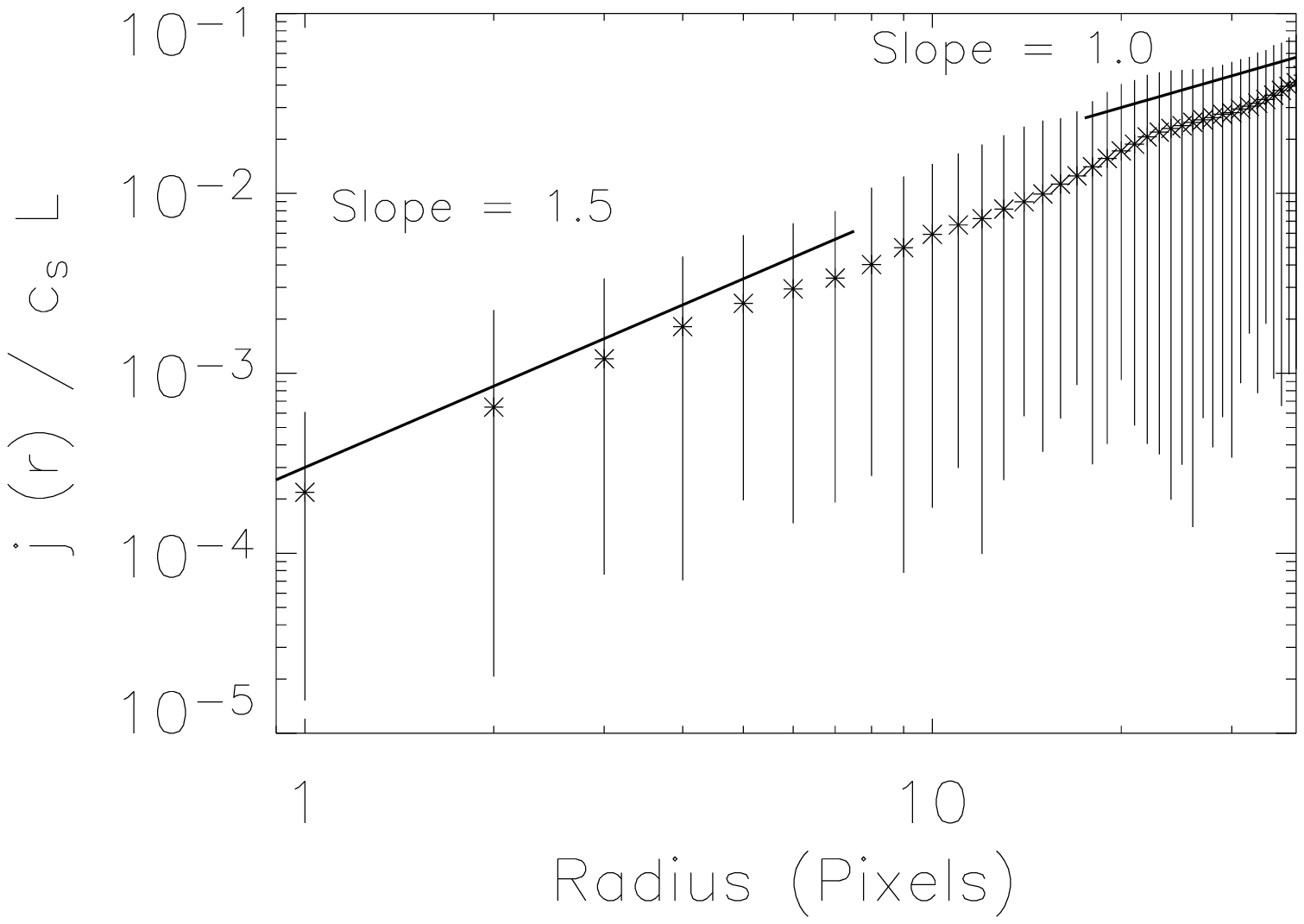}
\caption{Specific angular momentum profiles of the three most massive cores in each run.  The cores plotted in the left-hand column are from Run A2, the cores plotted in the right-hand column are from Run A5.  The abscissa is in units of pixels; for a box size of $0.1$ pc (from Table \ref{table_sim_list}), one pixel corresponds to $81$ AU.\label{jprofiles1}}
\end{center}
\end{figure*}

\begin{figure*}
\begin{center}
\includegraphics[width=84mm]{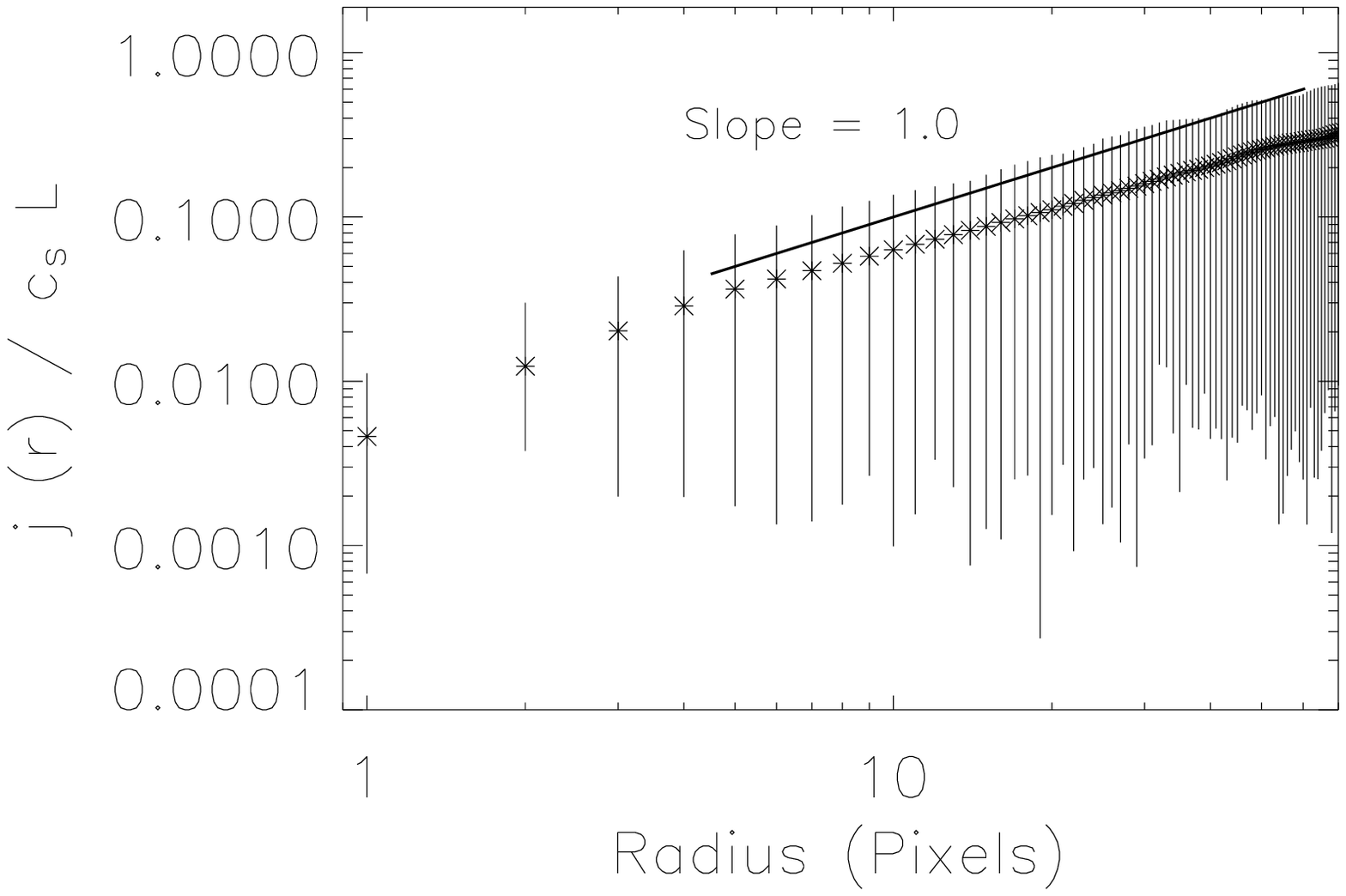}\includegraphics[width=84mm]{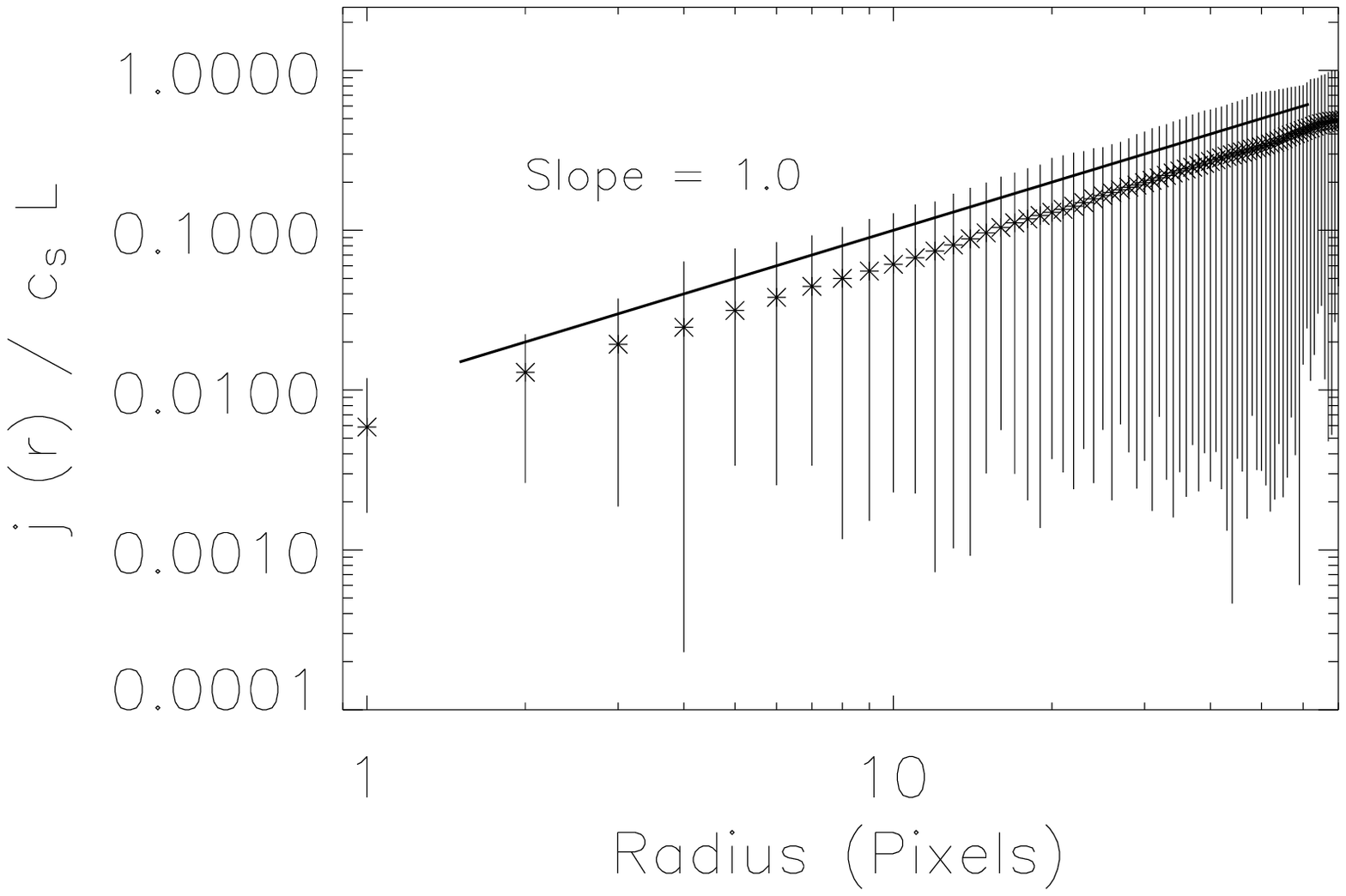} \\
\includegraphics[width=84mm]{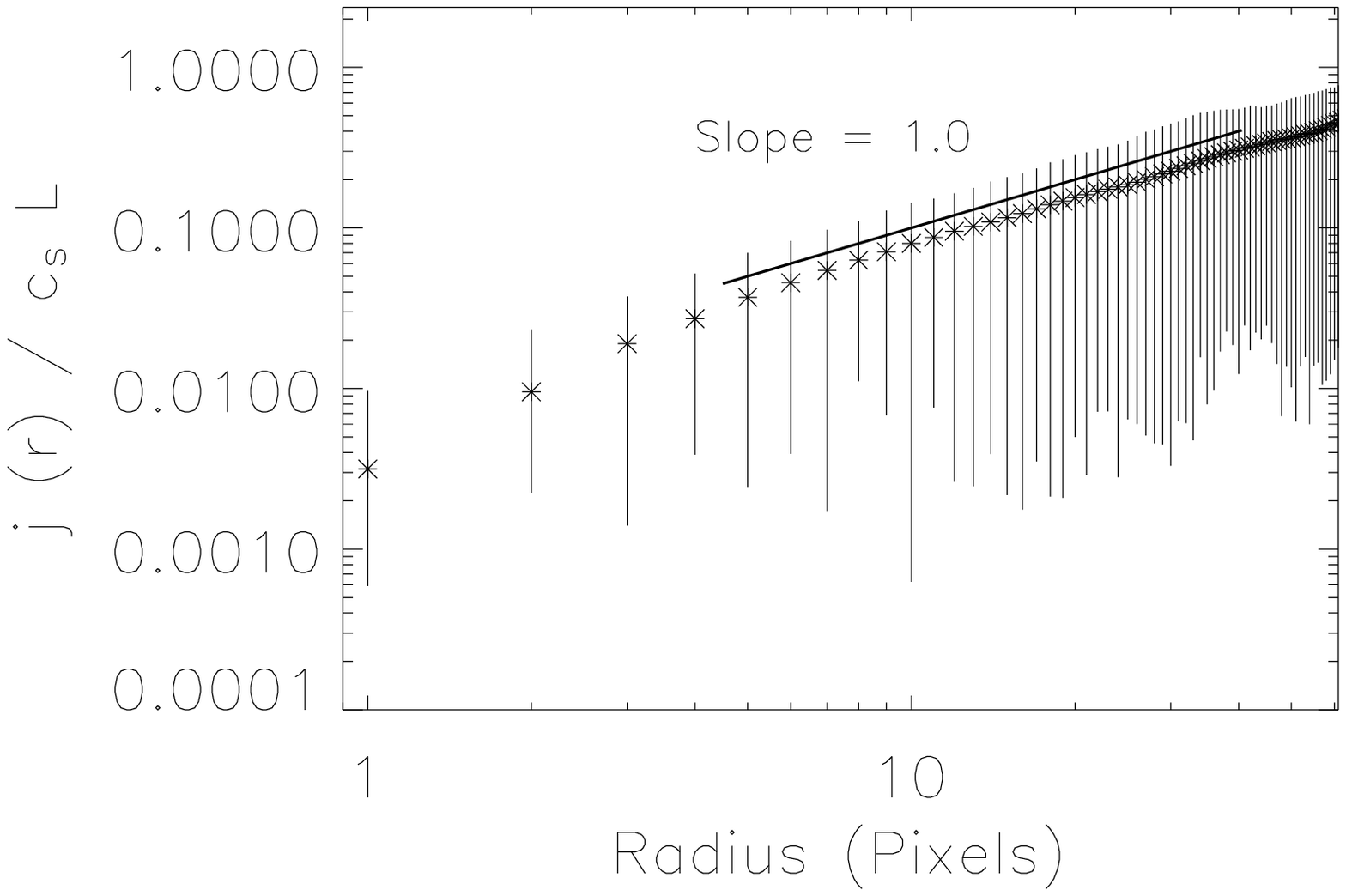}\includegraphics[width=84mm]{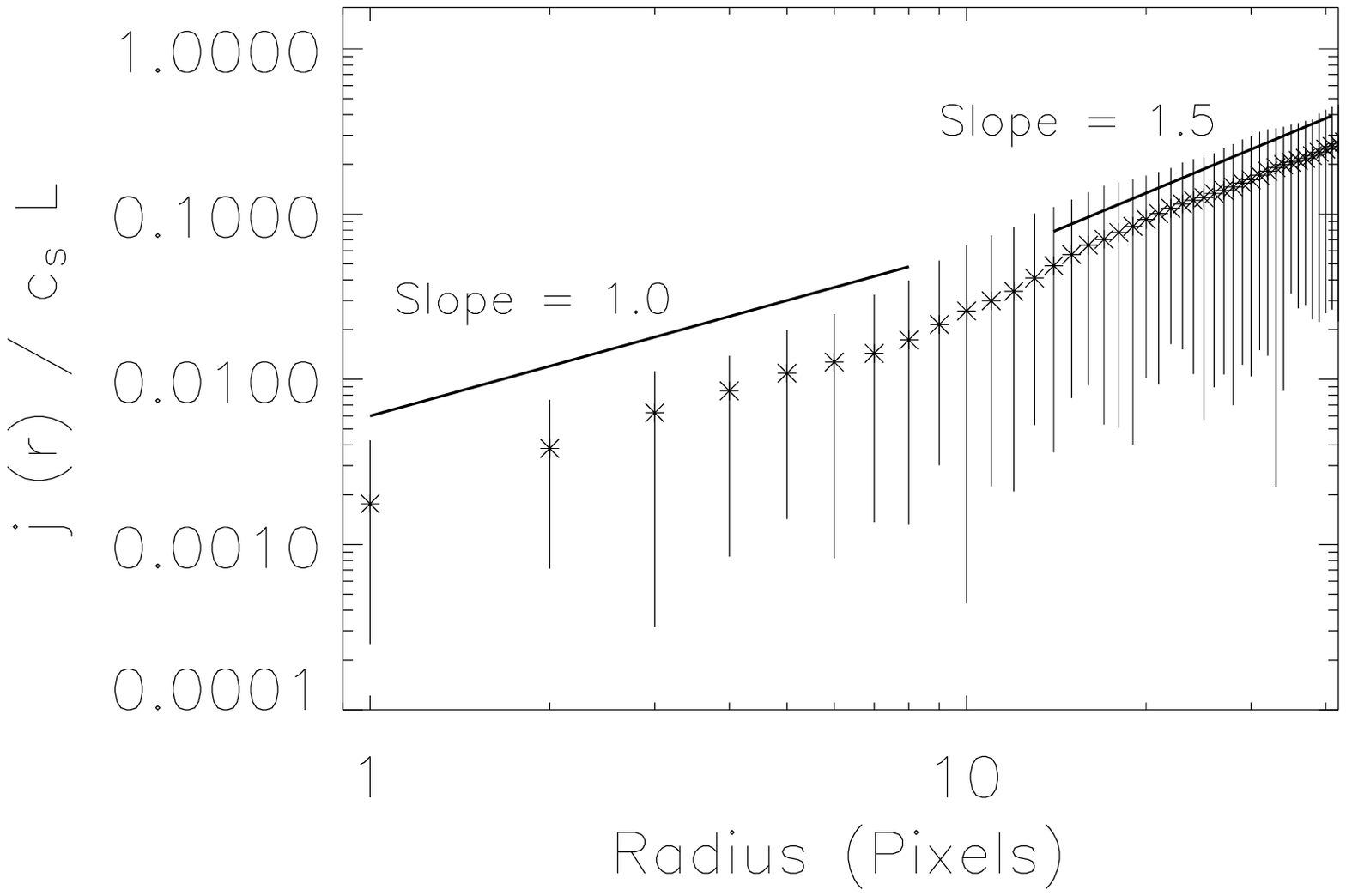} \\
\includegraphics[width=84mm]{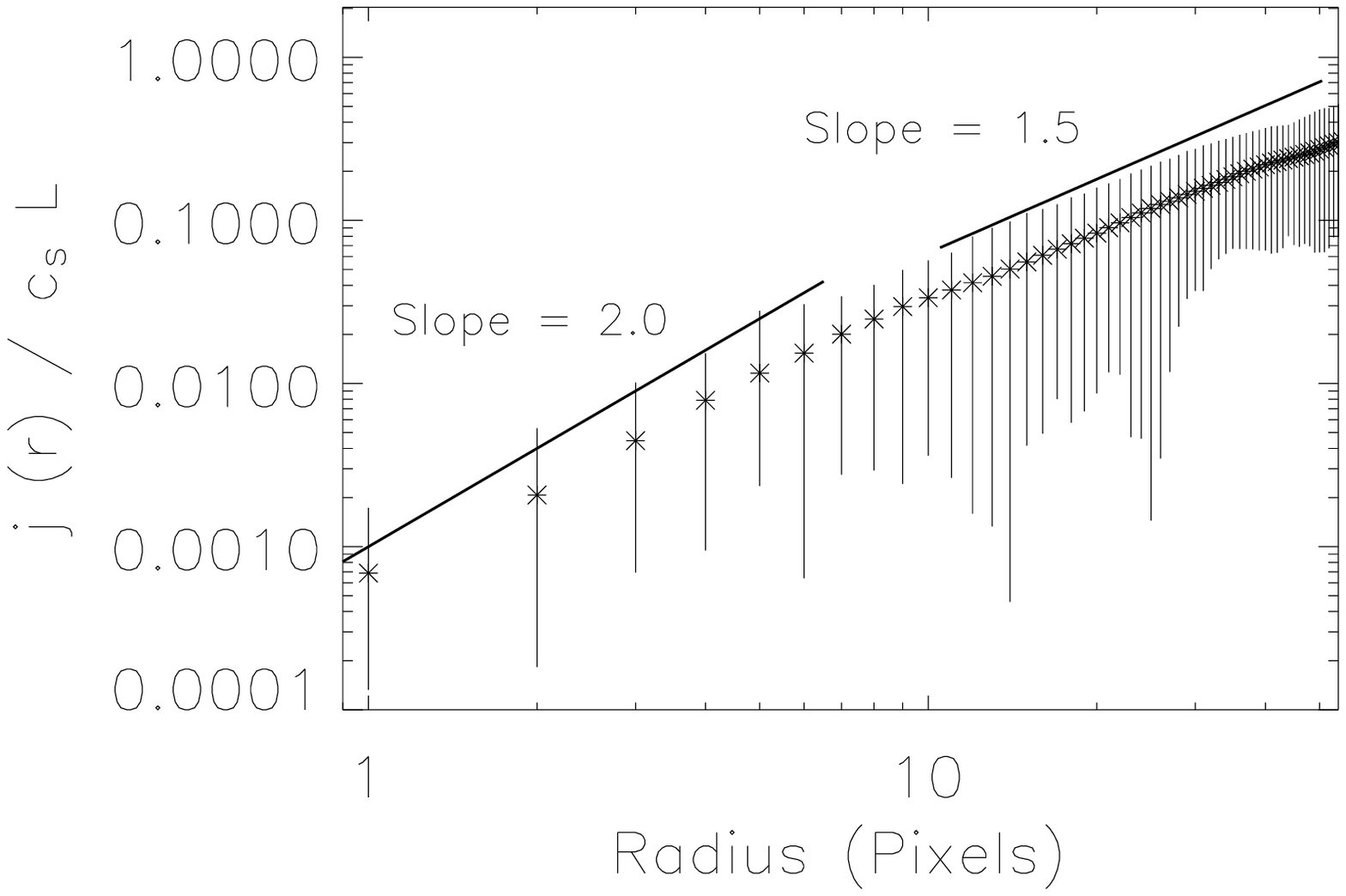}\includegraphics[width=84mm]{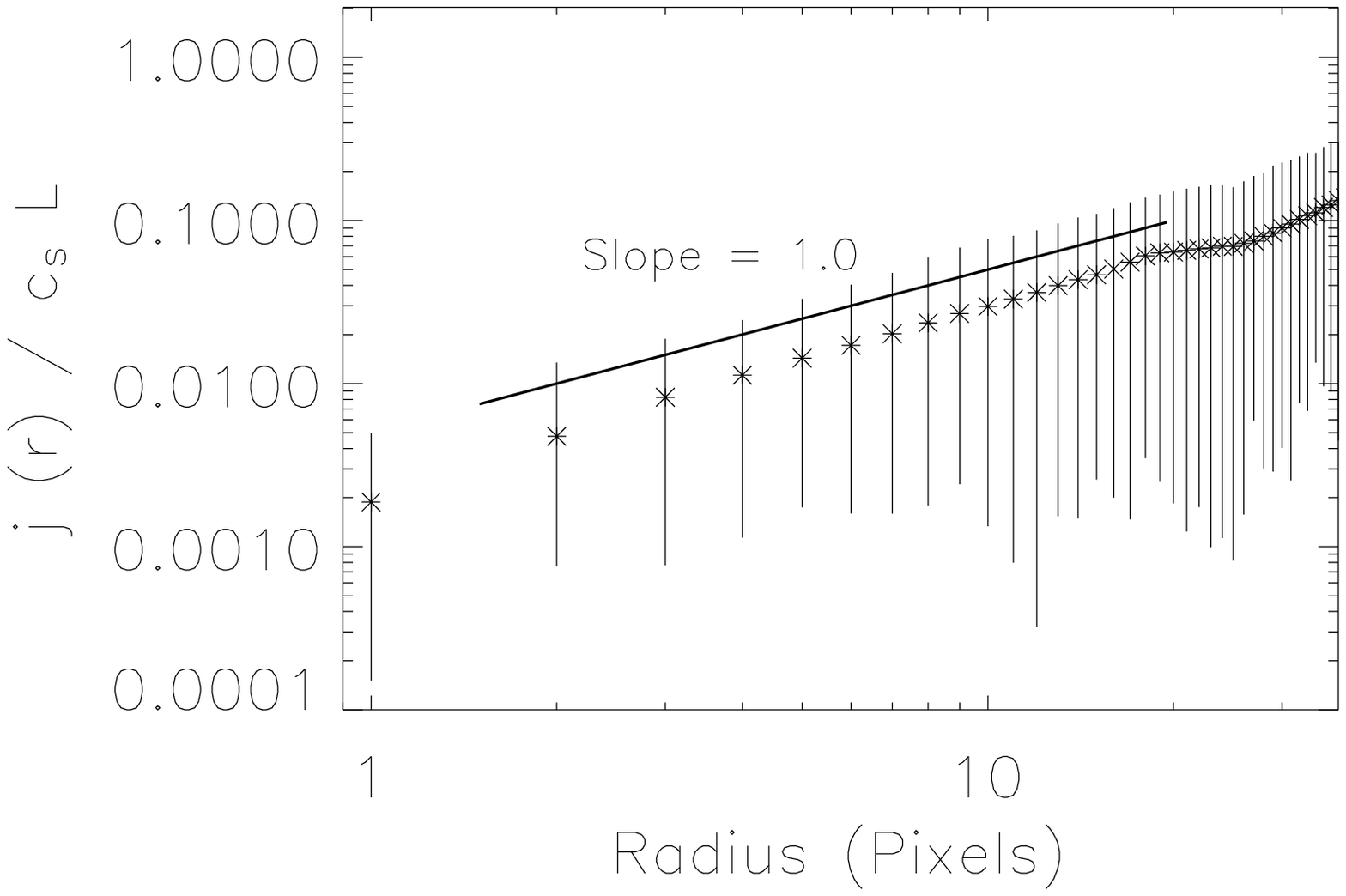}
\caption{Specific angular momentum profiles of the three most massive cores in each run.  The cores plotted in the left-hand column are from Run B2, the cores plotted in the right-hand column are from Run B5.  The abscissa is in units of pixels; for a box size of $0.32$ pc (from Table \ref{table_sim_list}), one pixel corresponds to $258$ AU.\label{jprofiles2}}
\end{center}
\end{figure*}

\begin{figure*}
\begin{center}
\includegraphics[width=84mm]{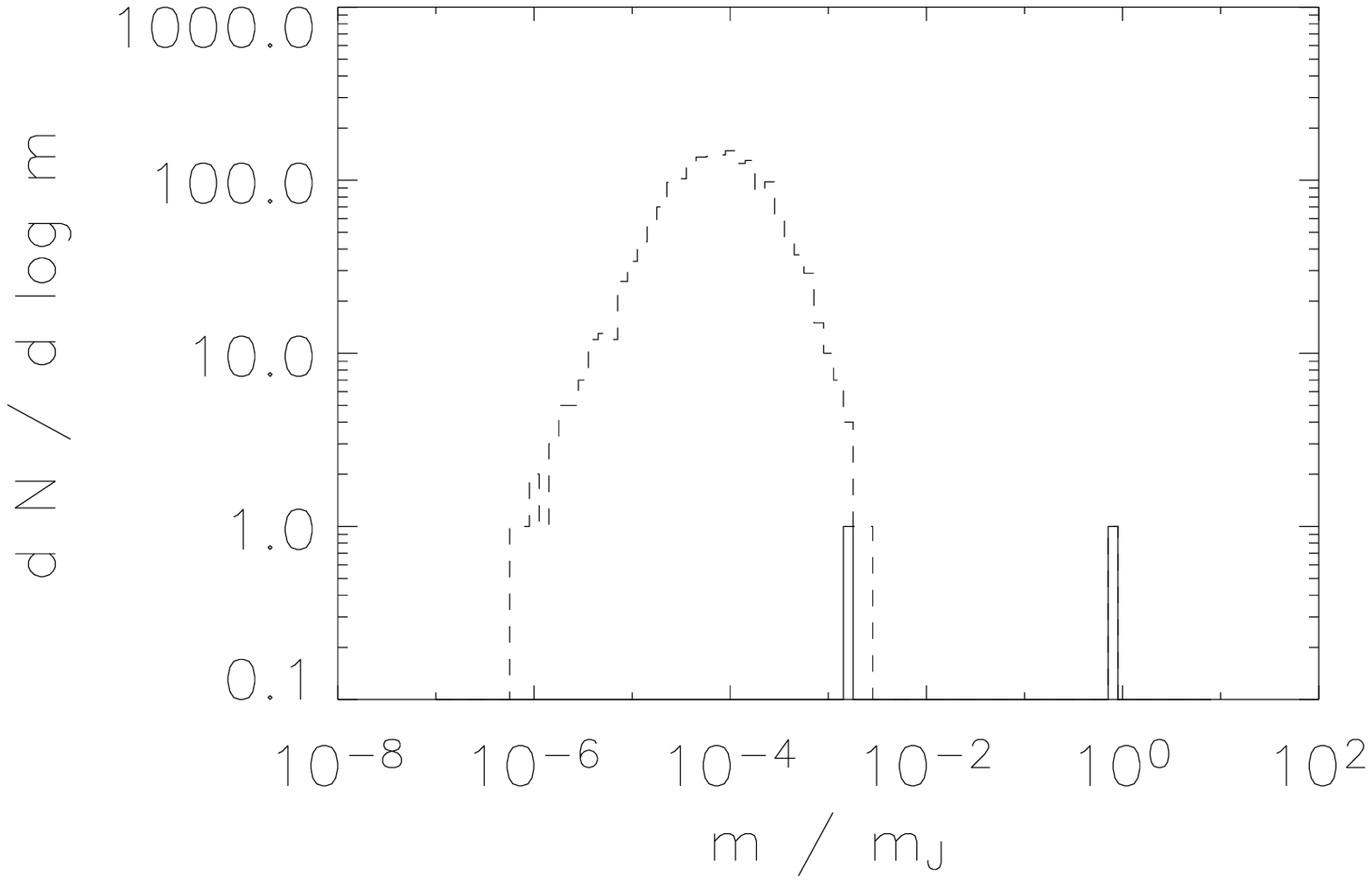}\hspace{5mm}\includegraphics[width=84mm]{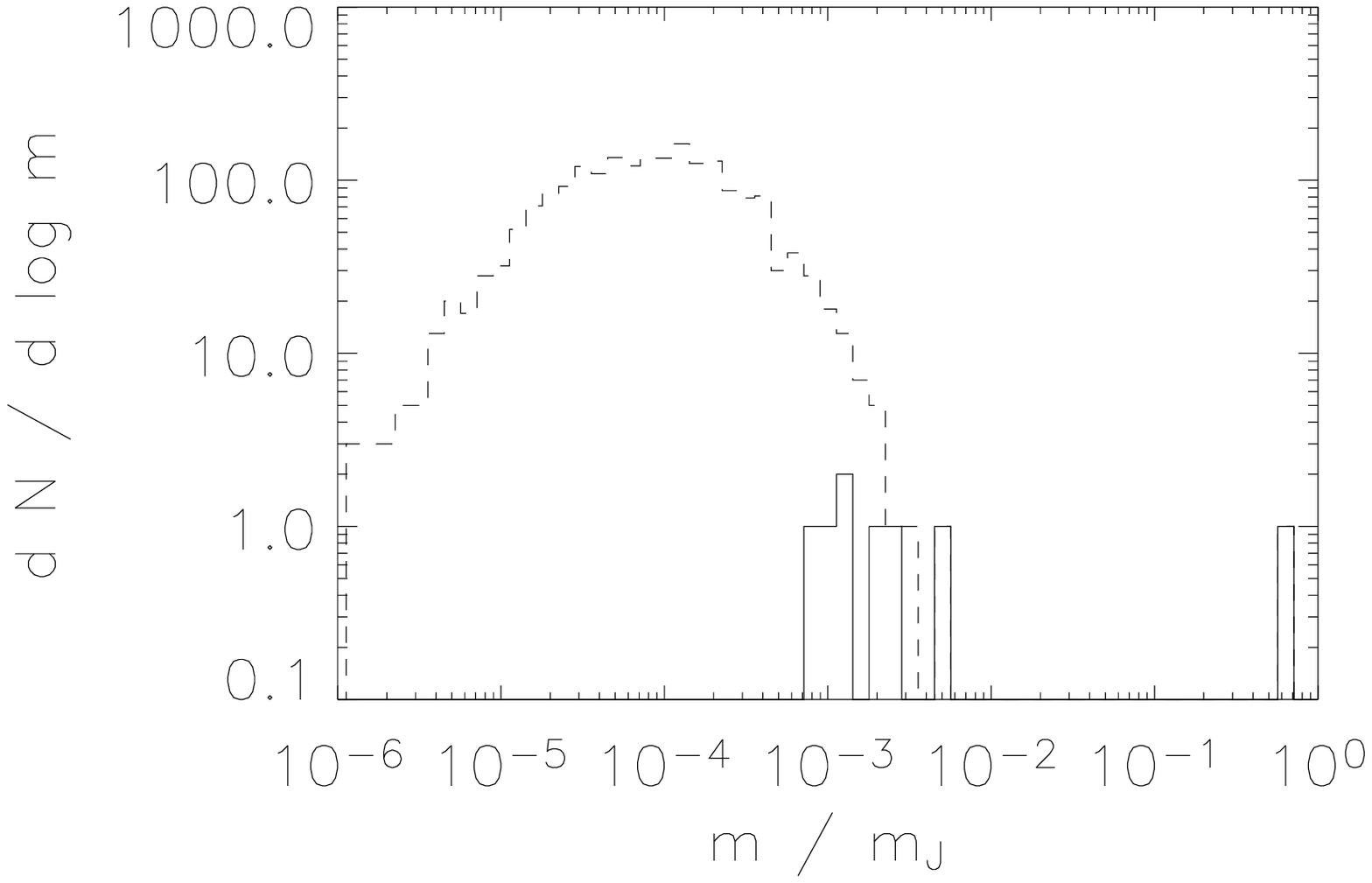}\\\vspace{10mm}
\includegraphics[width=84mm]{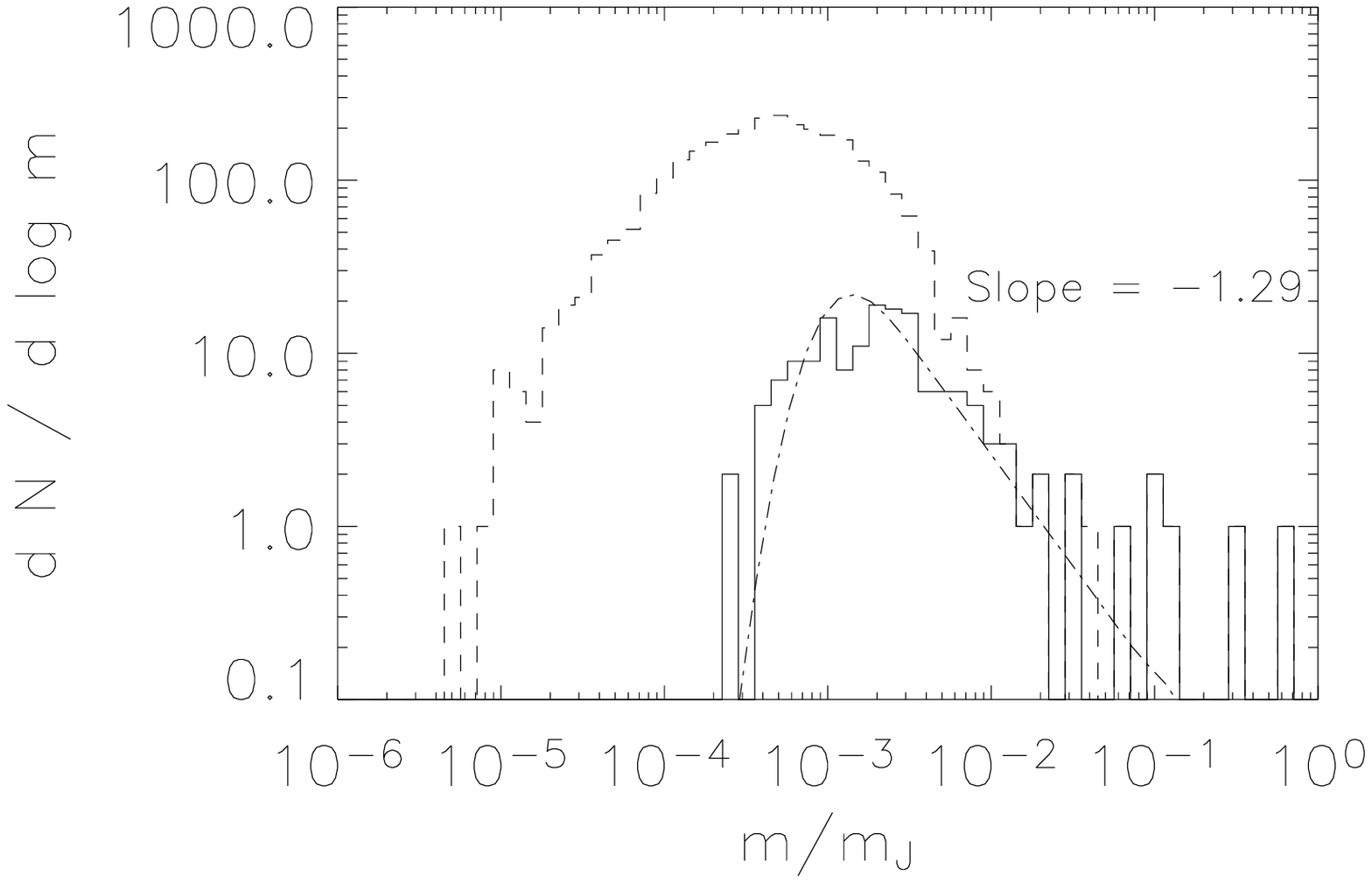}\hspace{5mm}\includegraphics[width=84mm]{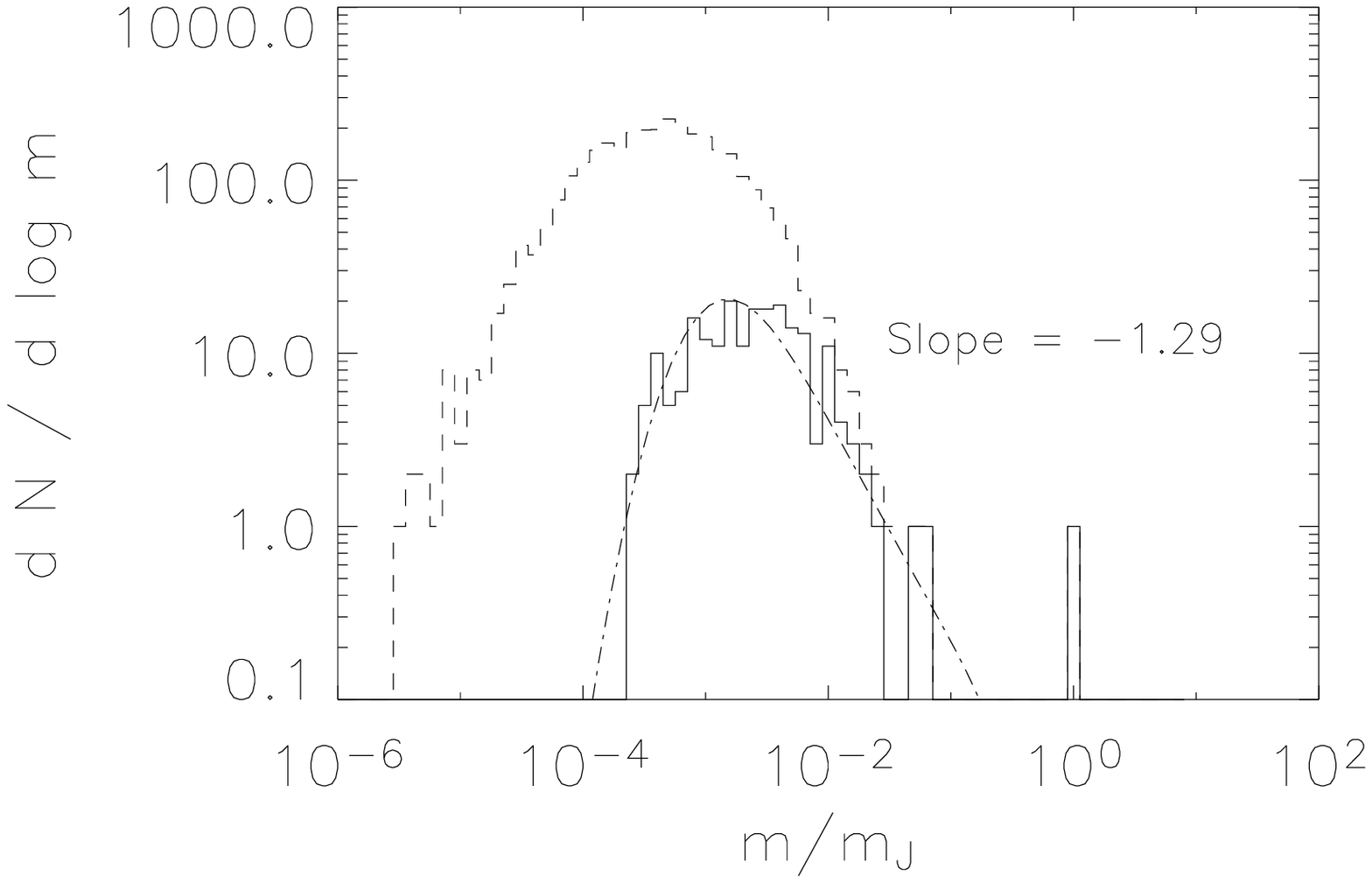}
\caption{Mass distribution of fluctuations for the various runs.  Runs A are on the top row, runs B are on the bottom row.  The Mach 2 runs are on the left, the Mach 5 run are on the right.  Solid lines represent the bound cores; the dashed lines represent the entire data set of fluctuations.  The fit is the theoretical power spectrum of \protect\citet{padoan02}.\label{massdistribution}}
\end{center}
\end{figure*}

\begin{figure*}
\begin{center}
\includegraphics[width=84mm]{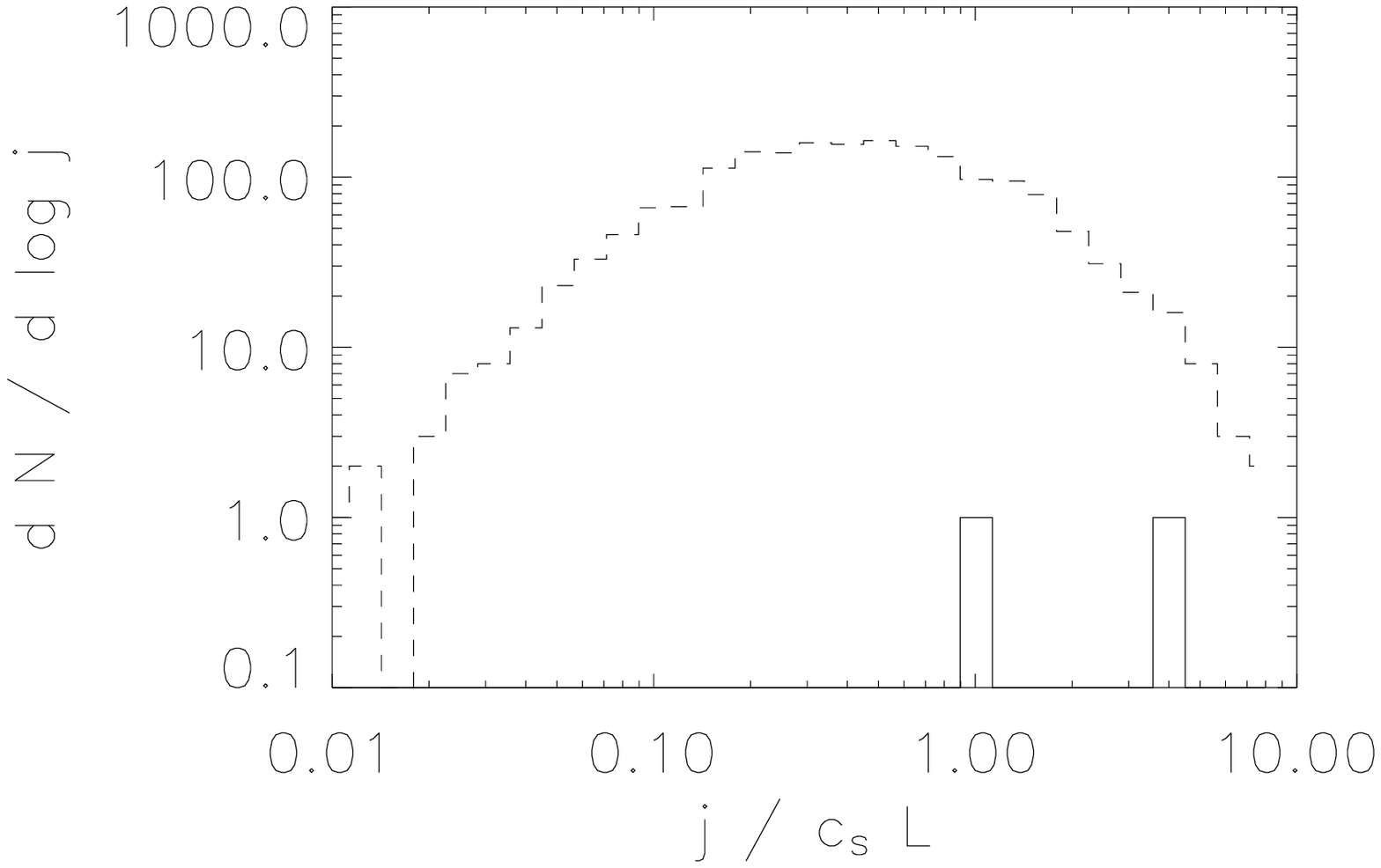}\hspace{5mm}\includegraphics[width=84mm]{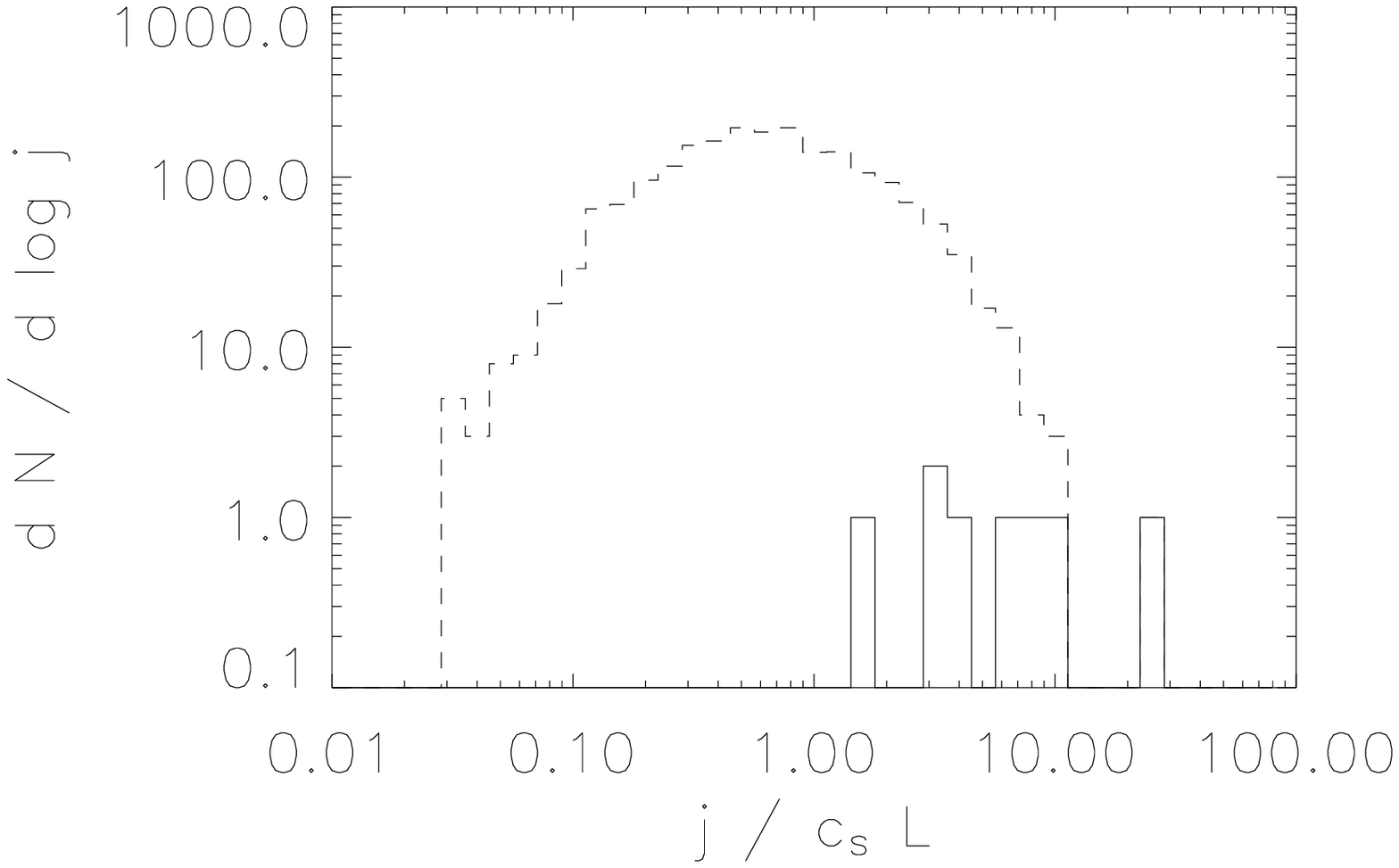}\\\vspace{10mm}
\includegraphics[width=84mm]{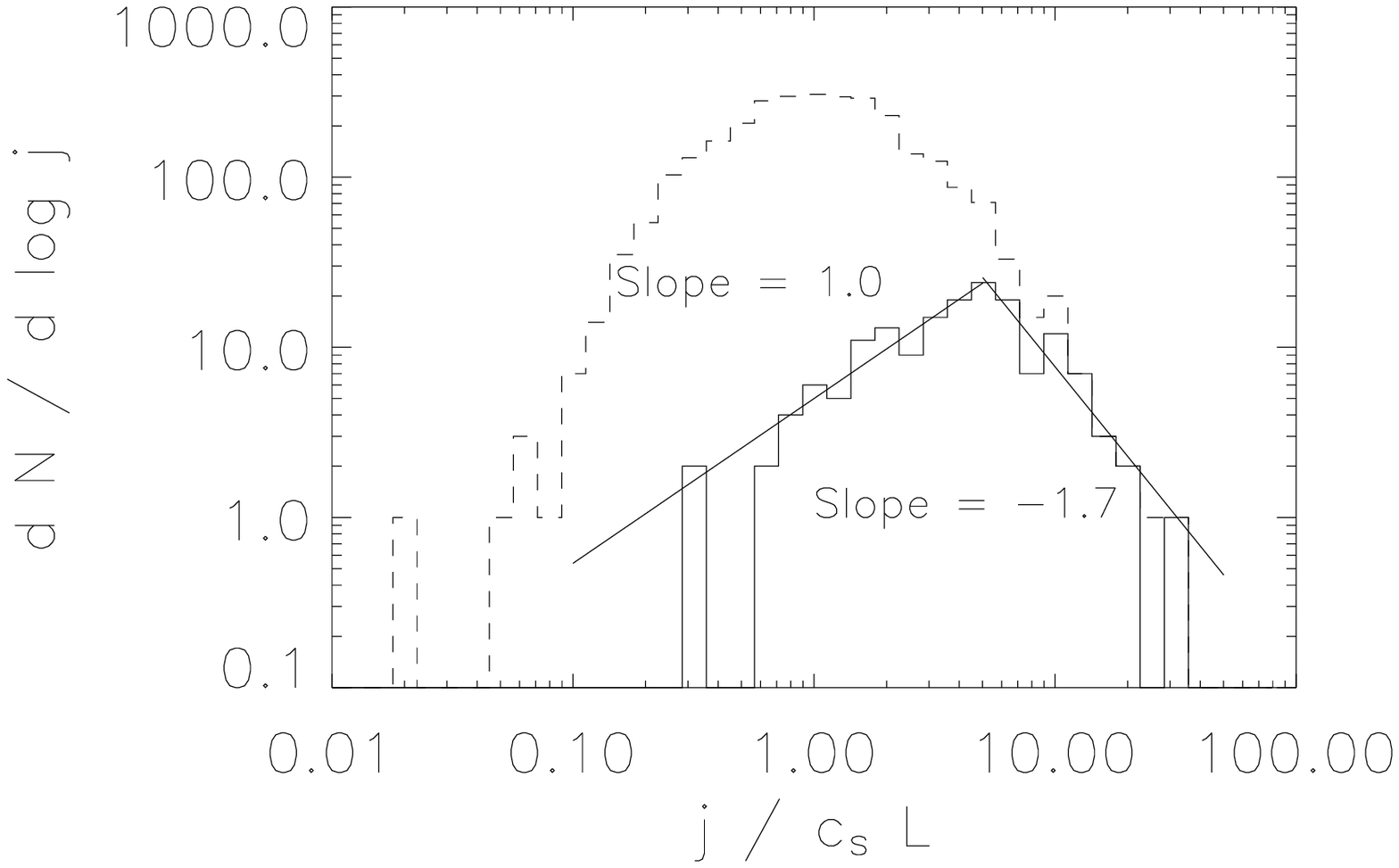}\hspace{5mm}\includegraphics[width=84mm]{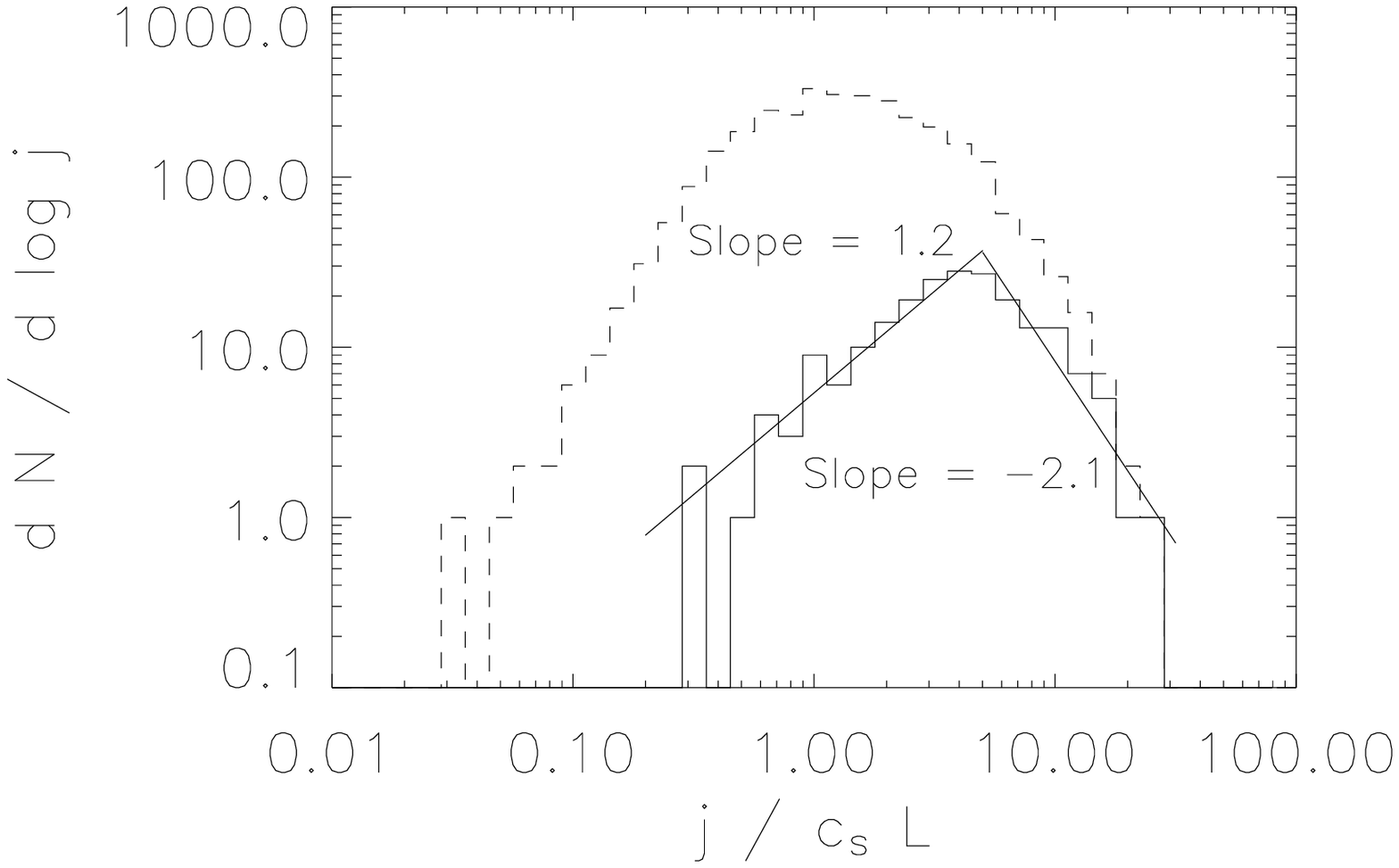}
\caption{Specific angular momentum distribution of fluctuations from the three-dimensional data sets.  The top row contains runs with 1.1 Jeans masses on the grid, the bottom row has 4.6 Jeans masses.  The left column starts with a Mach 2 RMS velocity, the right column with Mach 5.  Solid lines denote the specific angular momentum distribution of only the bound cores; dashed lines indicate the specific angular momentum distribution of all of the fluctuations.\label{jdistribution}}
\end{center}
\end{figure*}

A volume rendering image of a slab of our data cube from one of the runs we discuss (B5) is presented in Fig. \ref{volrend}.  This particular image occurs at 0.5 flow-crossing times, and is not centered on any particular feature as Figs. \ref{cdimages}---\ref{figslice} are.   As our fluctuation-finding algorithm associates any local density peak with an individual fluctuation, it is not surprising that we find many fluctuations in our simulations.  The fluctuations are typically arranged along the wispy filaments that are the result of interacting shocks.

	Observations of molecular cloud cores are unable to resolve the three-dimensional structures of the cores, as we are able to here.  Instead, they must resort to spectral line information, inferring the structure from the shape of the spectral lines.  We provide for comparison in Fig. \ref{cdimages} line maps of our simulations at their final time outputs, overlayed on simulated observations of the column density.  There are significant differences between the evolution of the A runs with 1.1 initial Jeans masses on the grid and the B runs with 4.6 initial Jeans masses.  There does not appear to be a significant difference between runs with different initial kinetic energies that have the same number of Jeans masses.

	The line maps are generated by treating the fluid as if it were in the optically thin limit.  For densities of $\sim 10^5 \;\mathrm{cm}^{-3}$ as found in our simulations, this might correspond to a tracer like CS.  We take lines-of-sight along one of the axes of the grid, and for each line-of-sight, we calculate the line profile via the equation (e.g. \citealt*{walker94})
\begin{eqnarray}
I(v_\mathrm{obs}) & \propto & \Sigma \rho e^{-(v_\mathrm{obs}-v_\mathrm{los})^2/c_s^2}\label{eq_line}
\end{eqnarray}
where $v_\mathrm{obs}$ is the velocity channel being observed, $v_\mathrm{los}$ is the true line-of-sight velocity in that cell, and the sum is over all of the pixels along the line-of-sight.  Equation (\ref{eq_line}) assumes that the line is optically thin, so that the details of line emission depend only on the amount of matter present to radiate.  The Gaussian term includes the effects of thermal broadening.  The resulting set of line profiles are then normalized with respect to the maximum line intensity in the map.  The lines-of-sight are then averaged into bins in x and y, and displayed over the column density map.  The area covered by the line profiles on the column density map indicates the size and location of each bin (generally 32x32 pixels).

The contour map in Fig. \ref{cdimages} is generated by convolving the integrated column density with a Gaussian filter, with a FWHM of 32 pixels.  The contour levels are linearly spaced in column density, with values of (10,20,30,40,50,60,70,80,90) per cent of the maximum density value.  The combination of the convolved column density map and line map simulates a typical observation of a molecular cloud clump, where the resolution of the data has been reduced by the resolution of the telescope, and the structure smeared out by the beam  (for comparison, the unconvolved column density map can be found in Fig. \ref{cdimages2}, with logarithmic contours).  To put these numbers on an observational basis, at a distance of 1000 pc, this ``beamwidth'' would correspond to an angular beamwidth of 2.6 arcsec for the A runs, and 8.3 arcsec for the B runs, if we use the size scaling found in Table \ref{table_sim_list} (for a cloud like Orion at $\sim 500\;\mathrm{pc}$, these numbers work out to 5.2 arcsec for the A runs, and 16.6 arcsec for the B runs).  The Very Large Array, for comparison, has a FWHM of 8.4 arcsec at 3.6 cm wavelengths in D-array, and 2.8 arcsec at 1.3 cm wavelengths.  Thus, the maps in Fig. \ref{cdimages} are analogous to the type of line map produced by these radio telescopes.

	In the lower-Jeans-mass runs, the initial shocks created by the turbulence can not accumulate enough mass for gravity to overcome thermal and turbulent pressure.  As a consequence, the shocks must dissipate before gravity becomes significant.  Gravitational collapse only occurs once the turbulent energy can decay sufficiently for the self-gravity to overcome the turbulent pressure.  Only a few core-like structures can be identified, as the fluid motions are dominated by the self-gravity of the main core, rather than being broken up by turbulent motions.  The line profiles in this map tend to be smooth and symmetric, although the line profile through the centre of the large core for both of the A runs is quite wide.

	In the higher-Jeans-mass simulations, as in the low-mass runs, the initial shocks are not strong enough to induce gravitational collapse.  However, it takes significantly less time for the turbulent energies to decay to the point where gravitational collapse can occur, because $t_\mathrm{ff}$ is shorter.  As a result, significant turbulent motions still remain, and the fluid is broken up into many fragments.  We can thus identify many potential cores in these simulations.  The line profiles are significantly more asymmetric for the B runs than the A runs.  This reflects the importance of turbulent motions on the line profiles.

	Due to the stochastic nature of the turbulence, some dense core will collapse before any other fluctuation can evolve to a significant degree.  As our criterion for stopping the simulations is based on a density threshold, the simulation will end before many regions of the simulation can collapse.  This is not a problem in runs A, as the turbulent effects have damped out and a single core dominates the simulation.  Stochastic effects are a concern in runs B, as turbulent fluctuations may cause additional cores to undergo runaway growth in the same manner as the first significant core that forms.  This could be a potential bias when interpreting statistical information about the cores, such as the mass distributions.

\subsection{Internal structure of cores}\label{internalstructure}

The fluctuations and cores in our simulations can have complicated structures.  We show slices through one plane of the largest core in each of the runs in Fig. \ref{figslice}, that show both the density structure and velocity field in that slice.  The A runs exhibit the fairly symmetric structure of the largest core that dominates the image.  The B runs, however, show an intricate density and velocity field as the fluid is channeled into the larger cores primarily along filamentary structures.

In the absence of turbulence, a pressure-bounded isothermal sphere that is gravitationally unstable will collapse.  The details are developed analytically by \citet{larson69} and \citet{penston69}, and have a self-similar form.  The Larson/Penston solution consists of a flat inner region, transitioning to a $r^{-2}$ profile on the outside.  The collapse proceeds asymptotically towards a density singularity (in  a real system, collapse would be halted by opacity and thermonuclear fusion as a protostar).  The infall velocity reaches a maximum of $3.3$ times the sound speed just outside of the flat central density region.  Numerical calculations by \citet{foster93} confirm this behaviour.

  The presence of turbulent motions complicates this analysis.  A phenomenological attempt to explain turbulent linewidths is a logatropic equation of state, $P/P_0 = 1 + A\ln \rho/\rho_0$, which has the effect of providing relatively more pressure support in low density regions than in the high-density regions, in comparison to an isothermal equation of state  ($A\sim 0.2$ is an adjustable parameter).  \citet{mclaughlin97} performed the analogous calculation to the Larson-Penston self-similar solution, using this logatropic equation of state rather than the isothermal equation of state.  As with the Larson-Penston solution, the logatropic collapse consists of a flat inner part, but the outer parts follow a $r^{-1}$ profile.  Three-dimensional collapse calculations of a logatropic fluid reproduce these results \citep*{reid02}.

  We plot the density and velocity profiles of the three most massive cores in each simulation in Figs. \ref{dvrprof_A2}-\ref{dvrprof_B5}.  These plots are generated by finding the average value of the density and radial velocity within each shell of radius $r$; the vertical bars indicate the total range of values that the density or radial velocity has at that particular radius.
The most massive cores in each run have a $\rho \propto r^{-2}$ profile everywhere (corresponding to the singularity formation of the Larson-Penston collapse).  The lower-mass cores can have considerably shallower density profiles that could indicate that the core is at an earlier stage in the Larson-Penston solution, when only the flat central part of the self-similar solution is contained within the truncation radius, or that the core is behaving as the logatropic McLaughlin-Pudritz solution.

The degree of asymmetry of each core at a given density level is reflected by the scatter at that density.  We can see that this scatter is typically a factor of $\sim 5$ in radius, with most of the points within $\sim 3$.  This implies that the cores in our simulation are elongated.  The largest cores in our B runs are less spherical then the most massive cores in the A runs -- as turbulence is significantly damped in the A runs at this time, this is a natural result of the balance of self-gravity and thermal pressure (as opposed to the B runs, which are still closely tied to the planar nature of the parent turbulent flow)

  Our results indicate that the more developed cores have structure that is dominated by gravitational effects -- they exhibit the $r^{-2}$ density profile of a collapsing isothermal sphere.  The less-developed cores and unbounded fluctuations have significant non-thermal support (remember, we are forced to stop when the most advanced core violates the \citet{truelove97} criterion), and so exhibit density profiles that can be much shallower, resembling that of a logatropic sphere or shallower.  This reinforces the idea that these fluctuations will collapse only when their internal turbulent motions are sufficiently damped in order that gravity can dominate.

  The behaviour of the infall velocity profiles indicate that, whatever their shape, the cores are not collapsing spherically -- in some directions, they are not collapsing at all.  This results in the large spread of $v_r$ at any given radius.    We will also note that a few of the lower-mass cores in each of our runs appear to have no infall at all in their centres.  This suggests that their interiors are in hydrostatic balance, at least temporarily.

	We examine the rotational behaviour of fluctuations and cores by plotting the magnitude of the specific angular momentum $j=J/M$ (where J is the total angular momentum of a parcel of fluid and M is that parcel's mass) as a function of radius in Figs. \ref{jprofiles1}-\ref{jprofiles2}.  As with the radial velocities, the specific angular momentum profile consists of an envelope, with a range of values at smaller $\mathit{\bf j}$.  
If the core were to undergo bulk rotation, a similar distribution of $|\mathit{\bf j}|$ would be expected, as the fluid elements in the plane of rotation will have maximal values of $|\mathit{\bf j}|$, while the elements along the axis of rotation will have $j \sim 0$.  The envelopes typically follow a $j \propto r^\eta$ profile, with $\eta$ between 1 and 2.   The most massive cores tend to have a central region of $\sim 10$ pixels with $j \propto r^{1.5}$, with a slightly flatter profile in the outer regions of $r^{1}$.  This $r^{1}$ profile is consistent with fluid elements in orbit around a $\rho \propto r^{-2}$ density profile (since $v_\phi \sim \mathrm{constant}$), as we see in the more massive cores.  As the smaller cores tend to have more turbulent support in their envelopes, the density profile is shallower, as shown previously.  As $j\sim v_\mathrm{rot} r\sim r^2\rho^{1/2}$, this allows the fluid to support a specific angular momentum profile that can be as steep as $r^{1.5}$ to $r^2$, as can be seen in Fig. \ref{jprofiles1}.  

	The maximum magnitude of $j$ is $\sim c_s L/4$, where $c_s$ is the sound speed and $L$ is the size of our simulation box ($L/4$ is the maximum scale of the turbulence, which by virtue of our turbulent spectrum, is also the scale that contains the most turbulent kinetic energy).  This is roughly the order-of-magnitude one would expect for cores collapsing out of oblique shocks in a turbulent fluid (the thickness of the shock $r\sim L/4M$, where M is the RMS Mach number of the fluid, is roughly the same size as the core that forms out of the shock; the velocity transverse to the shock is going to be on the order of $v \sim c_s M$, resulting in $|\mathit{\bf{j}}| \sim v r \sim c_s L/4$.  Note that this is independent of M).  

\section{STATISTICAL PROPERTIES OF CORES} \label{statistics}

\subsection{Mass distribution}

	The mass distribution for the fluctuations in our runs are plotted in Fig. \ref{massdistribution}.  We have plotted the entire data set of fluctuations for each of our run with a dotted line, and the subset of bound cores (see Section \ref{subsection_virial_stability}) with a solid line.  We are using the differential mass distribution, such that $dN(m)/d\log(m)$ counts the number of fluctuations with mass between $\log(m)$ and $\log(m)+\dif\log(m)$.

  The entire fluctuation set results in a distribution that is similar to a log-normal form at the peak mass and lower, but with a Salpeter power-law form at higher masses.  The bound core subset consists of most of the massive cores, with a peak at $\sim 2\times 10^{-3}m_J$. As an example, for a clump with a size $L= 0.32\;\mathrm{pc}$, $T=20\;\mathrm{K}$, $n_\mathrm{J} = 4.6$ (from Table \ref{table_sim_list}), Equation (\ref{massScaling}) states that $m_\mathrm{TOT}= 105.1 \;m_\odot$, so that the peak in the IMF is at $\sim 0.05\; m_\odot$.  Above this peak, the distribution is a power law with index $\sim -1.3$ for the B runs (the Salpeter value for the IMF of stars is $-1.35$); the A runs only have a few cores that are bound.  While there are not sufficient bound cores in the A runs to perform a meaningful statistical analysis, we note that observations of isolated star-forming regions such as Taurus can have IMFs that are approximately flat \citep{luhman03}.

  The fits to the mass spectrum in Fig. \ref{massdistribution} are calculated using the analysis of \citet{padoan02}.  Their calculation posits that the cores out of which stars form are created as the result of the fragmentation of shocks.  In this scenario, the high-mass slope $\alpha = -3/(4-\beta)$ is determined by turbulent power spectrum $\beta$ of the fluid.  Here, $\beta$ is index of the one-dimensional power spectrum, calculated by integrating the three-dimensional power spectrum over a sphere in $k$-space.  The three-dimensional power spectra of our simulations are given by $(\delta \mathbf{v}_k)^2$, where $\delta \mathbf{v}_k$ is given by Equation (\ref{pwrspectrumeq}).  Thus, $\beta = n - 2$, where $n$ is our three-dimensional power-law index from Equation (\ref{pwrspectrumeq}).  For $n=11/3$, $\beta=5/3$ and $\alpha = -9/7 = -1.29$.  In the Padoan \& Nordlund formalism, the width of the mass spectrum is determined primarily by the turbulent Mach number; the theoretical curves we present in Fig. \ref{massdistribution} are calculated using the initial RMS Mach number of each simulation.  We scale this theoretical mass distribution such that the total number of cores are the same as the measured distribution for the bound cores from the simulation, and that the total mass contained within cores with mass less than $0.1 m_J$ are equal.  The width of the theoretical distribution and the high-mass slope match the simulation mass distributions quite well.

\subsection{Star formation efficiency}

  The fraction of the total mass contained within the bound subset of cores is quite large -- $75$ per cent for run A2, $67$ per cent for run A5, $49$ per cent for run B2, and $42$ per cent for run B5.  In order to turn this into a star formation efficiency, we also need to consider that not all of the mass within these cores will necessarily end up within the star -- it is not clear what fraction of the core mass will actually be accreted.  This is something that our isothermal equation of state fails to take into account, as the gas in our simulations can cool quicker than an ideal gas can.  If most of the mass in our bound cores goes into stars then we have found an upper limit on the SFE of a clump.

  A lower limit for the efficiency of turning gas into stars in molecular cloud clumps (the star formation efficiency, or SFE) is estimated to be $10-30$ per cent using infrared observations \citep{lada03}.  The fraction of gas within the bound cores of our B simulations is not significantly greater than this.  As outflows and stellar winds remove some of the mass of these cores as a star forms within, the resulting SFE of our B simulations will agree with the observed values for clustered star-forming regions.  The A runs would have a SFE significantly higher than this, although they might be appropriate for isolated star-forming regions.

\subsection{Specific angular momentum distribution}

  In Fig. \ref{jdistribution} we plot the distribution of fluctuation mean specific angular momenta in the same manner as we did for the mass distribution.  $dN(j)/d\log(j)$ is the number of fluctuations with specific angular momentum between $\log(j)$ and $\log(j)+\dif\log(j)$.  Here, the specific angular momentum is defined as $j=J/m$, where $J$ is the total angular momentum of the fluctuation and $m$ is the mass of the fluctuation.

The peak of this distribution for the entire data set is at $\sim c_s L/4$ (For sound speeds of $0.4 \;\mathrm{km s}^{-1}$ and clump length scales of $0.1 \;\mathrm{pc}$, this works out to $3.1\times 10^{21}\;\mathrm{cm}^2\mathrm{s}^{-1}$).  As we have shown in Section \ref{internalstructure}, this is the order-of-magnitude that one would expect from a network of oblique shocks.  For the bound cores, the peak occurs at $\sim 5 c_s L$.  The overall distribution appears to be symmetric, with a width of approximately one order-of-magnitude.  The bound core distribution is more asymmetric, with slightly more cores at lower values of $j$.  We can fit the bound core distributions with a broken power law for the B runs, which qualitatively gives a good fit at both low-$j$ (with a slope of $\sim 1$) and high-$j$ (with a slope of $\sim -2$) values.


\section{DISCUSSION AND CONCLUSIONS}

        Our set of simulations support the idea that molecular cloud cores, in which stars can form, arise from the turbulent fragmentation of the parent cloud.  We posit that the cores are forming out of oblique shocks that impart angular momentum, and thus a net rotation, to these cores.  We have demonstrated that the evolution of the cores that form from molecular cloud turbulence is driven by the degree of turbulence within the cores.  The mass distribution of cores that we find appears to be well-fit by the turbulent mass distribution of \citet{padoan02}, suggesting that these cores were initially formed from shocks.  Gravitational collapse can only occur when the turbulent energy has damped sufficiently.  There are also many cores that are bound through turbulent surface pressure, although these are not in a state of dynamical collapse.  The cores in general are not in virial equilibrium; indeed, many are quite far from virial equilibrium.  The most massive cores are supercritical in the sense of Bonnor and Ebert.

	Our simulations show that structure formation in a turbulent, self-gravitating clump depends upon the value of the ratio of the turbulent damping time to the free-fall time $t_\mathrm{damp}/t_\mathrm{ff}$.  If $t_\mathrm{damp}/t_\mathrm{ff} > 1$, the turbulence fragments the gas into self-gravitating cores with an IMF-like mass spectrum.  These simulations have a star formation efficiency of $\approx 45$ per cent, that is likely an upper limit to the true star formation efficiency of the gas.  These simulations represent a clustered mode of star formation.  Conversely, if $t_\mathrm{damp}/t_\mathrm{ff} \le 1$, a few large self-gravitating objects arise.  The star formation efficiency is quite high, $\approx 70$ per cent.  While feedback from star formation could possibly reduce this efficiency, it is difficult to see how outflows could unbind dense critical cores.  These simulations are thus more likely to represent the conditions of isolated star formation.

	The oblique shocks out of which the cores form impart angular momentum to the nascent cores.  The magnitude of this specific angular momentum is $\sim c_s L/4$, the product of the sound speed and the wavelength of the most energetic turbulent mode.  For our simulations, this works out to $\sim 10^{21} \mathrm{cm}^2 \mathrm{s}^{-1}$ -- the same order of magnitude as seen in observations of cloud cores \citep{goodman93}, but $\sim 3$ orders of magnitude above the values for the specific angular momenta of stars \citep{stassun99}.  The specific angular momentum distributions that we measure for our simulated cores peaks roughly at this value of $c_s L/4$, and spans an order of magnitude in specific angular momentum above and below this value.

	The mass distribution of bound cores is well-matched by the \citet{padoan02} mass spectrum for cores forming out of shocks, in terms of both the width of the distribution (related to the current root-mean-square Mach number of the fluid motions) and high-mass slope $\alpha = -1.29$ (related only to the power spectrum of the turbulence).  The latter in particular also agrees well with the observed Salpeter mass distribution $\alpha = -1.35$ for molecular cloud cores.  When scaled to appropriate values, the peak is located at $\sim 0.1 m_\odot$, approximately where observations suggest a peak in the mass spectrum occurs.

  We obtain density profiles for the most massive and dynamically evolved cores in our simulations of $\rho \propto r^{-2}$.  The less-developed cores have considerably shallower profiles, as significant turbulent motions remain that allow the core to support more mass than isothermal pressure along can provide.  We observe that the collapse proceeds in an aspherical manner, preferentially occurring in one direction.  This is to be expected for a core that is rotating, as rotational support will inhibit collapse motions.  In a real system, rotational support can be bypassed through viscous and magnetic forces, but we do not include these effects in our simulations.  The specific angular momenta of these cores has a similar directional departure from spherical symmetry.

  The success of purely hydrodynamic models in determining many of the key features of star-forming regions brings into question the role of magnetic fields.  As magnetic energies tend to be of the same order of magnitude as turbulent energies in molecular clouds \citep{myers88,bertoldi92}, magnetic pressure support will likely reduce the SFE of these clumps by making cores sub-critical that would otherwise be slightly bound.  Magnetic fields will also allow the cores to transport angular momentum more efficiently, reducing the total specific angular momentum of the protostar.  Related to this is the fact that magnetic fields, coupled with the rotation of the protoplanetary disk, will drive jets that can inject energy back into the cloud.  Finally, magnetic fields may serve to increase the overall lifetime of the molecular cloud.

  The authors would like to acknowledge the useful discussions we had with Bruce Elmegreen, Richard Klein, Phil Myers, James Wadsley and Christine Wilson.  We would like to thank an anonymous referee for useful comments on our manuscript.  We are also grateful for the skilled visualisation programming carried out by Weiguang Guan of the McMaster University RHPCS group, in providing Fig. \ref{volrend}.  Our computations were carried out on a 128 CPU AlphaServer SC, which is the McMaster University node of the SHARCNET HPC Consortium.  D.A.T. is supported by an Ontario Graduate Scholarship.  R.E.P. is supported by the Natural Sciences and Engineering Research Council of Canada.

\appendix
\section{Comparison of Watershed Fluctuation-Finding Algorithm with \textsc{clumpfind}}\label{appendixa}

\begin{figure*}
\begin{center}
\includegraphics[width=84mm]{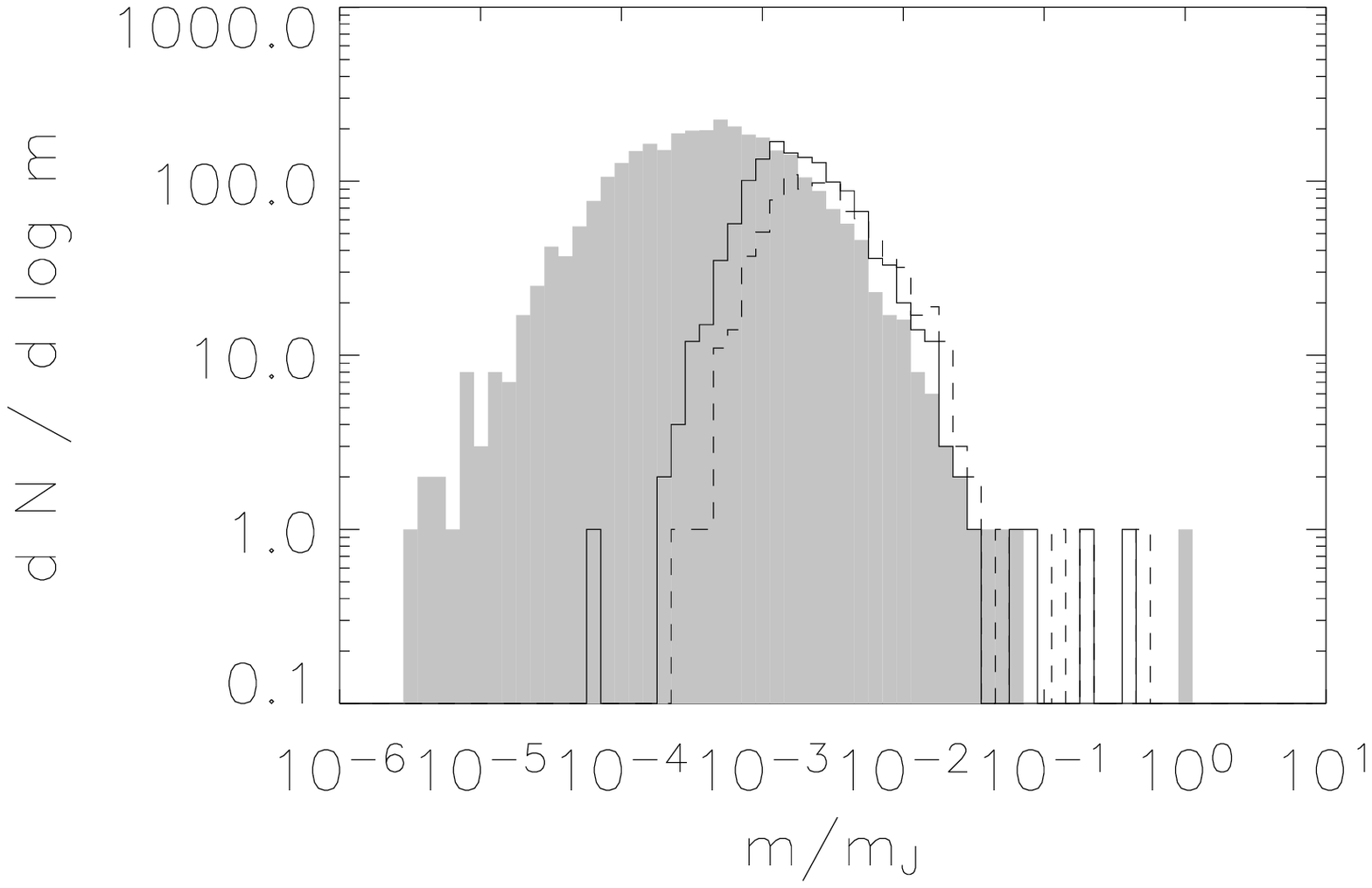}\hspace{5mm}\includegraphics[width=84mm]{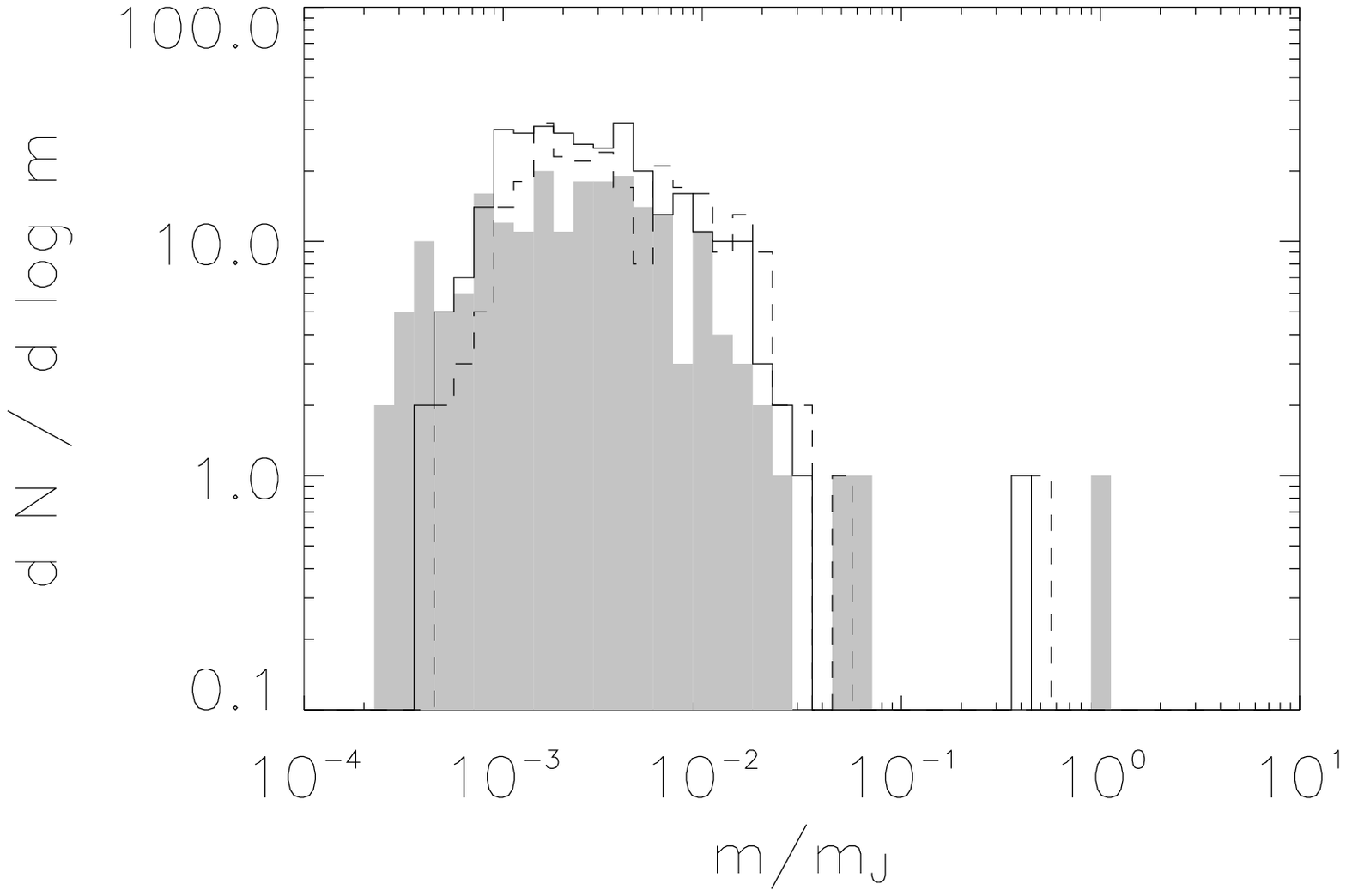}
\caption{Comparison of the results of the fluctuations found using the watershed algorithm described in Section \ref{watershedalgorithm} (filled region) and the \textsc{clumpfind} algorithm with a resolution of 0.05 dex (solid lines) and 0.1 dex (dashed lines).  The plot on the left compares the mass distributions for all fluctuations found by the algorithms; the plot on the right compares the mass distributions for only the bound cores.\label{clfindcomparison}}
\end{center}
\end{figure*}

The algorithm we use to find fluctuations, described in Section \ref{watershedalgorithm}, finds all of the local maxima in the density and all of the grid cells whose local gradient vector points towards that maxima.  Thus, it identifies both gravitationally bound cores as well as unbound density fluctuations.

The \textsc{clumpfind} algorithm \citep*{williams94} is commonly used to find clumps or cores in observational maps of star formation regions.  This algorithm works by finding all of the grid cells with intensity values between two contour levels.  The algorithm then determines if these regions are isolated from any other region or already-found core; if the region is isolated, it is labelled a new core.  If not, the algorithm determines which of the existing cores each cell is assigned to.

We compare the results of these two core-finding codes in order to provide a reference point for our watershed algorithm, to highlight its strengths and weaknesses.  We modified the \textsc{clumpfind} algorithm to account for a periodic grid, by allowing clumps to wrap around the boundaries.  We ran the \textsc{clumpfind} algorithm on Run B5 (binned to a $128^3$ grid) with a resolution of 0.05 and 0.1 dex in density (so that each contour level is $10^{0.05} = 1.12$ or $10^{0.1} = 1.26$ times the previous level).  For the 0.1 dex resolution search, we find a total of 888 fluctuations, of which 259 are bound.  These bound cores contain a total of 45 per cent of the mass.  For the 0.05 dex resolution search, we find a total of 1319 fluctuations, of which 317 are bound (containing 41 per cent of the total mass).  For comparison, our watershed algorithm found 2886 fluctuations, of which 207 where bound; 42 per cent of the mass is contained in bound cores.  The watershed algorithm finds significantly more objects that are not destined to collapse, as it is far more sensitive to small density fluctuations (although only 513 objects are found that have density contrasts between central density and mean surface density that is less than 1.25, and only 63 have density contrasts less than 1.12; none of these are bound.

The mass distributions for the two \textsc{clumpfind} searches are compared with the results of the watershed algorithm in Fig. \ref{clfindcomparison}.  For both \textsc{clumpfind} resolutions, we find good agreement with the watershed results for $m/m_j > 0.002$.  We also find excellent agreement between the distributions for the bound cores, which in turn agree very well with the Padoan-Nordlund mass distribution.  However, the watershed algorithm finds significantly more unbound fluctuations.  As our goals are to compare the nature of the bound cores with the unbound fluctuations, we favour the watershed algorithm.

We have also found that our watershed algorithm executes significantly faster than \textsc{clumpfind}.  The \textsc{clumpfind} algorithm at 0.1 dex took a comparable amount of time to execute as the original simulations did using \textsc{zeusmp}; at 0.05 dex it took nearly 5 times as long to execute as the \textsc{zeusmp} code.  The watershed algorithm, for comparison, only took about 0.1 times as long.  Coupled with the fact that we prefer the watershed algorithm's definition of a fluctuation, we feel that the watershed algorithm is the superior method for finding fluctuations and cores in our turbulent simulations.  However, \textsc{clumpfind} is probably better suited for finding cores in observations, as it can more naturally account for noise, simply by specifying an appropriate contour level.

\bibliographystyle{mn2e}
\bibliography{ref}

\begin{thebibliography}{}

\bibitem[\protect\citeauthoryear{{Bertoldi} \& {McKee}}{{Bertoldi} \&
  {McKee}}{1992}]{bertoldi92}
{Bertoldi} F.,  {McKee} C.~F.,  1992, ApJ, 395, 140

\bibitem[\protect\citeauthoryear{{Burkert} \& {Bodenheimer}}{{Burkert} \&
  {Bodenheimer}}{2000}]{burkert00}
{Burkert} A.,  {Bodenheimer} P.,  2000, ApJ, 543, 822

\bibitem[\protect\citeauthoryear{{Christensson}, {Hindmarsh} \&
  {Brandenburg}}{{Christensson} et~al.}{2001}]{christensson01}
{Christensson} M.,  {Hindmarsh} M.,    {Brandenburg} A.,  2001, Phys. Rev. E,
  64, 56405

\bibitem[\protect\citeauthoryear{{Ebert}}{{Ebert}}{1955}]{ebert55}
{Ebert} R.,  1955, Zeitschrift Astrophysics, 37, 217+

\bibitem[\protect\citeauthoryear{{Foster} \& {Chevalier}}{{Foster} \&
  {Chevalier}}{1993}]{foster93}
{Foster} P.~N.,  {Chevalier} R.~A.,  1993, ApJ, 416, 303+

\bibitem[\protect\citeauthoryear{{Frigo} \& {Johnson}}{{Frigo} \&
  {Johnson}}{1998}]{frigo98}
{Frigo} M.,  {Johnson} S.~G.,  1998, in IEEE International Conference on
  Acoustics, Speech and Signal Processing, vol. 3 {FFTW: An Adaptive Software
  Architecture for the FFT}.
pp 1381--1384

\bibitem[\protect\citeauthoryear{{Goodman}, {Benson}, {Fuller} \&
  {Myers}}{{Goodman} et~al.}{1993}]{goodman93}
{Goodman} A.~A.,  {Benson} P.~J.,  {Fuller} G.~A.,    {Myers} P.~C.,  1993,
  ApJ, 406, 528

\bibitem[\protect\citeauthoryear{{Jijina}, {Myers} \& {Adams}}{{Jijina}
  et~al.}{1999}]{jijina99}
{Jijina} J.,  {Myers} P.~C.,    {Adams} F.~C.,  1999, ApJS, 125, 161

\bibitem[\protect\citeauthoryear{{Johnstone}, {Fich}, {Mitchell} \&
  {Moriarty-Schieven}}{{Johnstone} et~al.}{2001}]{johnstone01}
{Johnstone} D.,  {Fich} M.,  {Mitchell} G.~F.,    {Moriarty-Schieven} G.,
  2001, ApJ, 559, 307

\bibitem[\protect\citeauthoryear{{Johnstone}, {Wilson}, {Moriarty-Schieven},
  {Joncas}, {Smith}, {Gregersen} \& {Fich}}{{Johnstone}
  et~al.}{2000}]{johnstone00}
{Johnstone} D.,  {Wilson} C.~D.,  {Moriarty-Schieven} G.,  {Joncas} G.,
  {Smith} G.,  {Gregersen} E.,    {Fich} M.,  2000, ApJ, 545, 327

\bibitem[\protect\citeauthoryear{{Klessen} \& {Burkert}}{{Klessen} \&
  {Burkert}}{2000}]{klessen00a}
{Klessen} R.~S.,  {Burkert} A.,  2000, ApJS, 128, 287

\bibitem[\protect\citeauthoryear{{Klessen} \& {Burkert}}{{Klessen} \&
  {Burkert}}{2001}]{klessen01a}
{Klessen} R.~S.,  {Burkert} A.,  2001, ApJ, 549, 386

\bibitem[\protect\citeauthoryear{{Kroupa}}{{Kroupa}}{2002}]{kroupa02}
{Kroupa} P.,  2002, Science, 295, 82

\bibitem[\protect\citeauthoryear{{Lada} \& {Lada}}{{Lada} \&
  {Lada}}{2003}]{lada03}
{Lada} C.~J.,  {Lada} E.~A.,  2003, ARA\&A, 41, 57

\bibitem[\protect\citeauthoryear{{Lada}, {Bally} \& {Stark}}{{Lada}
  et~al.}{1991}]{lada91}
{Lada} E.~A.,  {Bally} J.,    {Stark} A.~A.,  1991, ApJ, 368, 432

\bibitem[\protect\citeauthoryear{{Larson}}{{Larson}}{1969}]{larson69}
{Larson} R.~B.,  1969, MNRAS, 145, 271+

\bibitem[\protect\citeauthoryear{{Luhman}, {Brice{\~ n}o}, {Stauffer},
  {Hartmann}, {Barrado y Navascu{\' e}s} \& {Caldwell}}{{Luhman}
  et~al.}{2003}]{luhman03}
{Luhman} K.~L.,  {Brice{\~ n}o} C.,  {Stauffer} J.~R.,  {Hartmann} L.,
  {Barrado y Navascu{\' e}s} D.,    {Caldwell} N.,  2003, ApJ, 590, 348

\bibitem[\protect\citeauthoryear{{Mac Low}, {Klessen}, {Burkert} \&
  {Smith}}{{Mac Low} et~al.}{1998}]{maclow98}
{Mac Low} M.,  {Klessen} R.~S.,  {Burkert} A.,    {Smith} M.~D.,  1998,
  Physical Review Letters, 80, 2754

\bibitem[\protect\citeauthoryear{{Mangan} \& {Whitaker}}{{Mangan} \&
  {Whitaker}}{1999}]{mangan99}
{Mangan} A.~P.,  {Whitaker} R.~T.,  1999, IEEE Transactions on Visualization
  and Computer Graphics, 5, 308

\bibitem[\protect\citeauthoryear{{Matzner} \& {McKee}}{{Matzner} \&
  {McKee}}{1999}]{matzner99}
{Matzner} C.~D.,  {McKee} C.~F.,  1999, ApJL, 526, L109

\bibitem[\protect\citeauthoryear{{McKee} \& {Tan}}{{McKee} \&
  {Tan}}{2003}]{mckee03}
{McKee} C.~F.,  {Tan} J.~C.,  2003, ApJ, 585, 850

\bibitem[\protect\citeauthoryear{{McKee} \& {Zweibel}}{{McKee} \&
  {Zweibel}}{1992}]{mckee92}
{McKee} C.~F.,  {Zweibel} E.~G.,  1992, ApJ, 399, 551

\bibitem[\protect\citeauthoryear{{McLaughlin} \& {Pudritz}}{{McLaughlin} \&
  {Pudritz}}{1996}]{mclaughlin96}
{McLaughlin} D.~E.,  {Pudritz} R.~E.,  1996, ApJ, 469, 194

\bibitem[\protect\citeauthoryear{{McLaughlin} \& {Pudritz}}{{McLaughlin} \&
  {Pudritz}}{1997}]{mclaughlin97}
{McLaughlin} D.~E.,  {Pudritz} R.~E.,  1997, ApJ, 476, 750

\bibitem[\protect\citeauthoryear{{Motte}, {Andre} \& {Neri}}{{Motte}
  et~al.}{1998}]{motte98}
{Motte} F.,  {Andre} P.,    {Neri} R.,  1998, A\&A, 336, 150

\bibitem[\protect\citeauthoryear{{Myers}, {Evans} \& {Ohashi}}{{Myers}
  et~al.}{2000}]{myers00}
{Myers} P.~C.,  {Evans} N.~J.,    {Ohashi} N.,  2000, Protostars and Planets
  IV, pp 217--+

\bibitem[\protect\citeauthoryear{{Myers} \& {Goodman}}{{Myers} \&
  {Goodman}}{1988}]{myers88}
{Myers} P.~C.,  {Goodman} A.~A.,  1988, ApJL, 326, L27

\bibitem[\protect\citeauthoryear{{Ostriker}, {Gammie} \& {Stone}}{{Ostriker}
  et~al.}{1999}]{ostriker99}
{Ostriker} E.~C.,  {Gammie} C.~F.,    {Stone} J.~M.,  1999, ApJ, 513, 259

\bibitem[\protect\citeauthoryear{{Ostriker}, {Stone} \& {Gammie}}{{Ostriker}
  et~al.}{2001}]{ostriker01}
{Ostriker} E.~C.,  {Stone} J.~M.,    {Gammie} C.~F.,  2001, ApJ, 546, 980

\bibitem[\protect\citeauthoryear{{Padoan}, {Juvela}, {Goodman} \&
  {Nordlund}}{{Padoan} et~al.}{2001}]{padoan01}
{Padoan} P.,  {Juvela} M.,  {Goodman} A.~A.,    {Nordlund} {\AA}.,  2001, ApJ,
  553, 227

\bibitem[\protect\citeauthoryear{{Padoan} \& {Nordlund}}{{Padoan} \&
  {Nordlund}}{2002}]{padoan02}
{Padoan} P.,  {Nordlund} {\AA}.,  2002, ApJ, 576, 870

\bibitem[\protect\citeauthoryear{{Penston}}{{Penston}}{1969}]{penston69}
{Penston} M.~V.,  1969, MNRAS, 144, 425+

\bibitem[\protect\citeauthoryear{{Porter}, {Pouquet} \& {Woodward}}{{Porter}
  et~al.}{1994}]{porter94}
{Porter} D.~H.,  {Pouquet} A.,    {Woodward} P.~R.,  1994, Phys. Fluids, 6,
  2133

\bibitem[\protect\citeauthoryear{{Pudritz}}{{Pudritz}}{2002}]{pudritz02}
{Pudritz} R.~E.,  2002, Science, 295, 68

\bibitem[\protect\citeauthoryear{{Reid}, {Pudritz} \& {Wadsley}}{{Reid}
  et~al.}{2002}]{reid02}
{Reid} M.~A.,  {Pudritz} R.~E.,    {Wadsley} J.,  2002, ApJ, 570, 231

\bibitem[\protect\citeauthoryear{{Shu}, {Adams} \& {Lizano}}{{Shu}
  et~al.}{1987}]{shu87}
{Shu} F.~H.,  {Adams} F.~C.,    {Lizano} S.,  1987, ARA\&A, 25, 23

\bibitem[\protect\citeauthoryear{{Stassun}, {Mathieu}, {Mazeh} \&
  {Vrba}}{{Stassun} et~al.}{1999}]{stassun99}
{Stassun} K.~G.,  {Mathieu} R.~D.,  {Mazeh} T.,    {Vrba} F.~J.,  1999, AJ,
  117, 2941

\bibitem[\protect\citeauthoryear{{Stone} \& {Norman}}{{Stone} \&
  {Norman}}{1992}]{stone92a}
{Stone} J.~M.,  {Norman} M.~L.,  1992, ApJS, 80, 753

\bibitem[\protect\citeauthoryear{{Testi} \& {Sargent}}{{Testi} \&
  {Sargent}}{1998}]{testi98}
{Testi} L.,  {Sargent} A.~I.,  1998, ApJL, 508, L91

\bibitem[\protect\citeauthoryear{{Truelove}, {Klein}, {McKee}, {Holliman},
  {Howell} \& {Greenough}}{{Truelove} et~al.}{1997}]{truelove97}
{Truelove} J.~K.,  {Klein} R.~I.,  {McKee} C.~F.,  {Holliman} J.~H.,  {Howell}
  L.~H.,    {Greenough} J.~A.,  1997, ApJL, 489, L179+

\bibitem[\protect\citeauthoryear{{Truelove}, {Klein}, {McKee}, {Holliman},
  {Howell}, {Greenough} \& {Woods}}{{Truelove} et~al.}{1998}]{truelove98}
{Truelove} J.~K.,  {Klein} R.~I.,  {McKee} C.~F.,  {Holliman} J.~H.,  {Howell}
  L.~H.,  {Greenough} J.~A.,    {Woods} D.~T.,  1998, ApJ, 495, 821

\bibitem[\protect\citeauthoryear{{van der Tak}, {van Dishoeck}, {Evans} \&
  {Blake}}{{van der Tak} et~al.}{2000}]{vandertak00}
{van der Tak} F.~F.~S.,  {van Dishoeck} E.~F.,  {Evans} N.~J.,    {Blake}
  G.~A.,  2000, ApJ, 537, 283

\bibitem[\protect\citeauthoryear{{Vazquez-Semadeni}, {Ostriker}, {Passot},
  {Gammie} \& {Stone}}{{Vazquez-Semadeni} et~al.}{2000}]{vazquez00}
{Vazquez-Semadeni} E.,  {Ostriker} E.~C.,  {Passot} T.,  {Gammie} C.~F.,
  {Stone} J.~M.,  2000, Protostars and Planets IV, pp~3--+

\bibitem[\protect\citeauthoryear{{Vincent} \& {Soille}}{{Vincent} \&
  {Soille}}{1991}]{vincent91}
{Vincent} L.,  {Soille} P.,  1991, IEEE Transactions on Pattern Analysis and
  Machine Intelligence, 13, 583

\bibitem[\protect\citeauthoryear{{Walker}, {Narayanan} \& {Boss}}{{Walker}
  et~al.}{1994}]{walker94}
{Walker} C.~K.,  {Narayanan} G.,    {Boss} A.~P.,  1994, ApJ, 431, 767

\bibitem[\protect\citeauthoryear{{Williams}, {de Geus} \& {Blitz}}{{Williams}
  et~al.}{1994}]{williams94}
{Williams} J.~P.,  {de Geus} E.~J.,    {Blitz} L.,  1994, ApJ, 428, 693

\end{thebibliography}

\end{document}